\documentclass[reprint,amsmath,amssymb,aps,superscriptaddress,prc,floatfix]{revtex4-1}
\usepackage{amsmath,amsfonts}
\usepackage{graphicx}
\usepackage{longtable}
\usepackage{bm}
\usepackage{hyperref}
\hypersetup{colorlinks=true,breaklinks,linkcolor=red,citecolor=blue}

\begin{document}
\title{Pairing in high-density neutron matter including short- and long-range correlations}

\author{D. Ding}
\affiliation{Department of Physics, Washington University, St. Louis, Missouri 63130, USA}

\author{A. Rios}
\affiliation{Department of Physics, Faculty of Engineering and Physical Sciences, University of Surrey, Guildford, Surrey GU2 7XH, United Kingdom}

\author{H. Dussan}
\affiliation{Department of Physics, Washington University, St. Louis, Missouri 63130, USA}

\author{W. H. Dickhoff}
\affiliation{Department of Physics, Washington University, St. Louis, Missouri 63130, USA}

\author{S. J. Witte}
\affiliation{Department of Physics, Washington University, St. Louis, Missouri 63130, USA}
\affiliation{Department of Physics and Astronomy, University of California, Los Angeles, CA 90095, USA}

\author{A. Carbone}
\affiliation{Institut f{\"u}r Kernphysik, Technische Universit{\"a}t Darmstadt, 64289 Darmstadt, Germany
and ExtreMe Matter Institute EMMI, GSI Helmholtzzentrum  f{\"u}r  Schwerionenforschung GmbH, 64291 Darmstadt, Germany}

\author{A. Polls}
\affiliation{Departament d'Estructura i Constituents de la Mat{\`e}ria and Institut de Ci{\`e}nces del Cosmos, Universitat de Barcelona, Avinguda Diagonal 647, E-8028 Barcelona, Spain}

\date{\today}

\begin{abstract}
Pairing gaps in neutron matter need to be computed in a wide range of densities to address open questions in neutron star phenomenology. Traditionally, the Bardeen-Cooper-Schrieffer approach has been used to compute gaps from bare nucleon-nucleon interactions. Here, we incorporate the influence of short- and long-range correlations in the pairing gaps. Short-range correlations are treated including the appropriate fragmentation of single-particle states, and substantially suppress the gaps. Long-range correlations dress the pairing interaction via density and spin modes, and provide a relatively small correction. We use different interactions, some with three-body forces, as a starting point to control for any systematic effects. Results are relevant for neutron-star cooling scenarios, in particular in view of the recent observational data on Cassiopeia A. 
\end{abstract}


\maketitle

\section{Introduction}

Superfluids play an important role in nuclear physics, from nuclear structure to neutron-star observations~\cite{Dean2003}. In the latter, superfluidity is a key ingredient in the description of several relevant astrophysical phenomena. In a sense, neutron stars are essential testing grounds for the pairing properties of dense systems. 
It is important that theoretical considerations on dense matter superfluids are guided by knowledge on another relevant testing ground of superfluidity: finite nuclei. Many-body theory can provide insight on  different aspects of singlet and triplet pairing in nuclear matter~\cite{Steiner2009,Leinson2010}. 

On the one hand, neutron stars cool predominantly by neutrino emission~\cite{Yakovlev2004}. This process depends sensitively on the neutrino weak rates in the dense nuclear medium~\cite{Reddy1997,Sedrakian2007}. The presence of superfluid pairs in the crust and the core suppresses some of these rates. A comparison with astrophysical observations, providing age and temperature estimates for several pulsars, can be used to test the validity of some assumptions on the core's pairing properties. Recently, observations of rapid cooling in Cassiopeia A have been interpreted as evidence of the onset of triplet-pairing-mediated cooling in the core~\cite{Page2011}. While there is some debate regarding these observations on the astrophysical community~\cite{Elshamouty2013,Ho2015}, it seems appropriate to review the status of many-body calculations of the pairing gaps in infinite neutron matter. 

On the other hand, the accepted model for glitching in pulsar requires the presence of superfluids in the crusts of neutron stars~\cite{Anderson1975,Pines1985,Haskell2015}. Neutron superfluid vortices are not free to move, but rather remain pinned through the interaction with lattice nuclei or defects. While the bulk of the star slows down by the emission of electromagnetic radiation, the superfluid component can only change its angular momentum via a substantial reconfiguration of the vortex network. This occurs in a single violent, short rearrangement event, which gives rise to the glitch. The strength and periodicity of these glitches provide constraints on the amount of the star's momentum of inertia which is stored in the superfluid~\cite{Link1992,Andersson2012,Chamel2013}. Traditional models consider the superfluid as pinned to the crust, but a recent analysis has suggested that glitch phenomena require superfluids in the outer core of the star, too~\cite{Ho2015b}. In this picture, the gap closure density of singlet superfluids becomes a sensitive probe of neutron star interior physics. While the effect of triplet pairing has not  been studied in this context, its strength and density dependence are likely to be relevant as well. 

In nuclei, a well-known experimental property of pairing is its isovector nature. Neutron-neutron and proton-proton pairs dominate to a large extent nuclear pairing~\cite{Afanasjev2015}. The possibility of $np$ pairing is less likely to occur in $N \ne Z$ nuclei, but even in $N \approx Z$ nuclei isoscalar $np$ pairing is elusive~\cite{Frauendorf2014}. In stark contrast, Bardeen--Cooper--Schrieffer (BCS) pairing calculations in sub-saturation infinite matter with bare nucleon-nucleon (NN) forces predict a dominant pairing gap, of the order of $10$ MeV, in the $np$ sector~ \cite{Baldo1992,Dean2003,Muther2005,Maurizio2014}. Mechanisms have been proposed to explain why spin triplet pairing is particularly suppressed in nuclei~\cite{Frauendorf2014,Gezerlis2011}. One needs to call upon many-body theory, particularly beyond BCS pairing, to reconcile empirical observations with theory. After all, the well-known BCS method works best for uncorrelated systems, and one can put into question its applicability in dense, correlated nuclear systems~\cite{Dickhoff08}. 

Other than dressing the in-medium interaction, correlations beyond the traditional mean-field and BCS frameworks have a large impact on the redistribution of single-particle (sp) strength~\cite{Dickhoff2004a}. Studies in electron knock-out experiments have unambiguously demonstrated that approximately $35 \, \%$ of the sp strength lies beyond the quasi-particle peak~\cite{Lapikas1993,Benhar2008,Arrington2011}. About one third of this sp strength depletion can be attributed to short-range correlations (SRC), that empty the nuclear Fermi sea and promote strength to the high-momentum region~\cite{Rohe2004,Rios2009,Rios2014}. SRC are very similar in different nuclei, which bodes well with the idea that SRC are caused to some extent by the local short-distance repulsion of the NN force ~\cite{Alvioli2013,Atti2015}. This idea has been exploited to study generic properties of short-range correlated pairs, including their spin and isospin content~\cite{Wiringa2014,Neff2015,Ryckebusch2015}. All theoretical approaches indicate that SRC are dominated by $np$ pairs in a region of the order of 50-100 MeV below the mean-field potential~\cite{Muther95}, in accordance to experimental findings~\cite{Sargsian2014,Hen2014}. 

Pairing is active in a momentum region close to the Fermi surface. The removal of sp strength in this momentum region due to SRC should impact the corresponding pairing gap~\cite{Bozek1999,Shen2003,Muther2005}. To go beyond BCS theory, the Gorkov-Green's functions diagrammatic approach can be used to include correlations systematically into the pairing properties~\cite{Soma2011}. Microscopic calculations taking into account SRC indicate a reduction of singlet pairing gaps in nuclear and neutron matter~\cite{Bozek2000,Muther2005}. A qualitatively similar answer is obtained when using a BCS-like approach that is quenched by $\mathcal{Z}$ factors~\cite{Shen2003,Dong2015}, although a realistic description requires a full account of the spectral function width~\cite{Bozek2003,Muther2005}. At finite temperature, SRC also correct the corresponding critical temperature of the pairing transition. We emphasize that a $\mathcal{T}$-matrix resummation that fully accounts for SRC is needed to describe pseudo-gap phenomena in the vicinity of the phase transition~\cite{Schnell1999}. Moreover, and according to the Thouless criterion, the onset of pairing is in one-to-one correspondence with the appearance of a pole in the $\mathcal{T}$-matrix~\cite{Thouless1960,Kadanoff1961,Alm1993}.

In finite nuclei, long range correlations (LRC) are responsible for about $20 \, \%$ of the sp fragmentation via the coupling of sp states to low-lying resonances and collective modes ~\cite{Dickhoff2004a}. For pairing properties, polarization effects should mostly renormalize the effective interaction~\cite{Wambach1993}, even at very low densities \cite{Schulze2001}. Density and spin collective motion is expected to dress the interaction of paired particles, which does not necessarily resemble the free space NN force~\cite{Shen2003}. While the properties of LRC can be different in finite and infinite nuclear matter, a screening mechanism that dresses the pairing interaction is expected to affect both singlet and triplet pairing in neutron-star matter~\cite{Shen2003,Shen2005,Cao2006,Dong2013}. In particular, corrections at the level of the effective pairing interaction occur at low energies and Fermi liquid theory (FLT) can be used to provide a phenomenological, but systematic, understanding of screening~\cite{Migdal1967,Wambach1993,Schwenk2003,Schwenk2004}. 

In the following, we use a theoretical approach that combines self-consistent Green's functions (SCGF) techniques and FLT to include consistently the effect of SRC and LRC in the singlet and triplet gaps of neutron matter. SRC are included by means of a well established finite-temperature ladder resummation scheme, which is also able to describe quantitatively the density and isospin dependence of high-momentum components and tensor-like correlations~\cite{Frick2003,Rios2009,Rios2014}. We extrapolate self-energies to zero temperature, and use them to generate spectral functions that provide a quantitative account of the removal of strength close to the Fermi surface~\cite{Muther2005}. Based on the Gorkov formalism, our approach provides a quantitative estimate of the effect of SRC on pairing gaps. The feedback from the superfluid phase into the normal Green's functions properties below the critical temperature is missing, but it is expected to be small. 

We extend the treatment of SRC presented in Ref.~\cite{Muther2005} in four directions for the astrophysically relevant case of neutron matter. First, we consider, in addition to the singlet $^1$S$_0$ case, the case of pairing in the coupled triplet wave, $^3$PF$_2$. The size and density regime in which this gap operates could have relevance for neutrino cooling~\cite{Page2011} and glitch phenomena in pulsars~\cite{Ho2015b}. Second, we have implemented a computational method to extrapolate self-energies, spectral functions and thermodynamical properties from finite-temperature calculations into zero temperature. 

Third, with the aim of quantifying any potential systematic uncertainties related to the underlying Hamiltonian, we work with three different NN interactions. We use two high-quality phase-shift equivalent potentials, the CDBonn~\cite{Machleidt1995} and Argonne v18 (Av18)~\cite{Wiringa1995} forces. In terms of SRC, the latter is traditionally considered to be harder than the former, in the sense that it induces larger high-momentum components in the many-body wave function~\cite{Rios2009,Rios2014}. Moreover, we also employ the next-to-next-to-next-to-leading-order (N3LO) Idaho potential, that has been derived in the context of chiral perturbation theory~\cite{Entem2003}. The cut-off associated to the chiral expansion is implemented in the form of a regulator function in relative momentum which sharply cuts the potential from $\Lambda=500$ MeV on. As a consequence, pairing calculations, which are directly sensitive to the relative momentum dependence of the matrix elements due to the BCS kinematics, become sensitive to the artificial regulator function for Fermi momenta above $k_F \approx 2.5$ fm$^{-1}$.  
We also present results which are computed by supplementing the Idaho NN potential with an N2LO three-neutron force (3NF), following an uncorrelated average over the third particle that is consistent with the use of the Green's function formalism \cite{Holt2010,Carbone2013a,Carbone2013b,Carbone2014}. 3NFs are included both at the level of the effective interaction and on the treatment of SRC, in this initial exploratory study. More sophisticated calculations including 3NFs are our priority for the near future.

Finally, we supplement the calculations of SRC in the gap equation with a physically relevant screening of the pairing matrix elements. To account for LRC in our infinite matter calculations, we screen the effective interaction using FLT as a starting point. Following Refs.~\cite{Shen2003,Cao2006,Dong2013}, we dress the interaction with successive particle-hole excitations. These are coupled to the paired nucleons by vertices that are self-consistently determined using the concept of the induced interaction~\cite{babu1973}. Collective modes on top of this are described using Fermi liquid theory, with Landau parameters obtained from Ref.~\cite{Cao2006}. As a general conclusion, we find that the effect of LRC on triplet pairing gaps is smaller than that of SRC. 

This paper is organized as follows. Section \ref{sec:method} provides an overview of the method, with specific subsections devoted to the discussion of our treatment of SRC, temperature extrapolations and LRC. Results are summarized in Sec.~\ref{sec:results}. Singlet pairing gaps are discussed in subsection~\ref{sec:results1S0} and triplet gaps, in subsection~\ref{sec:results3PF2}. A preliminary discussion on the effect of 3NF is provided in subsection~\ref{sec:results3NF}. We draw conclusions and provide an outlook of potential future work in Sec.~\ref{sec:conclusions}. The appendices provide a discussion of numerical aspects related to zero-temperature extrapolations and fits.

\section{Methods}
\label{sec:method}

\subsection{Short-range correlations}
\label{sec:SRC}

In the following, we describe a method to include SRC into the pairing properties of dense nuclear matter. More specifically, we look at the inclusion of fragmented sp states into the gap equation. Technically, our method is founded in the Gorkov-Green's functions theory for the description of condensed fermionic systems~\cite{Abrikosov1965,Dickhoff08,Soma2011}. A diagrammatic expansion in terms of self-consistent propagators exists in this case, and involves, in addition to the usual Green's functions, anomalous propagators. We do not provide details on the derivations of the equations here: they have been presented elsewhere in the nuclear physics literature~\cite{Bozek2000,Muther2005,Dickhoff08}. We also note that similar approaches exist in condensed matter, particularly in the context of the BCS-BEC crossover \cite{Haussmann2007}. In that field, our approach is reminiscent to the fluctuation-exchange (FLEX) scheme~\cite{Chen2005}. 

A specific formulation of Gorkov's theory allows for the resummation of correlations on the normal component of the propagator using the normal Dyson equation~\cite{Pines1962,Migdal1967,Muther2005,Dickhoff08}. This is particularly useful for strongly correlated nuclear systems, in which there is already a substantial fragmentation in the normal state sp propagator, $G^N$. Below the critical temperature, the Gorkov formalism couples the full superfluid sp propagator, $G$, to its normal component via an anomalous self-energy,
\begin{align}
G(k, \omega) &= G^N(k, \omega) - G^N(k, \omega) \Delta(k, \omega) G(k, \omega) \, .
\end{align}
In turn, the anomalous propagator, $F$, and self-energy (or superfluid gap), $\Delta$, are related to both the normal and the full propagators:
\begin{align}
F(k, \omega) &= G^N(-k, -\omega) G(k, \omega)  \Delta(k,\omega) \, .
\end{align}
In the lowest order diagrammatic approximation, $\Delta$ is an energy-independent quantity. We note, however, that this formulation goes beyond the BCS approach in that the fragmentation of states at the normal level is described in terms of a fully dressed sp normal propagator, $G^N$.

Working in a partial wave basis, and after a suitable angle-average procedure has been considered, the expression for the lowest-order dressed anomalous self-energy leads to the generalized gap equation,
\begin{align}
\Delta^{JST}_{L}(k) = - \sum_{L'} \int_0^\infty \frac{dk' \, k'^2}{\pi} \frac{ \langle k | \mathcal{V}^{JST}_{LL'} | k' \rangle }{\xi(k')} \Delta^{JST}_{L'}(k') \, .
\label{eq:gap}
\end{align}
$\Delta^{JST}_L(k)$ is the pairing gap for a given partial wave, $L$, in the channel of total angular momentum $J$, pair spin $S$ and pair isospin $T$. Pairs are in a BCS-like state of opposite sp momenta, $k_1=-k_2$. The effective pairing interaction, $ \mathcal{V}_{LL'}$, is therefore a function of relative momentum, $k=k_1$, and we work at zero pair centre-of-mass momentum. In BCS theory, $\mathcal{V}_{LL'}$ would simply be a bare NN interaction. However, the effect of the medium is important for the pairing interaction, even at very low densities~\cite{Abrikosov1965,Schulze2001,Dong2013}. We will therefore introduce a description of the screening of $\mathcal{V}_{LL'}$ with polarization effects in Sec.~\ref{sec:LRC}. In our nomenclature, the polarization effects in the effective interaction are equivalent to LRC effects. 

Our main emphasis is on quantifying the effect of correlations in pairing properties, and particularly in finding behaviours that are generic to all nuclear forces. We will therefore focus on calculations involving NN interactions with different short-range and tensor structures. The results that contain 3NF have been obtained with a suitable non-correlated average over the third particle. At the level of the effective two-body interaction, this involves an integration over the third particle that includes the full anti-symmetrization of the three-body matrix element \cite{Holt2010,Carbone2013a}. At the self-energy level, there is another, differently weighted, one-body contribution of 3NFs to the Hartree-Fock term. 

In addition to the effective interaction, the kernel of the gap equation is determined by an energy denominator, $\xi(k)$. In the Gorkov approach, the denominator is the double energy convolution~\cite{Muther2005}:
\begin{align}
\frac{1}{2 \xi(k)} = \int \frac{d \omega}{2\pi} \frac{d \omega'}{2\pi} &A^N (k,\omega) A (k,\omega') \times \nonumber \\
& \frac{1-f(\omega)-f(\omega')}{\omega+\omega'}  \, .
\label{eq:xik}
\end{align}
The temperature, $T$, is included in the Fermi-Dirac distribution, $f(\omega)=\left[ 1+\exp(\omega-\mu)/T) \right]^{-1}$. The spectral function $A^N(k,\omega)$ is related to the normal component of the self-energy,
\begin{align}
A^N(k,\omega) = \frac{-2 \text{Im} \Sigma^N(k,\omega)}
{ \left[ \omega - \frac{k^2}{2m} - \text{Re} \Sigma^N(k,\omega) \right]^2 + \text{Im} \Sigma^N(k,\omega)^2 } \, ,
\end{align}
and includes information related to sp fragmentation in the normal phase~\cite{Dickhoff08,Rios2009}. We work with self-energies that have been obtained within a finite-temperature $\mathcal{T}-$matrix SCGF approach, discussed in detail in Refs.~\cite{FrickPhD,RiosPhD,Soma2008}. At temperatures close to the pairing phase transition, $A^N$ develops a characteristic two-peak structure as a function of energy, which is an indication of a pseudo-gap phase~\cite{Schnell1999}. At and below the critical temperature, the method is not valid anymore, as evidenced by the appearance of the Thouless pole in the $\mathcal{T}-$matrix~\cite{Thouless1960,Kadanoff1961,Alm1993}. We therefore obtain the normal spectral function at zero temperature by extrapolating finite temperature results down to zero temperature, as explained in the following subsection. This is in agreement with the physical interpretation of $A^N$ as a normal state spectral function. 

In the lowest-order BCS approach at zero temperature, the spectral functions in the convolution of Eq.~(\ref{eq:xik}) become delta functions in energy. The normal self-energy has a single peak at the normal quasi-particle energy, whereas the superfluid spectral function shows two unequally weighted solutions for a given momentum~\cite{Bozek2000,Muther2005}. The energy denominator is a function of momentum, 
\begin{align}
\xi(k) = | E(k) |
\label{eq:xie}
\end{align}
with the effective sp energy, 
\begin{align}
E^2(k)= \chi^2(k) + \overline \Delta^2(k) \, .
\label{eq:edk2}
\end{align} 
To obtain this result, we have also assumed that there is no renormalisation of the sp peaks, $\mathcal{Z}(k)\approx1$. In a low-density BCS approximation, the effective sp energy corresponds to a kinetic spectrum, $\chi(k) = \frac{k^2}{2m} - \mu$, with $\mu$ the chemical potential of the system. One can also add a mean-field potential contribution to the spectrum, or describe its effect by means of an effective mass~\cite{Abrikosov1965,Dickhoff08}. The averaged gap, $\overline \Delta$ is associated with the partial wave that is active for a given Fermi momentum. In practice, different partial waves are active in different regions of Fermi momentum. Consequently, we consider  $\overline \Delta(k) \approx \Delta^{JST}_{L}(k)$ in the solution of the gap equation in a given $JST$ channel~\cite{Dean2003}. For coupled channels, we take 
$\overline \Delta^2(k) = \left( \Delta^{JST}_{L} (k) \right)^2 + \left( \Delta^{JST}_{L'} (k) \right)^2 $.

The gap equation is an integral non-linear equation for $\Delta^{JST}_{L}(k)$, which appears explicitly in the denominator via Eq.~(\ref{eq:edk2}). In the fully correlated theory, in contrast, the gap appears indirectly in the definition of the superfluid spectral function,
\begin{align}
A(k,\omega) &= -2 \text{Im} G(k,\omega) \, .
\end{align}
The full sp propagator, $G(k,\omega)$, computed in the superfluid phase, differs from the normal one by a factor that is proportional to the square of the gap,
\begin{align}
G(k,\omega) &= G^N(k,\omega) \left [ 1 -G^N(-k,-\omega) \Delta^2(k,\omega) G(k,\omega)\right] \, .
\end{align} 
Consequently, $A^N$ and $A$ only differ from each other close to the Fermi momentum and energy, where pairing effects are more prominent \cite{Bozek1999,Muther2005}. 

A fully self-consistent description of pairing requires an explicit iterative calculation of both $A^N$ and $A$~\cite{Bozek1999}. Here, we take a different approach, which is an initial step towards a fully self-consistent solution, including all the relevant correlations. First, because the normal and superfluid spectral functions are very similar in a wide energy and momentum domain, we compute the contribution to the effective denominator from the double convolution of \emph{normal} spectral functions, 
\begin{align}
\frac{1}{2 \chi(k)} = \int \frac{d \omega}{2\pi} \frac{d \omega'}{2\pi} &A^N (k,\omega) A^N (k,\omega') \times \nonumber \\
& \frac{1-f(\omega)-f(\omega')}{\omega+\omega'}  \, .
\label{eq:chi}
\end{align}
This will account explicitly for fragmentation effects on the gap equation. The superfluid gap is then generated by the gap equation, Eq.~(\ref{eq:gap}), with an energy denominator that has the same structure as in the BCS expression, Eqs.~(\ref{eq:xie}) and (\ref{eq:edk2}). One can indeed generate a superfluid spectral function from the resulting gap, see e.g. Ref.~\cite{Muther2005}.
Further calculations of the normal self-energy then in principle require the effect of the gap to be included in the normal propagators~\cite{Bozek1999}. We expect such feedback effects from the superfluid phase to be small in comparison to the relatively large fragmentation of strength associated with SRC which is captured effectively by Eq.~(\ref{eq:chi}).

This approach was first exploited in Ref.~\cite{Muther2005} to study the singlet pairing properties of neutron matter  and triplet pairing in symmetric nuclear matter. A major conclusion of that study was the large impact of sp fragmentation on pairing properties. Gaps in infinite matter are substantially quenched by the removal of strength mediated by SRC, generally by a factor of $\approx 20 \, \%$. In nuclear matter, the SRC effect alone precludes the formation of a $^3$SD$_1$ pairing gap in nuclear matter at saturation density. This provides support for the lack of experimental evidence of isoscalar $np$ pairing from an infinite-matter calculation. 

Finally, we would like to comment on quasi-particle approximations to the pairing problem. In a quasi-particle limit, the spectral functions in Eq.~(\ref{eq:xik}) and Eq.~(\ref{eq:chi}) are replaced by delta functions centered around quasi-particle energies. In both the normal and the superfluid case, these are weighted by the corresponding renormalizations of the quasi-particle poles, or $\mathcal{Z}$-factors \cite{Baldo2000,Bozek2000,Dong2015}. These will effectively account for a removal of strength in different regions of momentum space, including the Fermi surface. In general, one finds that the quasi-particle energy denominator of Eq.~(\ref{eq:xie}) is divided by a factor $Z^2(k)$. These quasi-particle approximations, however, include fragmentation in a very crude way and cannot reliably predict the effect of quenching due to SRC obtained with realistic spectral functions in the double convolution of Eq.~(\ref{eq:chi}) \cite{Muther2005}. 

\subsection{Zero temperature extrapolation}
\label{sec:T0}

The normal spectral function at zero temperature, $A^N(k,\omega)$, has been computed as an extrapolation of finite temperature self-energies. Numerical results of SCGF ladder approximation self-energies with microscopic NN interactions have been available in the literature for the last decade~\cite{Frick2003,RiosPhD,Soma2008}. The pairing instability, however, precludes a direct calculation within the ladder approximation and normal propagators below the critical pairing temperature, $T_c$~\cite{Alm1993,Muther2005}. Consequently, for a fixed density, we perform a series of  finite temperature calculations to determine the real and imaginary parts of the self-energy, and use these as input for an extrapolation to zero temperature. The fit is further constrained by the requirement that the macroscopic properties that are computed at $T=0$ provide a thermodynamically consistent description of the system. Numerical details are provided in Appendix~\ref{app:T0}. Below, we discuss the properties of the extrapolated self-energies. 

 \begin{figure}
\begin{center}
\includegraphics[width=\linewidth]{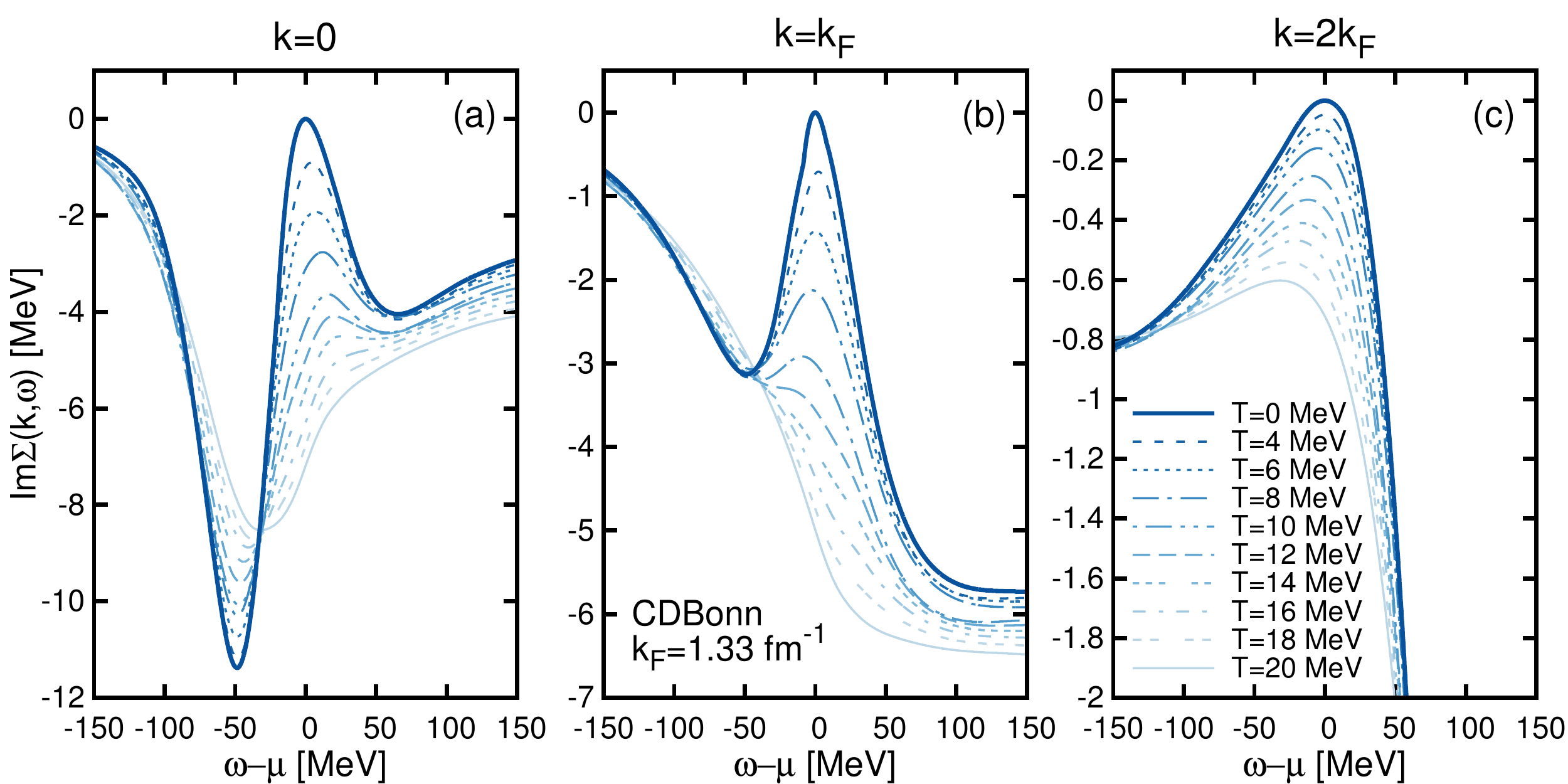}
\caption{Imaginary part of the self-energy around the Fermi energy at $k_F=1.33$ fm$^{-1}$ for the CDBonn interaction for several temperatures. Panels (a), (b) and (c) correspond to momenta $k=0$, $k_F$ and $2k_F$, respectively. The $T=0$ extrapolation is shown in a solid line. Note the different vertical scales of each panel.\label{fig:self}}
\end{center}
\end{figure}

The three panels of Fig.~\ref{fig:self} show the imaginary part of the self-energy as a function of energy, $\omega$, for three characteristic momenta. Panels (a) and (c) show self-energies well below and above the Fermi momentum, respectively. Panel (b), in contrast, shows the $k=k_F$ case. Results are displayed for the CDBonn interaction at $k_F=1.33$ fm$^{-1}$, but equivalent conclusions are found with other NN interactions and 3NFs in this density regime. At large temperatures, there is little (or no) distinction between the hole, $\omega < \mu$, and the particle $\omega>\mu$, parts of $\text{Im}\Sigma$. As temperature decreases, however, a structure develops close to $\omega \approx \mu$, with $\text{Im}\Sigma$ approaching zero in absolute value. This is the area where temperature plays the most important role, and where the extrapolation procedure is most critical. A momentum- and energy-dependent polynomial fit, described in Appendix~\ref{app:T0}, captures this temperature dependence, and provides an extrapolated self-energy which provides consistent results.

\begin{figure}
\begin{center}
\includegraphics[width=\linewidth]{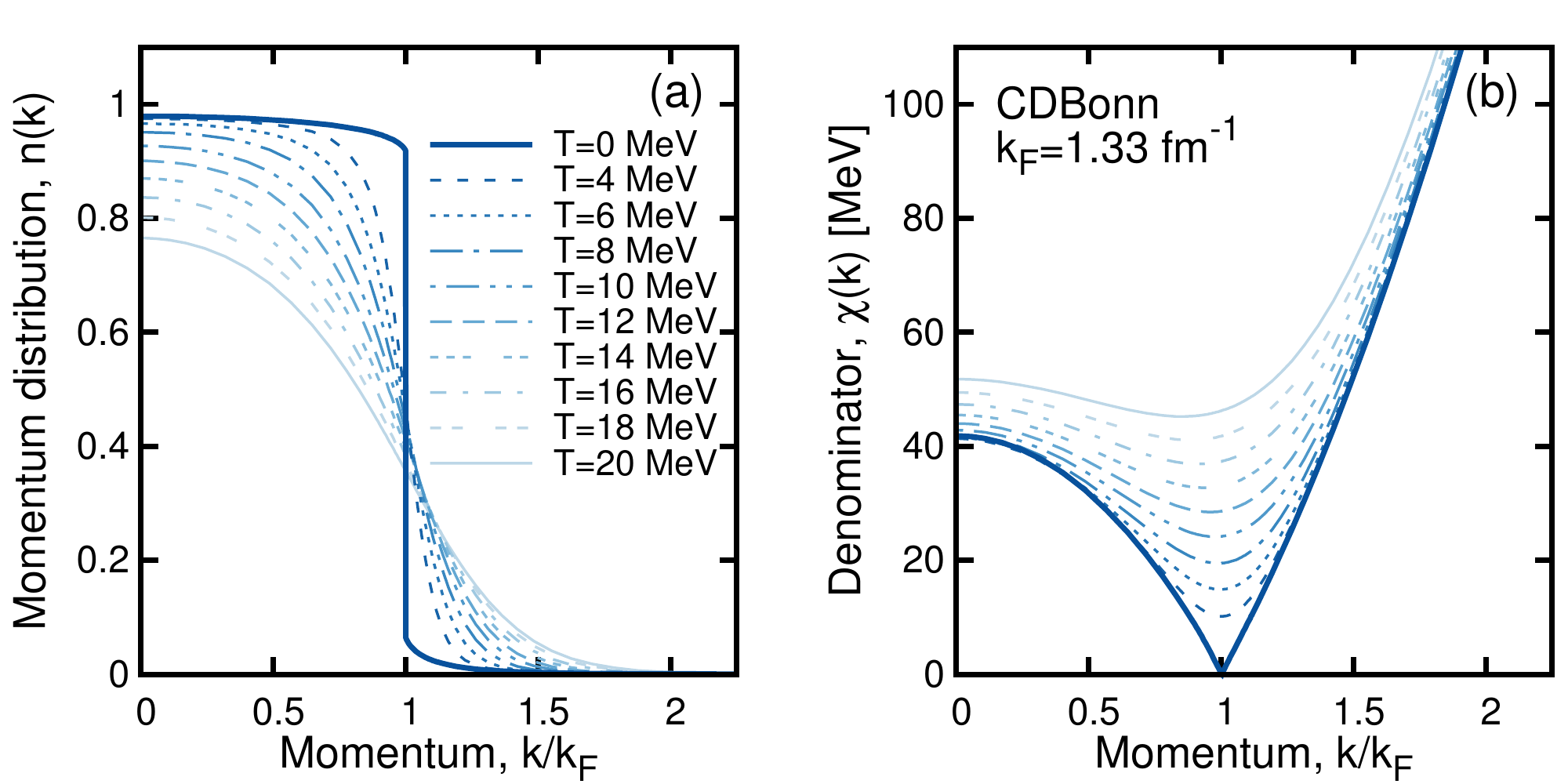}
\caption{Panel (a): momentum distribution at $k_F=1.33$ fm$^{-1}$ for the CDBonn interaction at different temperatures, including the $T=0$ extrapolation (solid line). Panel (b): effective sp denominator, $\chi(k)$, in the same conditions.\label{fig:mom_chi} }
\end{center}
\end{figure}

From the self-energy, one can obtain other relevant microscopic properties. Panel (a) in Fig.~\ref{fig:mom_chi} shows an example of the temperature extrapolation of the momentum distribution. Again, while this specific example is for the CDBonn interaction at $k_F=1.33$ fm$^{-1}$, very similar results are obtained with other forces in a wide density regime. As temperature decreases, one finds the expected behavior for the correlated momentum distribution: the Fermi surface becomes increasingly sharp, and low and high-momentum features build up. As expected, the zero temperature $n(k)$ has a sharp discontinuity across the Fermi surface. The exact shape of the momentum distribution for momenta within a few percent of $k_F$ is sensitive to the extrapolation procedure, particularly to the order of the extrapolating polynomial. However, the implementation of the thermodynamically consistent extrapolation procedure guarantees that, on average, the discontinuity of the Fermi surface is within a few percent of the derivative of the self-energy (i.e. the $\mathcal{Z}$-factor at $k=k_F$). We note that we have corrected the momentum distribution for missing strength effects, as discussed in Appendix~\ref{app:T0}.

\subsection{Pairing kernel with short range correlations}
\label{subsec:gap}
Pairing calculations require as input the double energy convolution of Eq.~(\ref{eq:chi}). This convolution is formally equivalent to the dressed but non-interacting two-body propagator, $\mathcal{G}_{II}^0$, in the in-medium $\mathcal{T}-$matrix equation at zero energy and centre-of-mass momentum~\cite{FrickPhD,Frick2003,Dickhoff08}. This is in correspondence to the well-known fact that Cooper pairing appears as a pole in the normal $\mathcal{T}-$matrix in these conditions. The experience gathered in performing the double convolution in SCGF finite temperature calculations is useful in computing the energy denominator~\cite{FrickPhD}. In particular, it is useful to keep track of the quasi-particle energies for each given sp momentum, so that the quasi-particle peak is well sampled in the double folding integrals~\cite{RiosPhD}.

Pairing calculations, particularly in the $^3$PF$_2$ channel, are very sensitive to the Fermi surface region, and inaccuracies on the double folding are amplified in final gap solutions. In particular, missing strength corrections, analogous to those discussed for $n(k)$ in Appendix~\ref{app:T0}, are essential to compute a continuous energy denominator in regions arbitrarily close to $k_F$. Panel (b) of Fig.~\ref{fig:mom_chi} shows the energy denominators for a CDBonn calculation at $k_F=1.33$ fm$^{-1}$ as a function of momenta for a variety of temperatures. Here, as it was the case with $n(k)$, the largest modifications due to temperature occur close to $k_F$. The low and the high momentum ends of $\chi(k)$ are less sensitive to temperature, and their details are well captured by finite temperature calculations. In contrast, the region around $k=k_F$ shows a non-negligible temperature dependence even in the lowest temperatures. In the quasi-particle approximation, the Sommerfeld expansions predicts a linear temperature dependence of the denominator in regions close to $k_F$. Panel (b) shows that a similar linear temperature dependence is generated in that region. 

\begin{figure}
\begin{center}
\includegraphics[width=\linewidth]{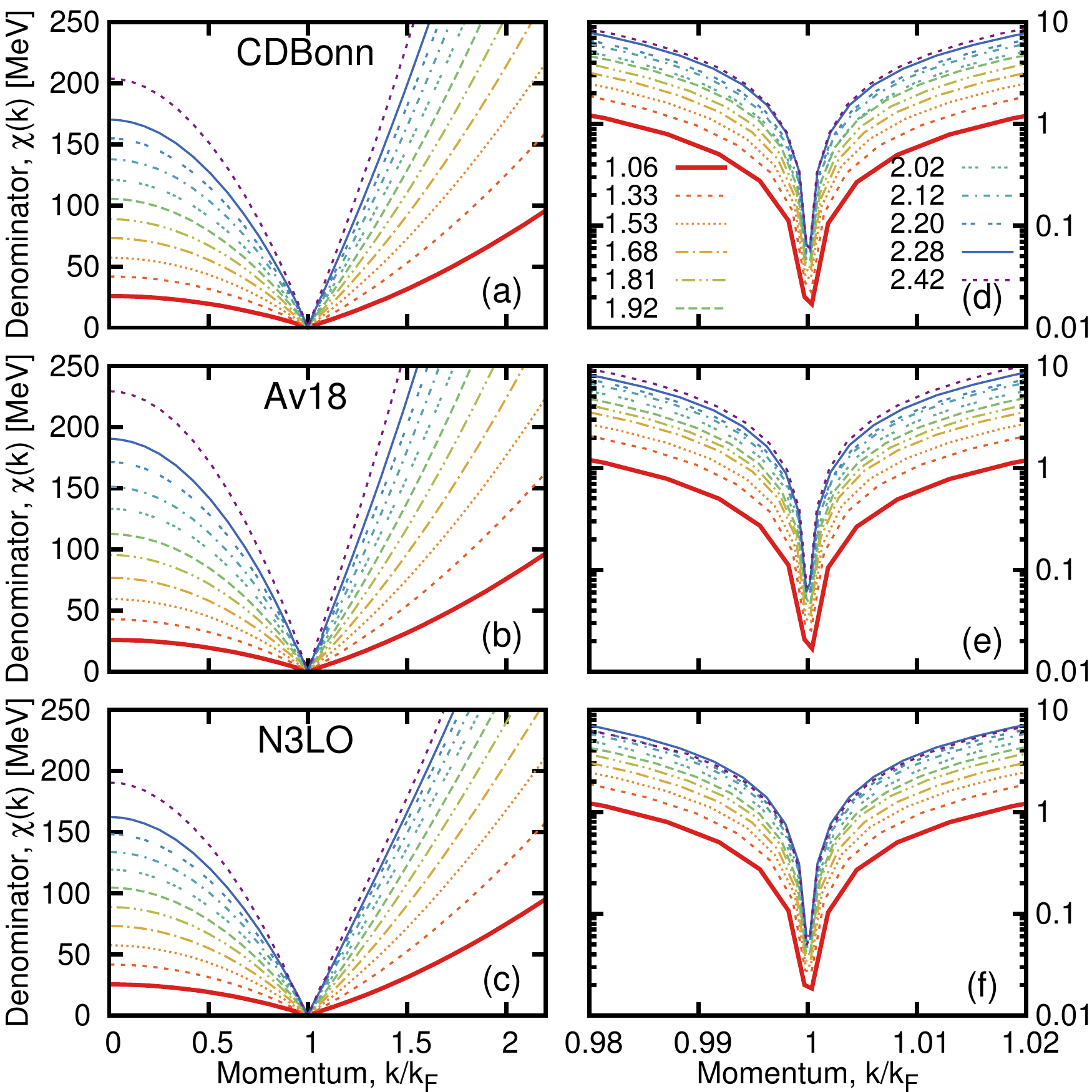}
\caption{Panels (a)-(c): energy denominator at $T=0$ as a function of momentum for different Fermi momenta, corresponding to the (a) CDBonn, (b) Av18 and (c) N3LO interactions. Panels (d)-(f): the same function, around the Fermi surface, plotted in a logarithmic scale. See text for details. \label{fig:chi_den} }
\end{center}
\end{figure}

The density dependence of the zero-temperature double convolution is displayed in panels (a) to (c) of Fig.~\ref{fig:chi_den}. Each panel represents the results obtained with a different NN interaction: (a) CDBonn \cite{Machleidt1995}, (b) Av18 \cite{Wiringa1995} and (c) the Entem-Machleidt N3LO potential \cite{Entem2003}. We note that the different NN interactions enter the denominator calculation via the convolution of different extrapolated spectral functions. The spectral functions of these three interactions are relatively dissimilar \cite{Rios2009}, but the integrated convolution smears out the differences to a certain extent. Consequently, the results obtained for $\chi(k)$ (and its density dependence) are relatively close for all the NN forces. 

In general, we find that the double convolution denominator increases with density. This density dependence occurs at all momenta, including the near vicinity of the Fermi surface that is displayed, in a logarithmic scale, in panels (d) to (f) of Fig.~\ref{fig:chi_den}. The gap equation is mostly determined by the energy denominator close to the Fermi surface, and hence it is important that this region is well sampled for pairing purposes. The denominator shows a linear behavior both below and above $k_F$, with a sharp minimum at the Fermi surface. The value of $\chi(k)$ at the Fermi surface, $k=k_F$, is relatively small, but non-zero. While the linear behavior is expected in a quasi-particle-type approach, the non-zero minimum is a direct consequence of the use of a double convolution beyond a quasi-particle picture. We note that, without the missing strength corrections discussed in Appendix \ref{app:T0}, the near-Fermi-surface behavior would be erratic. The corrected calculations, in contrast, provide a well-defined function of both momentum and density. 

\begin{figure}
\begin{center}
\includegraphics[width=0.56\linewidth]{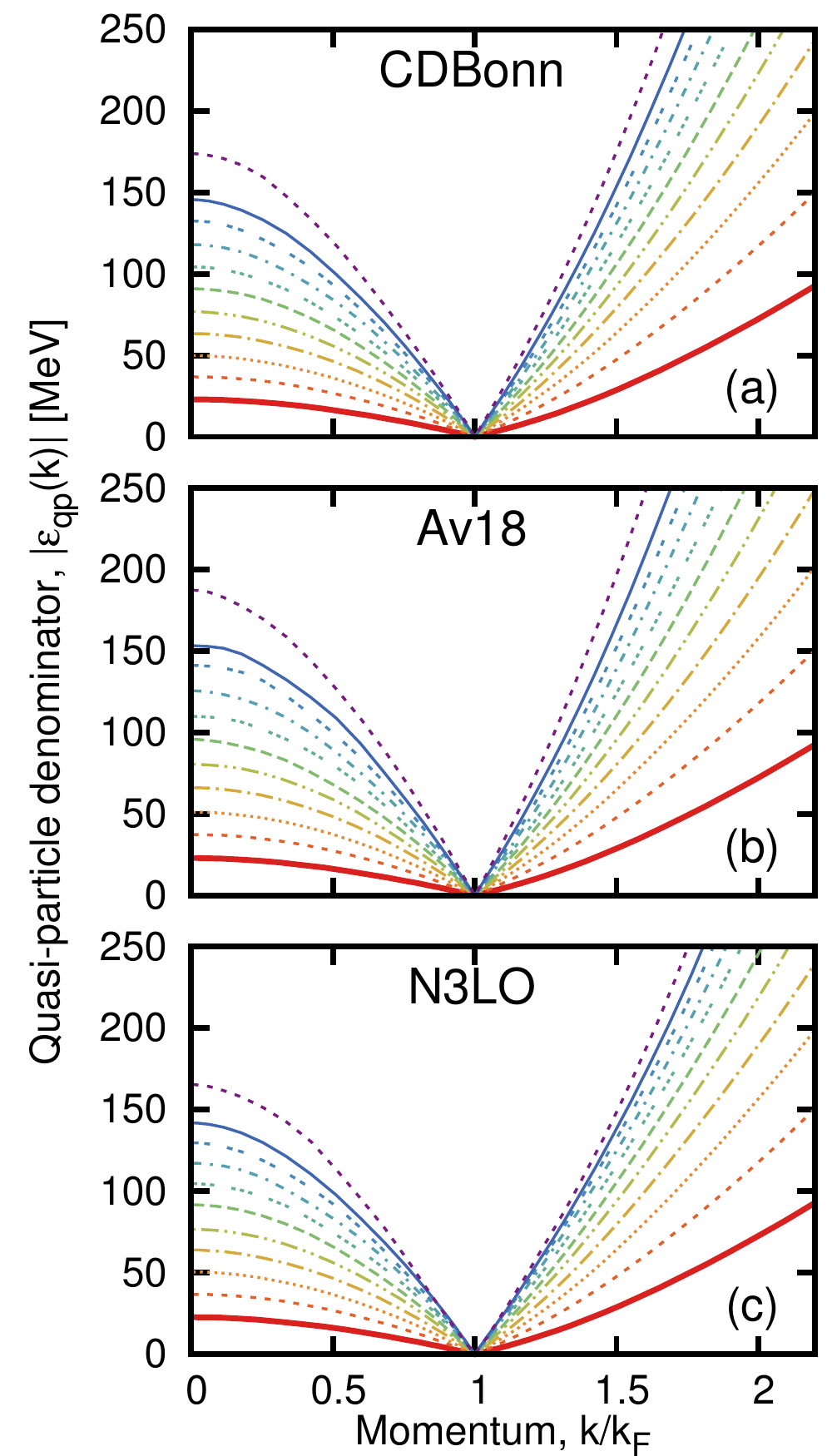}
\caption{Panels (a)-(c): energy denominator at $T=0$ in the quasi-particle limit as a function of momentum for different Fermi momenta, corresponding to the (a) CDBonn, (b) Av18 and (c) N3LO interactions. \label{fig:qp_den} }
\end{center}
\end{figure}

In the quasi-particle limit of Eq.~(\ref{eq:xie}), the denominator reflects the momentum and density dependence of the quasi-particle energy with respect to the chemical potential. We show the extrapolated zero-temperature quasi-particle denominator in Fig.~\ref{fig:qp_den}, in the same conditions and for the same NN forces as Fig.~\ref{fig:chi_den}. The quasi-particle energies are determined consistently by solving the corresponding implicit equation for the SCGF ladder self-energies,
\begin{align}
\varepsilon_\text{qp}(k) = \frac{ k^2 }{2m} + \text{Re} \Sigma^N(k,\varepsilon_\text{qp}(k) ) \, ,
\end{align}
and subtracting the chemical potential. 

The quasi-particle picture provides an intuitive understanding for the density dependence of the energy denominator. Broadly speaking, the quasi-particle spectrum is more stretched as density increases. A sign notwithstanding, this gives rise to an increase of the denominator as density rises at low momenta. In turn, when measured with respect to the Fermi momentum, the high-momentum quasi-particle energies become more repulsive as density increases, which gives rise to the $k>k_F$ behavior.

Two important conclusions can be drawn from a comparison of Figs.~\ref{fig:chi_den} and \ref{fig:qp_den}. On the one hand, the qualitative density and momentum dependence of the quasi-particle and the double convolution denominators are similar. In particular, both are increasing functions of density. As functions of momenta, the initial decrease below $k_F$ is followed by an increase above the Fermi surface. Furthermore, a linear behavior is found near the Fermi surface in both cases, as expected on general grounds. 

\begin{figure}
\begin{center}
\includegraphics[width=\linewidth]{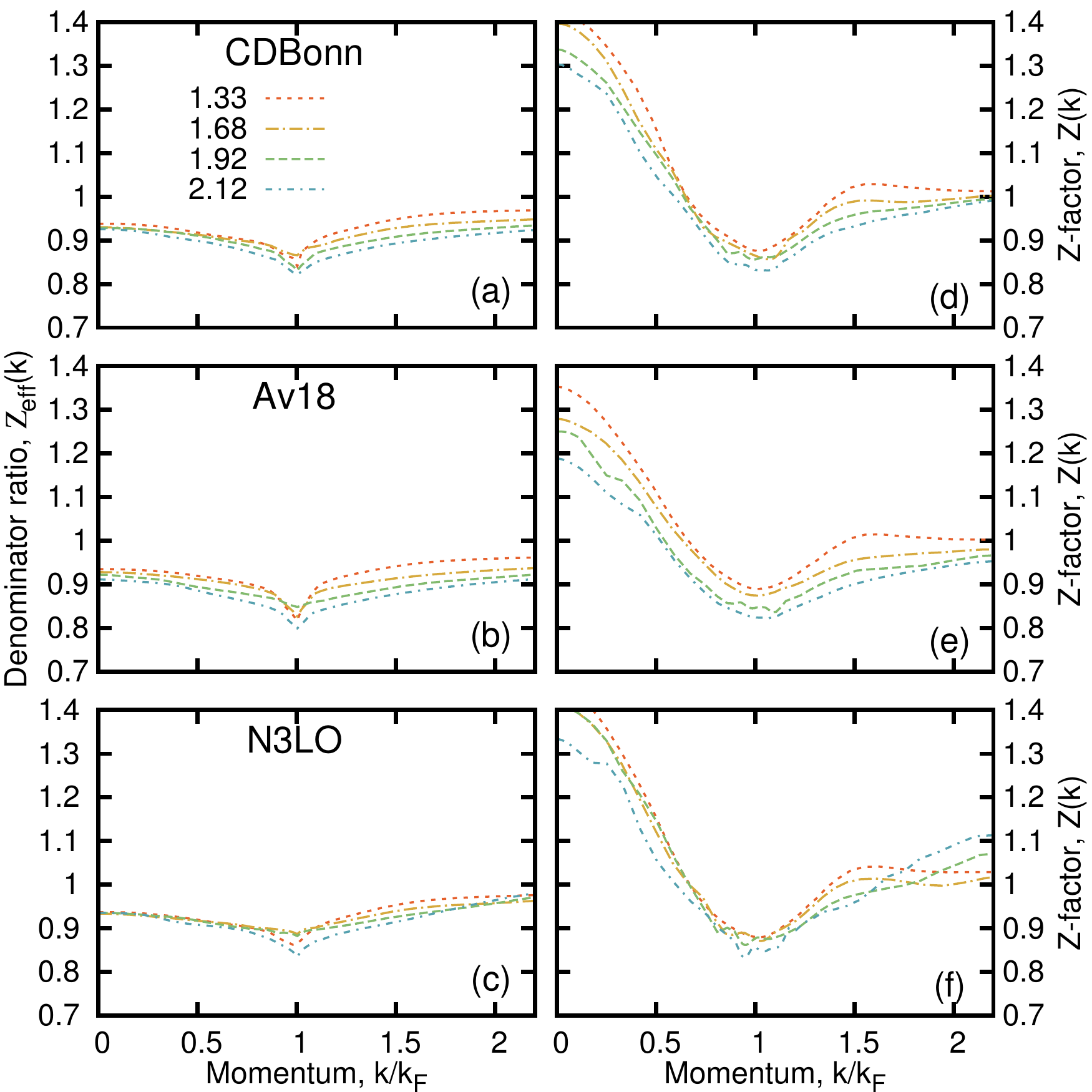}
\caption{Panels (a)-(c): ratio of denominators, Eq.~(\ref{eq:ratio}) at $T=0$ as a function of momentum for different Fermi momenta. The three panels correspond to the (a) CDBonn, (b) Av18 and (c) N3LO interactions. Panels (d)-(f): the actual $Z$-factor as a function of momentum in the same conditions. See text for details. \label{fig:zfacden} }
\end{center}
\end{figure}

On the other hand, there are quantitative differences between both denominators. The double convolution takes into account the fragmentation of quasi-particle states in the normal state. Since strength is removed from the full quasi-particle peak, the denominator becomes larger than the corresponding quasi-particle value. As a matter of fact, the difference between the two results can be parametrized in terms of an effective $\mathcal{Z}-$factor  \cite{Bozek2000,Muther2005},
\begin{align}
Z^2_\text{eff}(k) =\frac{ E(k) }{\chi(k)} \, .
\label{eq:ratio}
\end{align}
The ratio is displayed, for a subset of relevant densities and three NN forces, in panels (a)-(c) of Fig.~\ref{fig:zfacden}. $Z_\text{eff}$ is always in the range $\approx 0.8-0.9$, and shows a mild momentum dependence, with a minimum close to the Fermi surface. Our results suggest that the ratio decreases slowly with density. This is in accordance to an intuitive picture, where correlations, measured as a deviation from one in $Z_\text{eff}$, become more important at higher densities. 

It is important to stress that the effective denominator ratio, $Z_\text{eff}$, is different from the standard renormalization factor, 
\begin{align}
\mathcal{Z}(k) =\left.  \frac{ 1 }{ 1 - \partial_\omega \text{Re} \Sigma^N(k,\omega)}  \right |_{\omega=\varepsilon_\text{qp}(k) } \, .
\label{eq:zfactor}
\end{align}
We show this quantity, computed in the same conditions as  $Z_\text{eff}$, in panels (d)-(f) of Fig.~\ref{fig:zfacden}. Other than in the vicinity of the Fermi surface, the two renormalization factors provide very different results. $\mathcal{Z}(k)$ is generally well above $Z_\text{eff}(k)$. It peaks around $k=0$ at values of $\approx 1.3-1.5$, decreases to a minimum close to $k_F$ and subsequently raises again to $\approx 1$ at high momenta. Even though pairing properties are dominated by Fermi surface effects, where both renormalization factors are relatively similar, this comparison shows that a realistic description of the missing strength for pairing purposes can only be achieved approximately by a renormalization-corrected BCS-type approach~\cite{Dong2013,Dong2015}. In fact, because the removal of strength is underestimated in $\mathcal{Z}(k)$ with respect to $Z_\text{eff}$, the corresponding gap is larger in a $Z-$factor corrected BCS approach as compared to a fully correlated description~\cite{Bozek2000}. An additional difficulty is that it is unclear how particle number or density can be properly obtained from this approach.

\subsection{Long-range correlations}
\label{sec:LRC}

The most important effect of LRC on pairing properties will occur at the level of the effective pairing interaction when neutrons near the Fermi surface exchange possibly collective spin and density modes~ \cite{Wambach1993,Schulze2001,Schwenk2003,Schwenk2004,Sedrakian2006,Sedrakian2007}. Following Refs.~\cite{Shen2005,Cao2006,Dong2013}, we add to the interaction in the generalized gap equation, $\mathcal{V}^{JST}_{LL'}$, the corresponding contributions accounting for such fluctuations in a physically motivated way. 
We adopt the results of Ref.~\cite{Cao2006}, which incorporate an induced interaction that leads to a well-behaved particle-hole interaction that fulfills appropriate stability criteria, not obeyed by interactions that incorporate only the effect of SRC like \textit{e.g.} $\mathcal{G}$-matrices~\cite{dickhoff1987self}. The coupling to neutrons that are dressed by the full off-shell effect of SRC as described in the previous section is then governed by the exchange of both a density fluctuation and a spin mode. The collective features of these modes are controlled by self-consistently determined Landau parameters~\cite{Cao2006}.Their contribution to the pairing interaction requires a recoupling from the particle-hole channel to the particle-particle channel and is therefore different for spin singlet and spin triplet pairing. We note that this is a physically motivated approach to the treatment of LRC that has only been tested in the literature in extensions to BCS theory where SRC are included in terms of renormalization factors~\cite{Cao2006}. By restricting the effect of LRC to the effective interaction, we can test the effect of both SRC and LRC in pairing properties by turning either correlation effect on or off.

Following Ref.~\cite{Cao2006}, the interaction that treats LRC for the $^1$S$_0$ channel is given by 
\begin{align}
\mathcal{V}^{S=0}_\text{LRC} = 
{\textstyle{\frac{1}{2}}}  \mathcal{G}^0_\text{ph}  \mathcal{G}^0_\text{ph} \Lambda^{S=0}(q) - 
{\textstyle \frac{3}{2}} \mathcal{G}^1_\text{ph}  \mathcal{G}^1_\text{ph} \Lambda^{S=1}(q) \, ,
\label{eq:veffS0}
\end{align}
where $\mathcal{G}^S_\text{ph}$ represent the vertices that couple to the spin-$S$ excitation.
They can be thought of as particle-hole transformed $\mathcal{G}$-matrix elements averaged around the Fermi energy. As argued in Ref.~\cite{Cao2006}, these vertices are improved by employing the corresponding Landau parameters, as in the original work of Babu and Brown for liquid ${}^3$He~\cite{babu1973}. The iterated bubble series is then represented by:
\begin{align}
\Lambda^S(q) = \frac{\Lambda_0(q)}{1 - \Lambda_0(q) \mathcal{L}^S} \, ,
\label{eq:bubbles}
\end{align}
where $\mathcal{L}^S$ corresponds to the relevant Landau parameter. The density mode with total spin $0$ in the particle-hole channel is determined by $\mathcal{L}^0$  which is attractive at low density and usually denoted by $F_0$. The spin mode with total spin $1$ is determined by  $\mathcal{L}^1$, which is repulsive but has similar magnitude and is often denoted by $G_0$.  The static Lindhard function, $\Lambda_0(q)$
\begin{align}
\Lambda_0(q)=\frac{N(0)}{\mathrm{g}}\frac{1}{2}\left[-1+\frac{1}{q}\left(1-\frac{q^2}{4}\right) \ln\left|\frac{1-q/2}{1+q/2}\right|\right]
\end{align}
is employed in Eq.~(\ref{eq:bubbles}), with the appropriate density of states $N(0)=\frac{8mk_F}{\hbar^2}$ and degeneracy factor $\mathrm{g}=2$. 
We assume as in Ref.~\cite{Cao2006} that for neutron matter the effective mass in the density of states can be approximated by the bare mass.
The static Lindhard function is iterated to all orders according to Eq.~(\ref{eq:bubbles}) and generates negative particle-hole propagators, $\Lambda^S$. Projecting Eq.~(\ref{eq:veffS0}) onto $L = 0$, the resulting interaction can then be included into the gap equation for ${}^1S_0$ pairing for a given density and appropriate values of the Landau parameters.

\begin{figure}
	\begin{center}
		\includegraphics[width=0.9\linewidth]{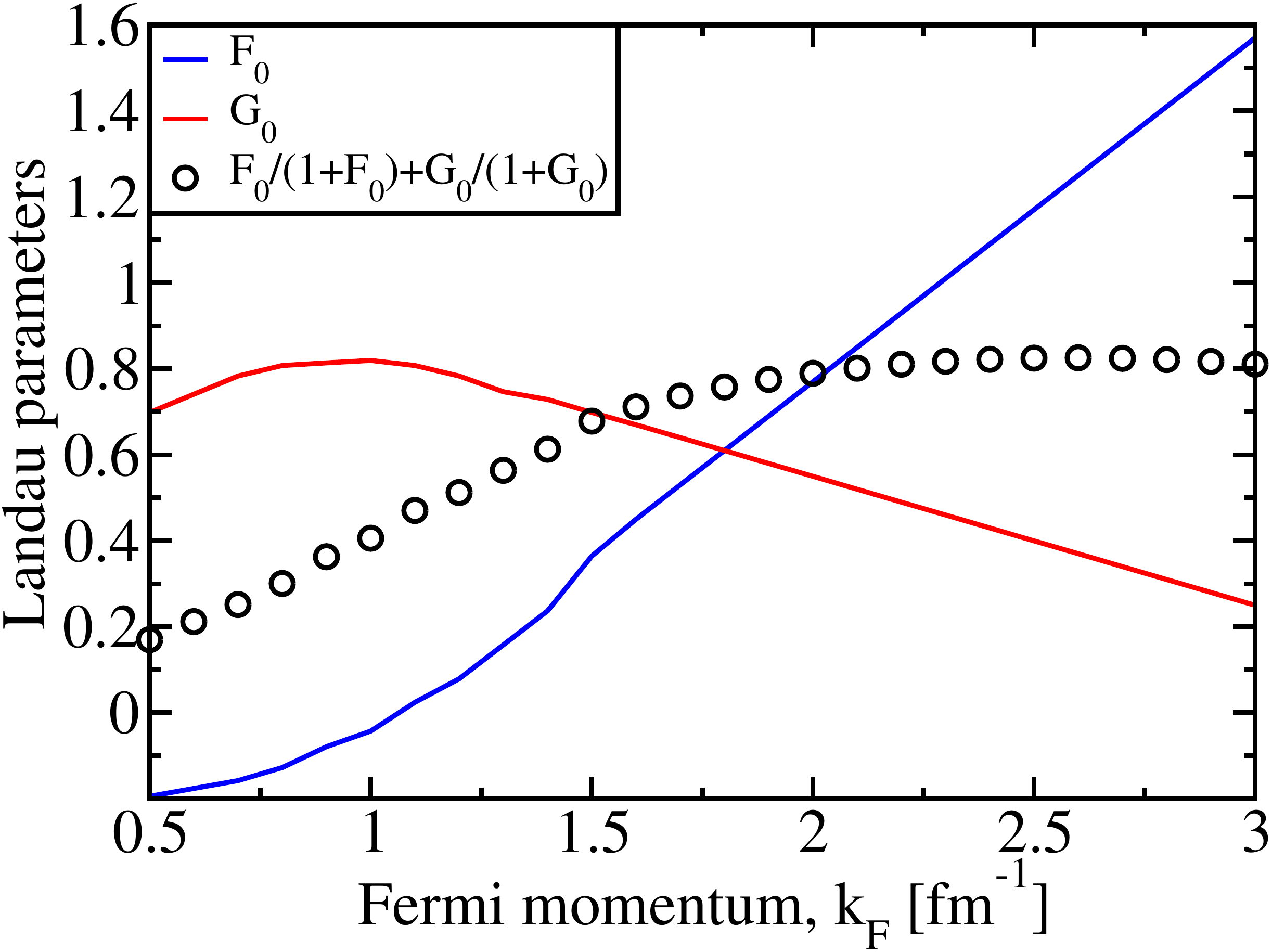}
	\end{center}
	\caption{Landau parameters $F_0$ and $G_0$ extrapolated to $k_F$ = 3.0 fm$^{-1}$ employing the results of Ref.~\cite{Cao2006}. The corresponding value of the forward scattering sum rule is indicated by the open circles. \label{fig:LDP} }
\end{figure}

\begin{figure}
	\begin{center}
		\includegraphics[width=\linewidth]{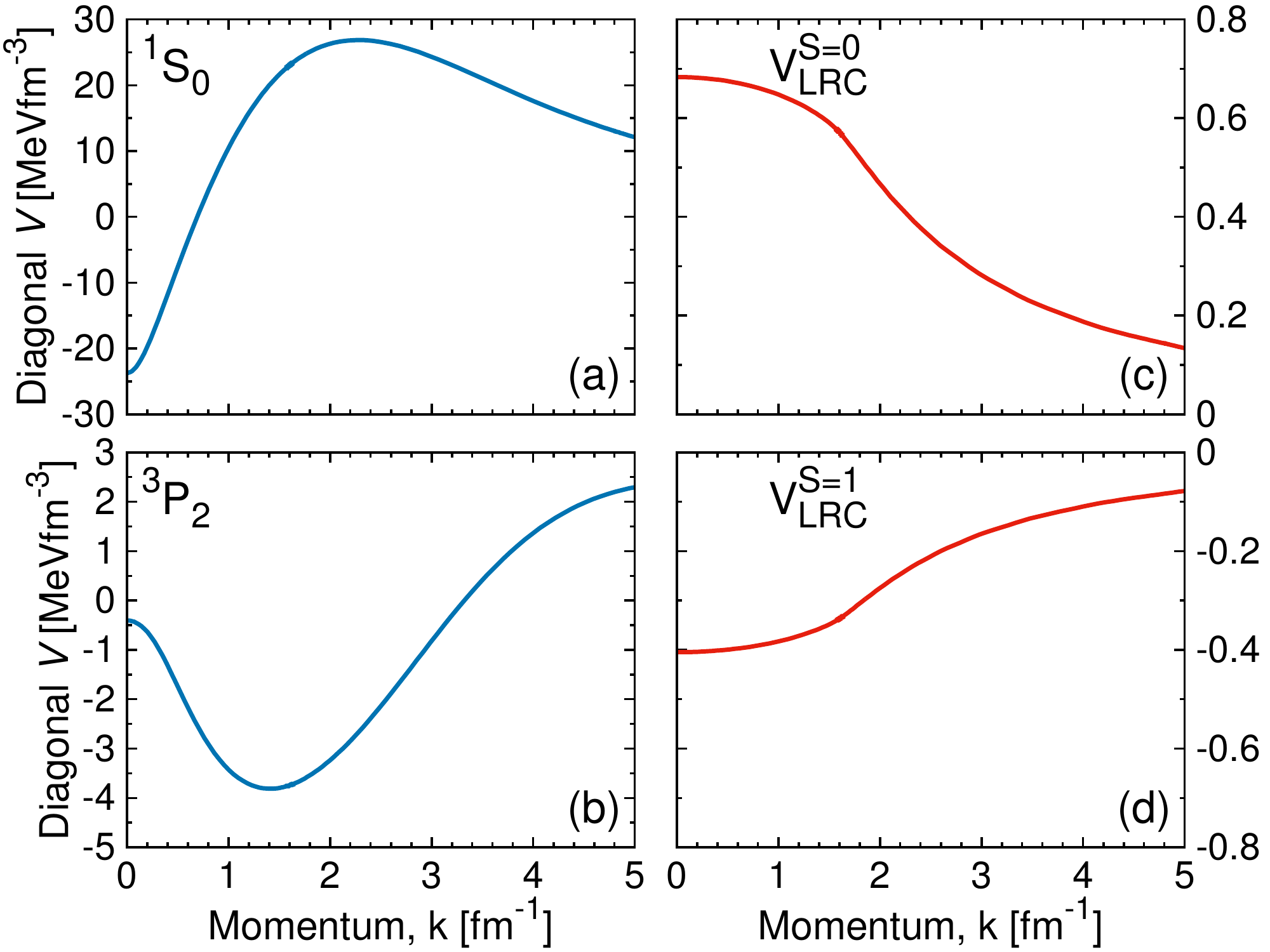}
	\end{center}
	\caption{Panels (a) and (b): diagonal matrix elements of the Av18 interaction in the (a) $^1$S$_0$ and (b) $^3$P$_2$ channels. Panels (c) and (d): diagonal matrix elements of the additional pairing interaction representing the low-energy medium polarization at a density corresponding to $k_F =1.6$ fm$^{-1}$ in the same channels. The scales are different for each panel. \label{fig:VLRC} }
\end{figure}

Both Landau parameters exhibit a modest density dependence in the domain relevant for singlet pairing ($k_F < 1.5$ fm$^{-1}$). The parameters are adopted from Ref.~\cite{Cao2006} and are plotted in Fig.~\ref{fig:LDP}.
The first term in Eq.~(\ref{eq:veffS0}) is attractive, whereas the second term is repulsive. In the density domain relevant for singlet pairing this repulsion dominates on account of the spin factor leading to an inevitable additional suppression of the gap in this channel. Figure~\ref{fig:VLRC} illustrates that the additional term, $\mathcal{V}^{S}_\text{LRC}$ [panels (c) and (d)], is relatively small compared to the bare interaction [panels (a) and (b)]. For the $^1$S$_0$ channel, LRC reduce the attraction of the bare Av18 interaction.

The procedure proposed in Ref.~\cite{Cao2006} is generalized here to the case of the $^3$PF$_2$ coupled channel. For the $^3$PF$_2$ channel which involves spin-$1$ pairs, the sampling over density and spin modes becomes:
\begin{align}
\mathcal{V}^{S=1}_\text{LRC} = 
\frac{1}{2} \mathcal{G}^0_\text{ph}  \mathcal{G}^0_\text{ph} \Lambda^{S=0}(q) + 
\frac{1}{2} \mathcal{G}^1_\text{ph}  \mathcal{G}^1_\text{ph} \Lambda^{S=1}(q) \, ,
\label{eq:veffS1}
\end{align}
with both terms yielding attraction. Contrary to the $^1$S$_0$ channel, this contribution will always lead to antiscreening of the gap, as it represents an attractive interaction. This point is illustrated in panel (d) of Fig.~\ref{fig:VLRC}, which shows the relatively small but nevertheless attractive contribution of the LRC interaction in the $S=1$ channel. This is to be compared to the bare interaction in the $^3$P$_2$ channel, shown in panel (b). 

The Landau parameters $F_0$ and $G_0$ from Ref. \cite{Cao2006} are extrapolated to higher densities in a smooth way as shown in Fig.~\ref{fig:LDP}. In Fig.~\ref{fig:LDP}, we also include the contribution to the forward scattering sum rule of the Landau parameters $F_0$ and $G_0$ indicated by the open circles (see \textit{e.g.} Refs.~\cite{Friman1979,Dickhoff1981x,dickhoff1987self}). While the extrapolated Landau parameters are both positive at higher density, the forward scattering sum rule is nevertheless approximately fulfilled when one allows for a negative contribution of the Landau parameter $F_1$ (for example of about $-0.5$~\cite{Dickhoff1981x}) to the sum rule, given by $F_1/(1+F_1/3)$. The extrapolation introduces some uncertainty in the effect of LRC for triplet pairing at higher density, but it should be emphasized that Eq.~(\ref{eq:veffS1}) leads to antiscreening whatever the numerical values or sign of the Landau parameters $F_0$ and $G_0$. Moreover, this small correction is motivated by well-explored many-body theory principles.

In future work, we intend to generate the Landau parameters from a consistent evaluation starting from the ladder-summed effective interaction. A proper inclusion of the induced interaction with this starting point is however considerably beyond the scope of the present work. Further, the possibility that the presence of the pion-exchange tensor interaction strongly influences the spin mode \cite{Dickhoff1981} should also be investigated (see also Ref.~\cite{Pankratov2015}). A proper treatment of retardation implied by the possibility of exchanging low-lying density and spin modes in principle generates a complex solution of the gap equation which should also be explored further (see \textit{e.g.} the work of Ref.~\cite{Sedrakian2003} for a calculation with a dynamic pion-exchange interaction).

\section{Results}
\label{sec:results}

\subsection{Singlet pairing}
\label{sec:results1S0}

\begin{figure}[t!]
\begin{center}
\includegraphics[width=0.6\linewidth]{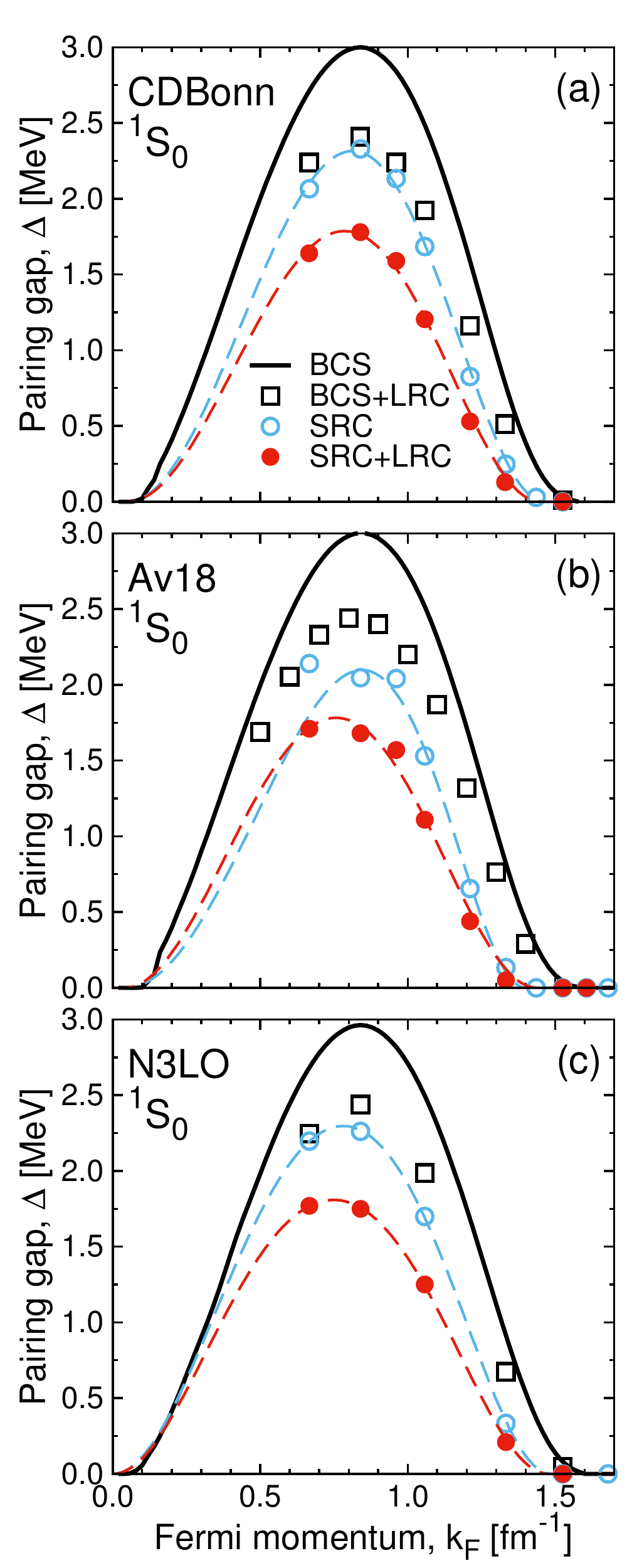}
\end{center}
\caption{Pairing gaps at the Fermi surface as a function of Fermi momentum in the $^1$S$_0$ channel. The three panels correspond to the (a) CDBonn, (b) Av18 and (c) N3LO interactions. Results for different approximations are presented: BCS (solid lines), beyond BCS with short-range correlations (empty circles), beyond BCS with long-range correlations (empty squares) and beyond BCS with both short- and long-range correlations included (solid circles). The dashed lines represent the fits provided in Table~\ref{table:1S0}. \label{fig:gap1s0} }
\end{figure}

Figure~\ref{fig:gap1s0} provides the pairing gaps in the $^1$S$_0$ channel computed at the corresponding Fermi surfaces in neutron matter for four different approximations. Results for the CDBonn, Av18 and N3LO NN forces are displayed in panels (a)-(c), respectively. Solid lines represent the standard BCS solution, computed using free sp spectra. The BCS+LRC results (empty squares) have been obtained by adding the dressed effective interactions of Eq.~(\ref{eq:veffS0}) to the bare NN forces in the gap equation, Eq.~(\ref{eq:gap}). SRC results (empty circles) are computed with bare NN forces, but double convolution denominators in the gap equation. Finally, the full circles are obtained from the full denominators and LRC-corrected effective interactions. 

The BCS results (solid line) are very similar for all forces, which confirms that phase-shift equivalence is enough to fix the value of the gap in this channel~\cite{Dean2003}. The BCS gap peaks at about $3$ MeV around $k_F=0.7$-$0.8$ fm$^{-1}$, and closes at $k_F \approx 1.5$ fm$^{-1}$. As mentioned above, LRC in this channel screen part of the attraction of the NN forces. Consequently, BCS+LRC gaps (squares) are generally smaller than BCS results. While the overall Fermi momentum dependence is similar, including a similar closure density, the maximum of $\Delta$ decreases to about $\approx 2.5$ MeV. It is important to stress that the screening is the same for all interactions. 

Including SRC within the Green's function formalism outlined above (empty circles), we find that the overall gap is reduced, with a maximum that now sits just above $2$ MeV. This result is expected: by removing strength from the Fermi surface, the pairing phase space is quenched and the corresponding pairing gap decreases by about $30 \%$. The mild density dependence of $Z_\text{eff}$ also explains why the BCS and the SRC results have similar density dependences. There is a tendency to have a slightly lower closure density for the SRC results, a feature we shall discuss further when we introduce numerical parametrizations below.

Finally, the complete results including both SRC and LRC are shown in full circles in Fig.~\ref{fig:gap1s0}. Screening effects in this channel are repulsive, and as a consequence the corresponding LRC+SRC gaps decrease in size by about $25 \%$ with respect to the SRC only data. These gaps peak at values of around $1.8$ MeV, for Fermi momenta close to $0.75$ fm$^{-1}$. While the overall density dependence is comparable to the previous results, we note a tendency to find a lower gap closure density. 

For a given channel, a convenient parametrization of the density dependence of the gap function is given by:
\begin{align}
\Delta^{JST}_L(k_F) = \Delta_0 \frac{ (k_F-k_0)^2}{(k_F-k_0)^2+k_1} \frac{ (k_F-k_2)^2}{(k_F-k_2)^2+k_3} \, ,
\label{eq:gapfit}
\end{align}
with $\Delta_0$, $k_0$, $k_1$, $k_2$ and $k_3$ numerical parameters \cite{Ho2015}.  In particular, $k_0$ and $k_2$ represent the Fermi momenta at which the gap opens and closes, respectively. Details on the numerical fit to this function are given in Appendix~\ref{app:fits}. We note that this parametrization is particularly sharp around the closure points, and that in the singlet channel we supplement the fit with a zero value at zero density. We show in Table \ref{table:1S0} the values of the parameters obtained for these fits. Further, we note that the dashed lines shown in Fig.~\ref{fig:gap1s0} correspond to the fit functions. 

The fit does reproduce the qualitative shape of the pairing gap. We take $k_2$,  displayed in column 5 of Table~\ref{table:1S0}, as a measure of the gap closure density. The confidence interval associated to the fit is within $0.12$ fm$^{-1}$ (for the worst fit) from the central value. We note that there is a robust agreement and for all forces and many-body approaches the gap closure sits between $1.4$ and $1.5$ fm$^{-1}$. 

This parametrization also allows a simple quantitative estimate of the gap maxima, and their location. For instance, the SRC maximum gap lies between $k_F=0.78$ (N3LO), $0.81$ (CDBonn) and $0.84$  fm$^{-1}$ (Av18), at a value between $\Delta_\text{max}^{^1S_0}=2.1$ MeV (Av18) and $2.3$ MeV (CDBonn and N3LO). Similarly, the SRC+LRC results peak between  $k_F=0.75$ (N3LO), $0.76$ (Av18) and $0.79$ fm$^{-1}$ (CDBonn) to maximum gaps of the order of $\Delta_\text{max}^{^1S_0}=1.8$ MeV for all three NN interactions.

We note that similar gaps have already been obtained in the literature. A comparison with the compilation of Ref.~\cite{Ho2015} shows that our results are close to the Cao-Lombardo-Schuck (CLS)~\cite{Cao2006} and Margueron-Sagawa-Hagino (MSH)~\cite{Margueron2008} singlet gaps. MSH is fit to the CLS results, so the agreement between the two is not surprising. Our results include LRC in a way that is similar to CLS, but we note that the SRC physics is considered only at the $\mathcal{Z}-$factor level in a Brueckner--Hartree--Fock calculation, and hence misses a complete description of hole-hole correlation effects.
 
\begin{table}[t!]
\caption{Parameters generated by a fit to the calculated gaps for the CDBonn, Av18, and N3LO interactions in the $^1$S$_0$ channel. For each interaction, the first line contains the results for the inclusion of SRC only, and the second the effect of both SRC and LRC. \label{table:1S0}}
\centering 
\begin{tabular}{cccccc} 
\hline \hline 
Singlet & $\Delta_0$ & $k_0$ & $k_1$ & $k_2$ & $k_3$ \\
 & [MeV] & [fm$^{-1}$] & [fm$^{-2}$] & [fm$^{-1}$] & [fm$^{-2}$]  \\
 \hline
CDBonn SRC & 26.59 & 0.05 & 1.79 & 1.46 & 0.76 \\ 
CDBonn SRC+P & 18.18 & 0.05 & 1.39 & 1.45 & 0.81 \\ 
Av18 SRC & 32.22 & 0.04 & 3.46 & 1.40 & 0.43 \\ 
Av18 SRC+P & 14.07 & 0.04 & 1.00 & 1.44 & 0.78 \\ 
N3LO SRC & 7.77 & 0.00 & 0.56 & 1.49 & 0.38 \\ 
N3LO SRC+P & 5.85 & 0.00 & 0.46 & 1.48 & 0.42 \\ 
\hline 
\end{tabular} 
\end{table}

All in all, the picture that arises for the singlet gaps provided by the three interactions is \emph{remarkably robust}. The small variation of the result among NN forces provides an insightful constraint on the model dependence of the gap properties. In spite of their different short-range and (less relevant for neutron matter) tensor components, the 3 interactions considered here predict singlet gaps which are very close to each other. More importantly, the many-body effects are very similar in all cases. SRC deplete the gap by about $25 \, \%$. When LRC are included on top of SRC, the gap that remains is around $60 \, \%$  of the original BCS result for all forces. The effect of the correlation-induced gap quench in pairing properties, like the Cooper pair coherence length~\cite{deBlasio1997}, or neutron star properties~\cite{Ho2015}, go beyond the scope of the present paper and will be studied elsewhere.

The robustness of the singlet gap results with and without correlation effects is one of the major conclusions of this work. We note, however, that this result is not necessarily easily anticipated. The gap itself is a function of both Fermi momentum, $k_F$, and sp momentum, $k$. So far, we have focused on the values at the Fermi surface, $\Delta^{JST}_L(k=k_F)$, but the momentum dependence provides useful information, too. In particular, as we are about to show, very different momentum dependences can lead to similar gaps at the Fermi surface. 

\begin{figure}[t!]
\begin{center}
\includegraphics[width=0.6\linewidth]{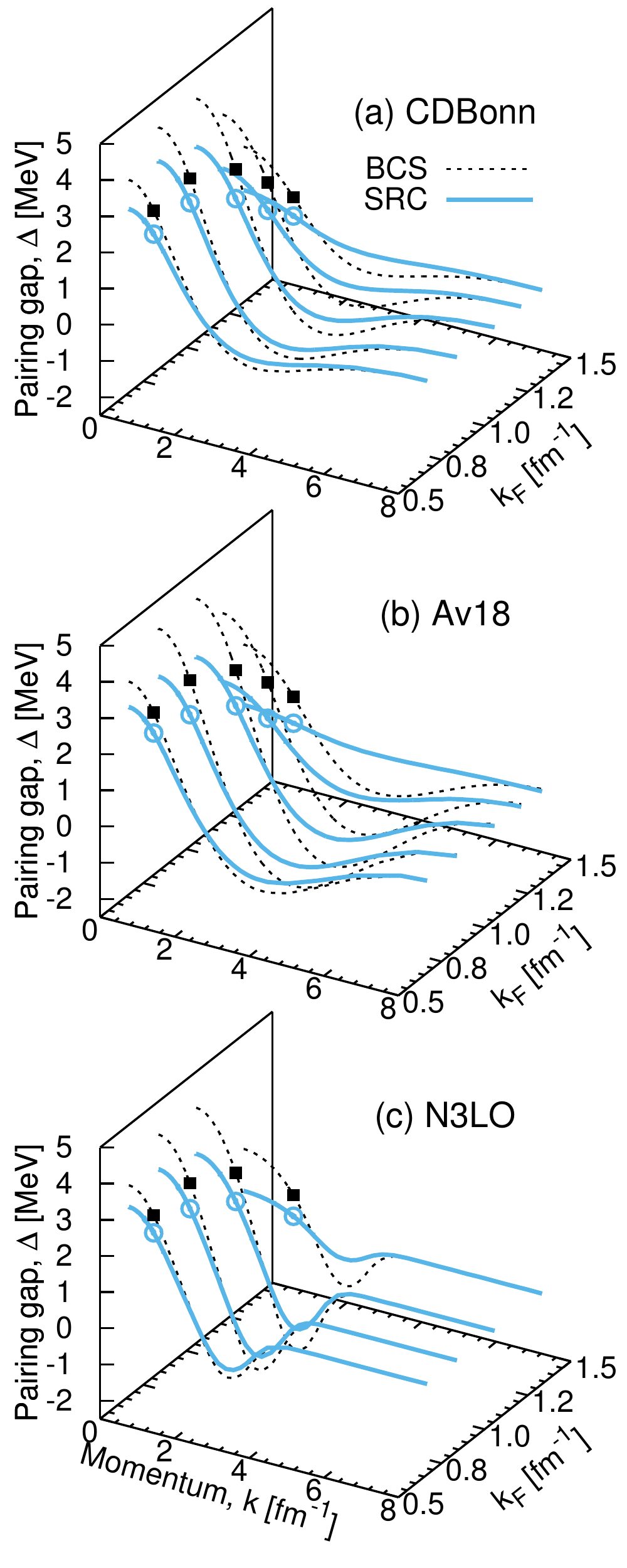}
\end{center}
\caption{Pairing gaps in the $^1$S$_0$ channel as a function of momentum for several Fermi momenta. The three panels correspond to the (a) CDBonn, (b) Av18 and (c) N3LO interactions. The BCS (dashed lines) and SRC (solid lines) results are presented. The symbols represent the corresponding gaps at the Fermi surface, as shown in Fig.~\ref{fig:gap1s0}. \label{fig:gap1s0_2d} }
\end{figure}

The pairing gap is shown as a function of the momentum, $k$, for a variety of Fermi momenta in the fence plots of Fig.~\ref{fig:gap1s0_2d}. Panels (a), (b) and (c) correspond to the three NN forces, CDBonn, Av18 and N3LO, respectively. We show results for both the BCS (dotted lines) and SRC-only (solid lines) approximations. In all cases, as expected in the singlet case, the gap peaks at zero momentum, subsequently decreases to negative values beyond the Fermi surface, and ultimately approaches zero asymptotically. In contrast to the results of N3LO in panel (c), both CDBonn and Av18 show a non-zero gap up to large values of momentum, $\approx 6-7$ fm$^{-1}$. In principle, N3LO is regularized at a momentum scale of $500$ MeV~\cite{Entem2003}, and hence it is not surprising to find that there is no support for a gap beyond $\approx 3$ fm$^{-1}$. 

The significant differences in high-momentum components do not alter the low-momentum gap. The values at the Fermi surface are close to each other, and indeed the momentum dependence is qualitatively similar for all three forces up to $k\approx2.5$ fm$^{-1}$. The general effect of SRC is to tame this momentum dependence. The maximum of the gap decreases, and its negative minimum becomes less negative. In accordance with some of the ideas mentioned above, the quench of the momentum dependence is relatively density- and momentum independent. 

\subsection{Triplet pairing}
\label{sec:results3PF2}

\begin{figure}[t!]
\begin{center}
\includegraphics[width=0.6\linewidth]{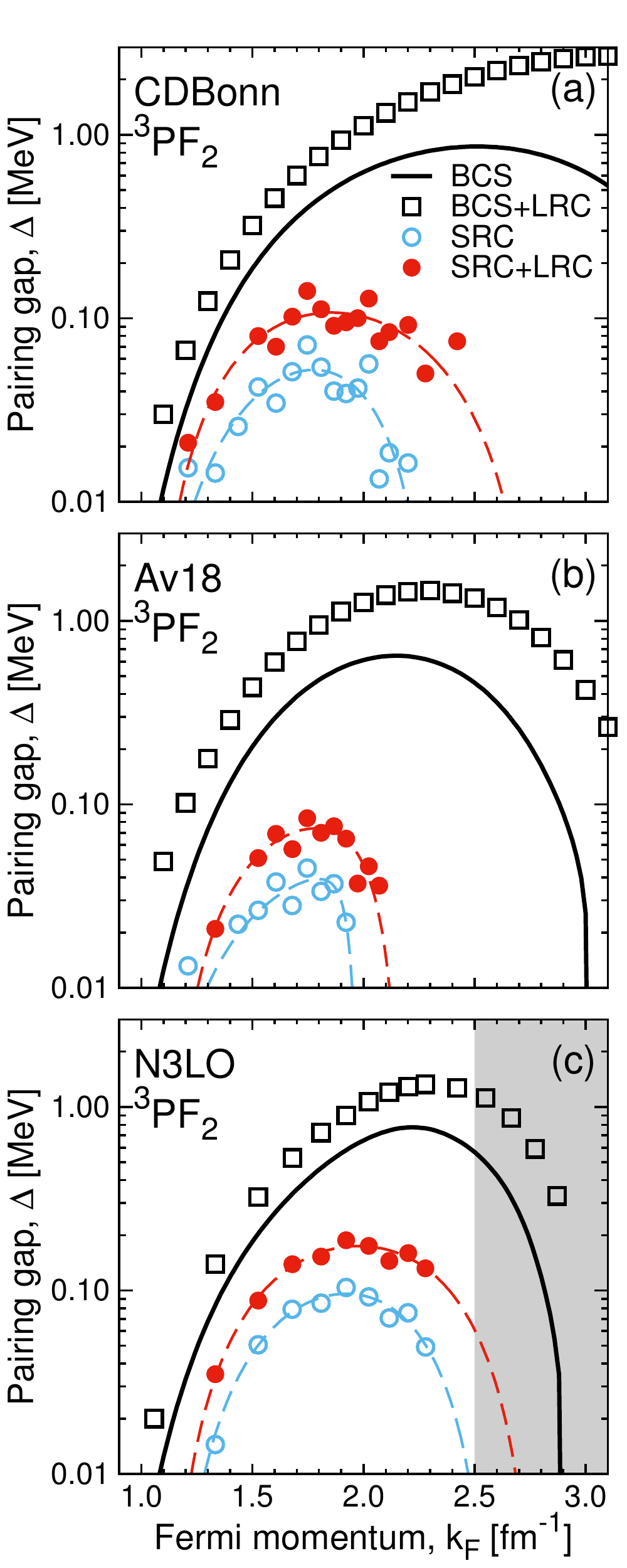}
\end{center}
\caption{The same as Fig.~\ref{fig:gap1s0} for the coupled $^3$PF$_2$ channel. The grey band on panel (c) indicates the region in which cut-off effects are relevant for the N3LO force. \label{fig:gap3pf2} }
\end{figure}

Whereas singlet pairing is active at relatively low densities and affects the dynamics of both the core and the crust, triplet pairing, concerning the coupled $^3$PF$_2$ channel, takes place within the neutron star core, at Fermi momenta  $k_F>1$ fm$^{-1}$~\cite{Ho2015,Page2011}. Higher Fermi momenta imply that higher relative momenta are explored in the bare (or the effective) interaction. Because phase-shift equivalent interactions are constrained only at low energies and relative momenta, it is not surprising that the corresponding gaps show a larger dependence on the NN force. We show the triplet gaps for three NN forces in panels (a)-(c) of Fig.~\ref{fig:gap3pf2}. We use a logarithmic scale to discriminate better the results of different many-body approximations. 

Differences between NN forces are already significant at the BCS level (solid lines). All gaps open at Fermi momenta around $k_F \approx 1.1$ fm$^{-1}$. Triplet gaps peak around $2$-$2.5$ fm$^{-1}$. CDBonn provides the largest maximum ($\Delta^{^3PF_2}_\textrm{max} = 0.86$ MeV at $k_F\approx2.5$ fm$^{-1}$), followed by N3LO ($\Delta^{^3PF_2}_\textrm{max} = 0.77$ MeV at $2.2$ fm$^{-1}$) and Av18 ($\Delta^{^3PF_2}_\textrm{max} \approx 0.64$ MeV at $k_F=2.1$ fm$^{-1}$). In turn, the gap closure happens at larger densities for interactions with larger gaps. For CDBonn, the closure occurs at a very high density, $k_F \approx 3.64$ fm$^{-1}$, beyond the limit of Fig.~\ref{fig:gap3pf2}. Av18 and N3LO, instead, provide BCS closure momenta between $2.9$ and $3$ fm$^{-1}$. 

\begin{figure}[t!]
\begin{center}
\includegraphics[width=\linewidth]{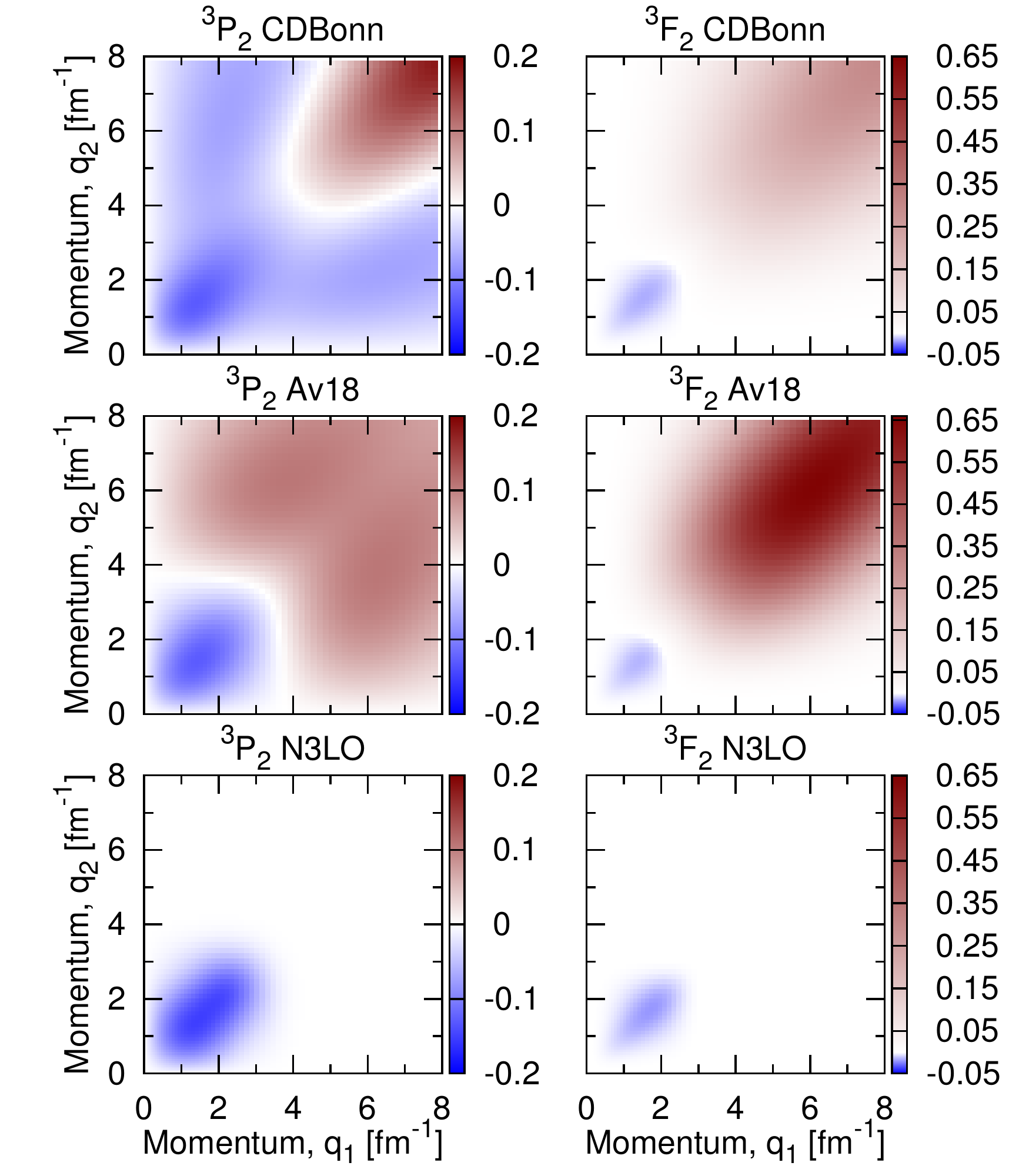}
\end{center}
\caption{Left panels: matrix elements of NN forces in fm for the $^3$P$_2$ partial wave. Right panels: the same for the $^3$F$_2$ wave. \label{fig:NNpots} }
\end{figure}

N3LO results are sensitive to the cut-off regularization of the NN force at large Fermi momenta \cite{Maurizio2014}. The grey band in panel (c) indicates the region in which regularization effects become observable. It is important to stress that the two similar gap closures for Av18 and N3LO are indicative of two very different physical effects. On the left (right) panels of Fig.~\ref{fig:NNpots} we show density plots for the $^3$P$_2$ ($^3$F$_2$) matrix elements of the three NN forces. A gap can only appear if attractive matrix elements of $\mathcal{V}$ are available. Consequently, N3LO can only sustain a gap up to about $\approx 3$ fm$^{-1}$ because its matrix elements are regularized, and hence tend to zero, beyond this momentum. In contrast, Av18 does have non-zero, rather repulsive matrix elements beyond about $\approx 4$ fm$^{-1}$. It is the appearance of these repulsive matrix elements that forbids pairing above the closure momentum for Av18. The top panels also illustrate why CDBonn sustains gaps up to larger Fermi momenta: the attractive nature of its $P-$wave matrix elements covers a large relative momentum region.

In contrast to the singlet case, the inclusion of LRC on top of the BCS result leads to higher gaps. This result is in line with the discussion of Sec.~\ref{sec:LRC}, because in this channel LRC are attractive and anti-screen the interaction. The effect is rather significant, with the Av18 BCS+LRC gap becoming more than a factor of $2$ larger than the original BCS result. The density dependence is also modified by LRC.

When SRC are considered (empty circles), all $^3$PF$_2$ gaps are strongly suppressed. All maximum triplet gaps fall below $0.15$ MeV: $0.04$ MeV for Av18, $0.05$ for CDBonn and about $0.1$ MeV for N3LO. The density dependence of these gaps is also different than the BCS prediction. We note that the data for the SRC and SRC+LRC is relatively noisy, due to the numerical limitations of the zero temperature extrapolation discussed in Appendix~\ref{app:T0}. Nevertheless, the gross features of the SRC effects are rather clear. 

If we take the $k_0$ parameter of the fits presented in Table~\ref{table:3PF2} as an indication of gap opening, triplet gaps start at $k_F \approx 1.1$ fm$^{-1}$ for all forces. The corresponding gap maxima occur at $k_F=1.77$, $1.79$ and $1.92$ fm$^{-1}$ for CDBonn, Av18 and N3LO, respectively. This is to be compared to the substantially larger BCS values of $\approx 2-2.5$ fm$^{-1}$. Finally, the gap closure occurs at lower Fermi momenta for the SRC gaps (see column 5 in Table~\ref{table:3PF2}) than the corresponding BCS results. N3LO predicts the larger closure at $k_F \approx 2.8$ fm$^{-1}$, whereas the lowest closure is given by Av18 at $2.0$  fm$^{-1}$. All in all, SRC triplet gaps are smaller and exhibit a smaller density range than their corresponding BCS counterparts. 

\begin{table}[t!]
\caption{Parameters generated by a fit to the calculated gaps for the CDBonn, Av18 and N3LO interactions in the $^3$PF$_2$ channel. For each interaction, the first line contains the results for the inclusion of SRC only, and the second the effect of both SRC and LRC.\label{table:3PF2}}
\centering 
\begin{tabular}{cccccc} 
\hline \hline 
Triplet & $\Delta_0$ & $k_0$ & $k_1$ & $k_2$ & $k_3$ \\
 & [MeV] & [fm$^{-1}$] & [fm$^{-2}$] & [fm$^{-1}$] & [fm$^{-2}$]  \\
 \hline
CDBonn SRC & 0.60 & 1.01 & 2.21 & 2.33 & 0.43 \\ 
CDBonn SRC+P & 0.41 & 1.03 & 0.56 & 2.81 & 1.00 \\ 
Av18 SRC & 0.09 & 1.01 & 0.64 & 1.98 & 0.005 \\ 
Av18 SRC+P & 0.17 & 1.10 & 0.35 & 2.18 & 0.05 \\ 
N3LO SRC & 0.43 & 1.13 & 0.83 & 2.59 & 0.41 \\ 
N3LO SRC+P & 0.60 & 1.11 & 0.69 & 2.79 & 0.53 \\ 
\hline 
\end{tabular} 
\end{table}

LRC, when considered in addition to SRC (solid circles), do not change the picture qualitatively. In this channel, spin-density fluctuations lead to a more attractive pairing interaction, and hence LRC increase the triplet pairing gap by a small percentage. LRC+SRC start at similar Fermi momenta than their SRC-only counterparts. The maximum gap that is produced, however, is almost twice as large: $0.17$ MeV for N3LO, $0.11$ for CDBonn and $0.07$ MeV for Av18. The corresponding Fermi momentum maxima are similar, $k_F=1.86$, $1.79$ and $1.98$ fm$^{-1}$ for CDBonn, Av18 and N3LO, respectively. In keeping with the larger maxima, gap closures also occur at larger Fermi momenta, with two interactions (CDBonn and N3LO) closing the gap at $k_F=2.8$ fm$^{-1}$, and the remaining one, Av18, at $2.2$ fm$^{-1}$. 

All in all, our prediction for triplet gaps are also reasonably robust, and independent of the NN force. Triplet gaps are always at the level of $10$s of keV, peaking at Fermi momenta in the region $1.7-2$ fm$^{-1}$ and closing earlier than the BCS predictions. To compare with previous literature in similar conditions, we consider the results of Dong \emph{et al.} in Ref.~\cite{Dong2013}. They use the Av18 interaction and parametrize SRC in terms of $\mathcal{Z}-$ factors. The maximum triplet gap in that calculation is $0.045$ MeV at $k_F\approx1.6$ fm$^{-1}$, in very good agreement to our SRC result ($0.04$ MeV at $1.79$ fm$^{-1}$). We note that none of the gaps considered in the recent astrophysically motivated compilation of Ref.~\cite{Ho2015} resemble our predictions. However, the maximum gaps that we produce compare well with the inferred value of triplet critical temperatures of Ref.~\cite{Page2011}, $T_c = 5 \times 10^8 \text{ K} \Rightarrow \Delta_\text{max}^{^3PF_2} \approx 0.08 \text{ MeV}$.

\begin{figure}[t!]
\begin{center}
\includegraphics[width=0.6\linewidth]{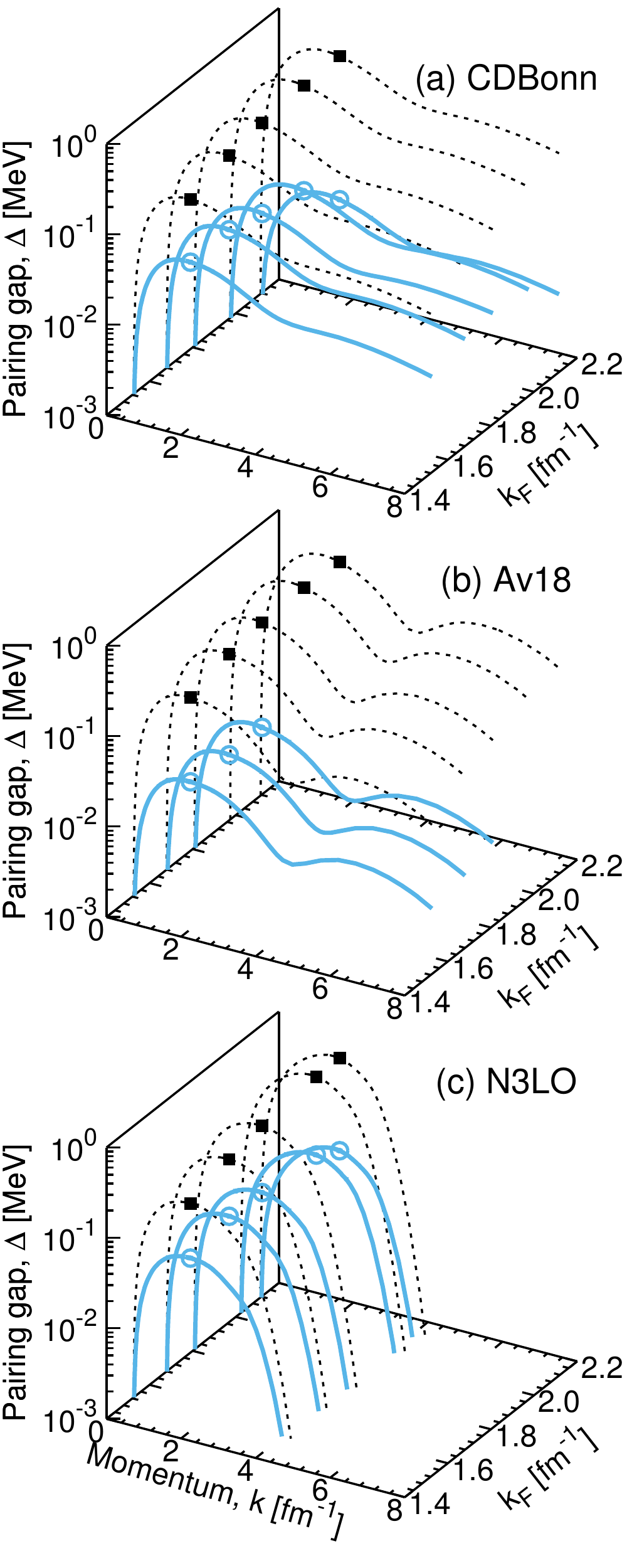}
\end{center}
\caption{The same as Fig.~\ref{fig:gap1s0_2d} for the pairing gaps in the  $^3$PF$_2$ channel. Note the logarithmic scale on the $z$ axis. \label{fig:gap3pf2_2d} }
\end{figure}

The agreement between the different NN interactions is not trivial, as we have discussed in the singlet case. We have already illustrated the very different momentum-space structure of the triplet components of the NN forces in Fig.~\ref{fig:NNpots}. Further, we present in Fig.~\ref{fig:gap3pf2_2d} the momentum dependence of the triplet gap components for several Fermi momenta. We use again a logarithmic plot to distinguish better all the presented results. The BCS predictions (dotted lines) begin at zero, as expected from non-$S$-wave pairing, peak close to the Fermi surface, and subsequently decay. While CDBonn and Av18 show an inflection point at $k\approx4$ fm$^{-1}$ and decay slowly with momentum, N3LO decays to zero for momenta well below this value. The corresponding SRC gaps show qualitatively similar behaviors and peaks, but are generally an order of magnitude smaller. There are differences in the density dependence, too. In any case, this figure illustrates the fact that, unlike the singlet case, triplet pairing gaps are more sensitive to the short-range (or, equivalently, high-momentum) components of the NN force.

\subsection{Three-neutron forces}
\label{sec:results3NF}

\begin{figure}
\begin{center}
\includegraphics[width=0.6\linewidth]{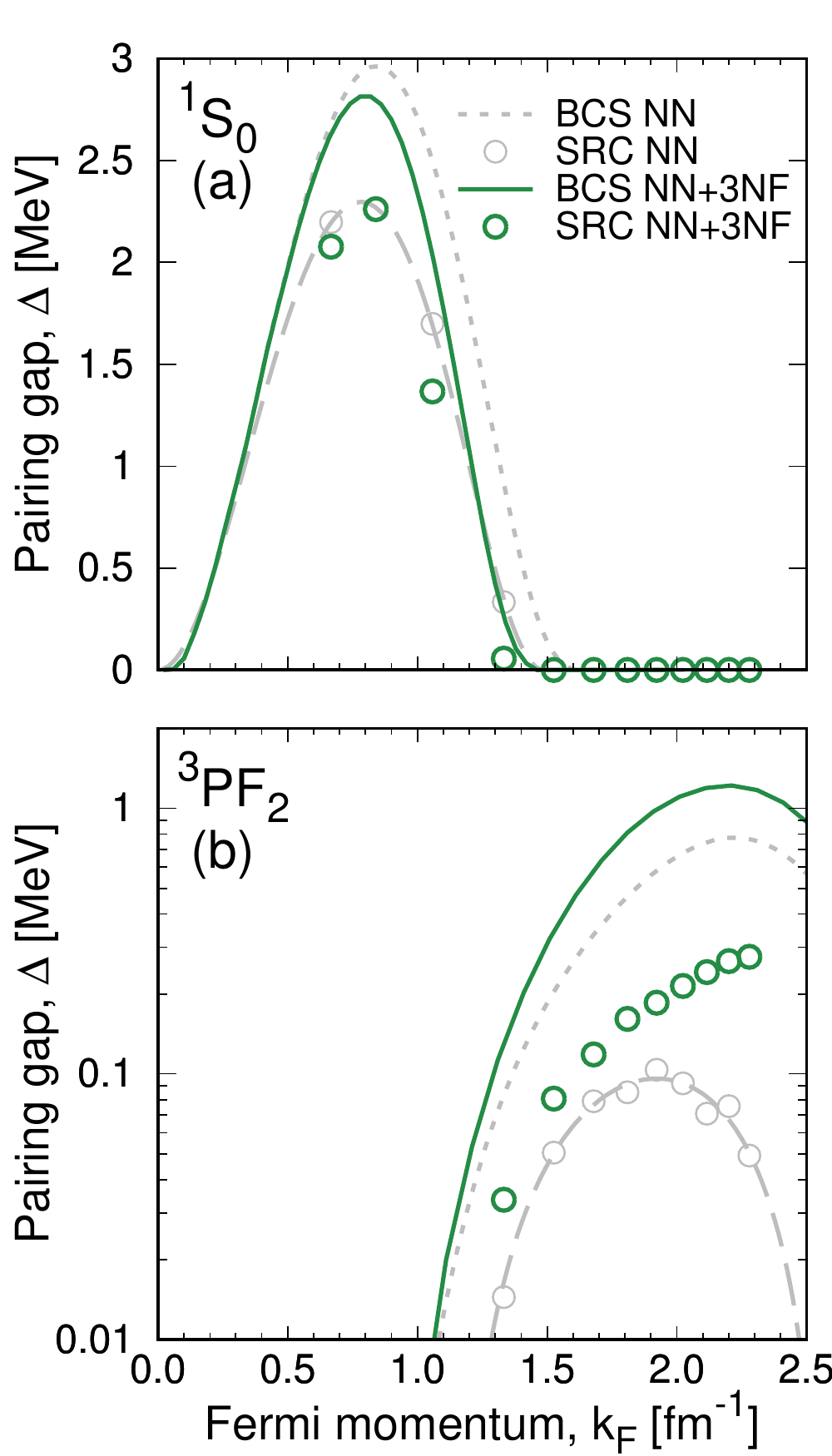}
\caption{Pairing gaps at the Fermi surface as a function of Fermi momentum in the $^1$S$_0$ [panel (a)] and $^3$PF$_2$ [panel (b)] channels. Results for different approximations are presented for the chiral N3LO Idaho NN force in the BCS (dotted lined) and BCS+SRC approximation (light circles). Results including 3NF (2N+3NF) are given in a solid (bold circles) line for the BCS (BCS+SRC) approximation. See text for details. \label{fig:gap_3NF}}
\end{center}
\end{figure}

The results discussed so far do not include any 3NF effects. We postpone a detailed discussion of specific 3NF effects to a future publication, but present here some preliminary results. These illustrate qualitatively the role played by 3NFs in pairing properties. We note that the effects are particularly small in the singlet channel. We use chiral N2LO 3NF, which at the neutron matter level, are predicted from the NN force without the need to fit any further low energy constants. Compared to the treatment in Ref.~\cite{Carbone2014}, the effective density-dependent two-body interaction is obtained from an uncorrelated average. The non-local regulator affects the integrated momentum variable, and has a 3NF cut-off $\Lambda_\text{3NF}=500$ MeV~\cite{Carbone2014}. Off-diagonal momentum matrix elements are obtained with the prescription introduced in Ref.~\cite{Holt2010}. 

3NF affect our calculations at two different levels. First, the effective pairing interaction itself is modified. At the singlet pairing level, one expects a repulsive effect that will reduce the gap \cite{Holt2010,Maurizio2014}. For triplet pairing, chiral N2LO forces produce attractive components that, in general, enhance the gap \cite{Carbone2014}. These aspects become particularly clear at the BCS level where, in our treatment, the only difference between NN and NN+3NF calculations are the effective interaction themselves.

The situation is illustrated in Fig.~\ref{fig:gap_3NF}. At the BCS level, the Fermi surface gap with NN forces only (dotted line) for the singlet [panel (a)] and triplet [panel (b)] channels is the same that has been presented in Figs.~\ref{fig:gap1s0} and ~\ref{fig:gap3pf2}, respectively. The solid lines, in contrast, are obtained including 3NFs within the BCS approach. For the singlet, one finds a decrease in the maximum gap of about $0.2$ MeV, and a narrower gap. Gap closure occurs around $k_F \approx 1.5$ fm$^{-1}$ when 3NFs are considered, instead of $1.6$ fm$^{-1}$. For the triplet, in contrast, the gap increases at all densities. The maximum gap, in this case, goes from $0.77$ MeV (NN only) to $1.21$ MeV (NN+3NF), reflecting the attractive nature of 3NFs in this channel. 

In addition to the pairing interaction, 3NFs affect our calculations via the the double convolution denominators. Changes in spectral functions from the self-consistent calculations will induce variations in gaps within the SRC approximation. We find that 3NFs modify quasi-particle energy peaks more than they modify the spectral functions widths \cite{Carbone2014}. These considerations are density-dependent, as expected. At the low densities relevant for singlet pairing, 3NFs are less important and the difference between NN and NN+3NF calculations should be small. At densities relevant for triplet pairing, but below the chiral cut-off, we find that the quasi-particle energies are shallower as a function of momentum, an effect that tends to enhance the gap. 

These intuitive features are reflected in the SRC results presented in Fig.~\ref{fig:gap_3NF}. For the singlet SRC gap, the results with and without 3NF agree well below $k_F \approx 1$ fm$^{-1}$. This indicates a small effect of 3NF. Above this density, the reduction of the gap associated to 3NF is similar in the BCS and in the SRC approximations. This suggests that, for the singlet channel, the gap reduction is due to the change in interaction. In contrast, SRC results for the triplet gap are already different close to gap opening. 3NF are active in the whole density range where triplet gaps are relevant. As density increases, we find that the SRC gap with 3NF is larger and broader than the NN-only counterpart. As seen in the BCS case, the effect of 3NF in the interaction should be to increase the gap slightly. Our SRC results, however, show no sign of gap closure at large densities ($k_F \approx 2.5$ fm$^{-1}$) as a consequence of the softening of the quasi-particle spectrum. We caution that at these high densities one approaches the limits of applicability of the employed chiral interactions. An implementation of these chiral forces with a larger cut-off could clarify the situation.

In both the singlet and triplet cases, SRC substantially reduce the pairing gap. The bulk of this effect, associated to the removal of strength around the Fermi surface, is universal and independent of NN or 3N forces. In the density regime where singlet gaps are relevant, the removal of strength is not affected by 3NFs. LRC, as implemented here, are independent of 3NF. They will produce an (anti-) screening effect in the (triplet) singlet which is of the same order of magnitude as in the NN-only case. As we have argued so far, the LRC effect is subdominant compared to SRC for NN forces, and we expect to find very similar results if LRC were implemented in the full 3NF framework. Full results including 3NFs will be presented in the near future.

\section{Conclusions}
\label{sec:conclusions}

We have proposed a method that combines the SCGF framework for the treatment of SRC and FLT for the incorporation of LRC to the neutron matter pairing problem with NN interactions. This approach has two major ingredients. On the one hand, extrapolated normal ladder self-energies provide access to zero temperature spectral functions and, in turn, these give rise to depleted energy denominators that quench the gap. On the other, the pairing interaction is treated beyond the static, bare level. Screening is provided by considering vertices represented by Landau parameters, that couple to spin and density oscillations, whose collectivity is also controlled by the same Landau parameters. 

The effect of SRC is to remove strength from the Fermi surface, thus necessarily reducing the gap. Compared to a quasi-particle, BCS-like picture, the energy denominators in the gap equation are quenched by a relatively momentum and density independent factor. In general, this is very different from the corresponding $\mathcal{Z}$-factors associated to the ladder SCGF self-energy. This indicates that pairing calculations with $\mathcal{Z}$-factors do not consider the full effect of SRC in a consistent way. 

We take three major conclusions from our work. First, the universal effect of SRC is to deplete the gap substantially with respect to its BCS value in the whole momentum range. In the singlet channel this translates into a decrease in the Fermi surface gap of about $10-15 \%$. In the triplet channel, the gaps were small, below $1$ MeV, at the BCS level. When SRC are considered, gaps decrease further to below $0.2$ MeV throughout a wide density regime. Second, whereas for the singlet case the effect of SRC is of the same order of the screening provided by LRC, in the triplet case LRC have an anti-screening effect that modestly increases the SRC determined gap for all three NN interactions. Third, the density dependence of triplet gaps is substantially modified by  SRC and LRC. We find gaps that open above $1.2$ fm$^{-1}$ and close below $2.6$ fm$^{-1}$ in all cases, with maxima that hardly reach $0.2$ MeV. Small triplet gaps of a similar size are commensurate with the Cassiopeia A rapid cooling scenario presented in Ref.~\cite{Page2011}. 

We have performed calculations with three very different, but phase-shift equivalent, interactions. We have also presented preliminary calculations including the effect of 3NFs. For the singlet channel, our conclusions are extremely robust and independent of the NN force. Triplet gaps, in contrast, depend on the specifics of the interaction itself. CDBonn in general provides the largest and widest triplet gaps, whereas Av18 provides small and narrow pairing gaps. Cutoff effects artificially cut the triplet gaps of N3LO above $2.5$ fm$^{-1}$. At the high densities involved in triplet gaps, 3NF are important, and tend to increase the gap at the BCS and SRC level. We want to stress, however, that the SRC and LRC effects are universal and independent of the nuclear force under consideration. Work on incorporating 3NFs in a consistent way and a subsequent detailed discussion of 3NF-induced effects, following Ref.~\cite{Carbone2014}, is our first priority in the near future.

The extension of the approach to asymmetric nuclear systems is also important~\cite{Alm1993,Lombardo2001}. There is a small admixture of protons inside neutron stars, and their pairing is relevant for neutron star matter. In-medium SRC effects should be similar for proton pairing, and the suppression in the proton channel might have consequences for neutron star cooling. Pairing at finite momentum is also a relevant physical phenomena, particularly since it can lead to different pairing phases~\cite{Stein2014}. The interplay of correlation and finite momentum effects will necessarily lead to a change in the phase diagram with respect to BCS results.

These calculations represent a first controlled step towards a full treatment of superfluidity within the Green's function formalism. At the SRC level, our treatment does not allow for the superfluid phase to feed back into the determination of the normal propagators. While feedback effects will be small, the reformulation of the problem in a Gorkov context would avoid the need of extrapolations from finite temperature. Such a self-consistent treatment of the ladder approximation in the pairing phase has never been implemented to our knowledge. At the LRC level, consistency at the Landau parameter level could provide small, quantitative differences in our results. Furthermore, the full spin dependence of the effective interaction, beyond the traditional Landau parameters~\cite{Schwenk2004}, could have an impact on pairing gaps. Finally, the inclusion of polarization effects beyond the low-momentum transfer limit is an interesting, if computationally expensive, possibility.

\appendix

\section{Numerical treatment of the temperature extrapolation}
\label{app:T0}
\begin{figure}[t!]
\begin{center}
\includegraphics[width=0.8\linewidth]{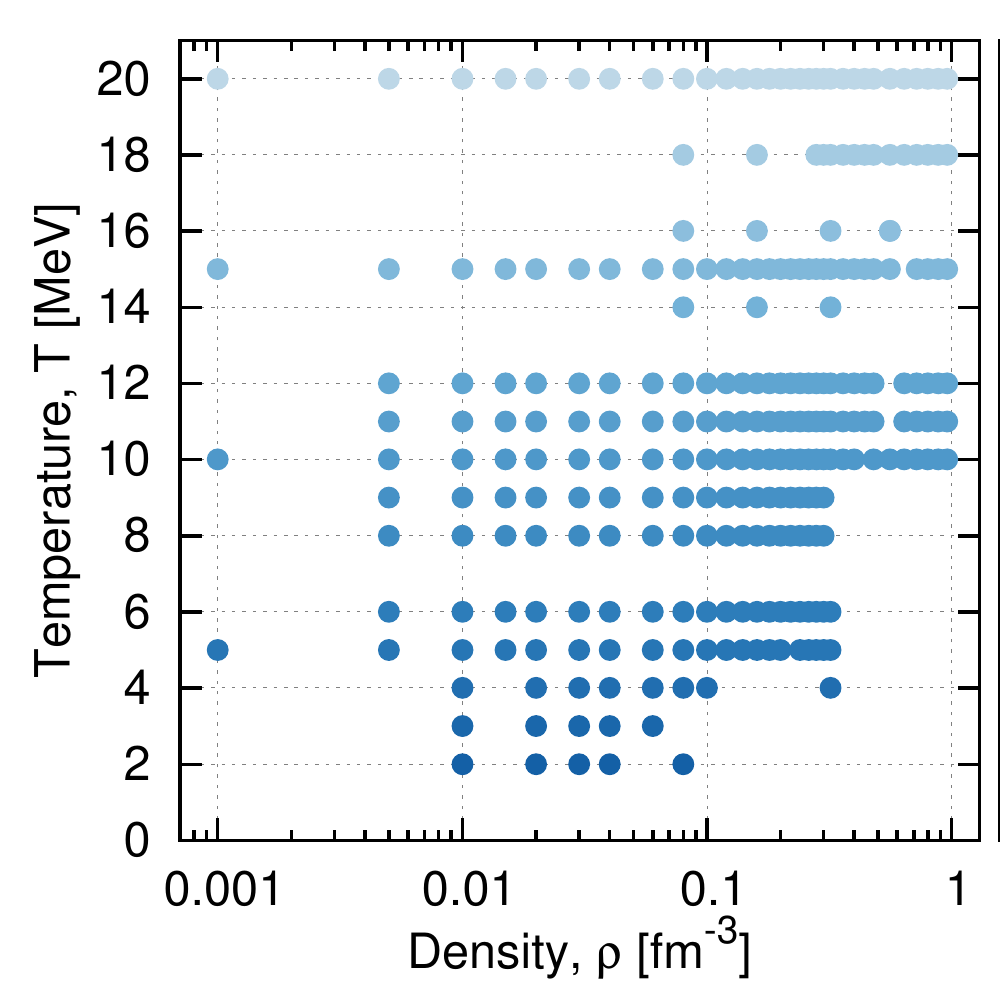}
\end{center}
\caption{\label{fig:data} Each point in this plot corresponds to a density and temperature where ladder self-energies have been computed. Finite temperature points are used to extrapolate to zero temperature.}
\end{figure}

For a given density, ladder self-energy calculations are typically performed for a set of $N_T\approx 3$ to $10$ temperatures. The degeneracy parameter, $\zeta = \frac{T}{\epsilon_F}$, with $\epsilon_F$ the non-interacting Fermi energy, is a proxy for temperature in Fermi gases and is a natural dimensionless extrapolation parameter, in accordance to the Sommerfeld expansion \cite{Ashcroft1976}. At each $\zeta$, the real and imaginary parts of the self energy are stored as arrays in energy and momentum space. Typically, between $4000$ to $11000$ energies are needed, whereas we work with a fixed set of $70$ points in the momentum mesh. The self-energy is fit by a polynomial function of $\zeta$,
$ \Sigma(k,\omega; \zeta) = \sum_{l=0}^{L} a_l(k,\omega) \zeta ^{2l} \, ,$
 in a window of $\zeta$ values. For a given density, we take an upper limit of $\zeta \approx 1$ and a lower limit of $\zeta  \gtrsim 0.07$ (as long as the pairing instability does not set in). This ensures that the finite temperature data is neither thermally dominated ($\zeta \gg 1$) nor insensitive to thermal effects ($\zeta \ll 1$). Fig.~\ref{fig:data} provides an illustration of the density and temperature mesh that we have used for the extrapolations with Av18.
 
The zero-temperature self-energy is the independent coefficient of the polynomial fit, $a_0(k,\omega)$, although in principle the fit can also be used to compute self-energies at arbitrary temperatures. For practical purposes, the interpolation involves only even powers, $2l$. Close to the Fermi surface, the temperature dependence of the self-energy is indeed expected to be quadratic \cite{Abrikosov1965}. However, numerically computed self-energies, particularly close to $\omega=\mu$, need not be soft, continuous functions of degeneracy, due to numerical noise. Consequently, a single fit might extrapolate quantities in an unphysical way. For this reason, we perform not only one, but a series of fits with different values of $L$, the maximum power of the polynomial. Generally, we go from $L=1$ (corresponding to a $T^2$ dependence) to about $L=4$, depending on the total number of temperatures available.  All polynomial fits are performed using a $\chi^2$ minimization procedure, which helps in evening out any numerical noise.
 
In the implementation, we extrapolate separately the imaginary and the real parts of the self-energy. Pairing is sensitive to the properties of $\Sigma$ close to the Fermi energy. This is where the temperature dependence is more difficult to capture with fits. For a given polynomial order, $L$, we therefore allow for two different options. We either take the extrapolated $\text{Im} \Sigma$ as face value ($\text{Im} \Sigma \le 0$ is imposed throughout, though) or we allow for a second option, where we match $\text{Im} \Sigma$ to the analytic function:
\begin{align}
\text{Im} \Sigma(k,\omega\approx\mu;T=0) \approx a_k (\omega - \mu)^2 e^{ -b_k (\omega - \mu) }
\label{eq:match}
\end{align}
in the vicinity of the Fermi energy. This has the correct quadratic dependence in  energy of a normal Fermi liquid \cite{Luttinger1961}. The exponential factor allows for a certain degree of asymmetry below and above the Fermi surface, which we find to be essential in order to match the numerical self-energies. This function is matched to the self-energy in a given range of energies, which is different for every momentum, $k$. 

With access to the $T=0$ self-energies and spectral functions, one can compute several zero temperature properties from the microscopic properties. For instance, the energy per particle, $\frac{E}{A}$, is obtained from the Koltun sum-rule at zero temperature \cite{Dickhoff08}. Alternatively, the finite temperature SCGF calculations yield a set of energies which can independently be extrapolated to zero temperature using a polynomial fit. For simplicity, we take the same $L$ in the fit of this macroscopically determined data and in that of the associated self-energy. 

A good extrapolation procedure for the self-energy should ensure consistency between the micro- and the macroscopic results. In the example above, we would like the Koltun sum-rule at zero temperature to yield the same energy per particle than the extrapolated value from finite temperature data. We therefore construct a quality measure that quantifies the distance between extrapolations of micro- and macroscopic evaluations for some relevant data. The measure is built from a weighted sum of the relative differences between microscopic and macroscopic determinations of density, chemical potentials, energies, kinetic energies and $Z$-factors. The latter is determined independently from the discontinuity of the momentum distribution at the Fermi surface and from the on-shell derivatives of self-energies. The polynomial with $L$ that minimizes the quality measure, whether matched or not according to Eq.~(\ref{eq:match}), is used in the extrapolation. This guarantees that the associated self-energy is consistent with both the microscopic and the macroscopic pseudo-data. The extrapolation procedure is automated, in an effort to avoid biases.

Below $T_c$ and as the temperature approaches zero asymptotically, the normal spectral function becomes an increasingly sharp function of energy close to the Fermi surface. It is important to keep track of these narrow structures in the calculation of the momentum distribution, $n(k)$. For a given momentum, $k$, the missing strength due to uncaptured narrow peaks can be estimated from the energy spectral function sum-rule \cite{Dickhoff08}. Deviations from $1$ indicate missing strength contributions, which we use to correct the momentum distribution. We include a quasi-particle term that is weighted to account for the missing strength. We have tested this procedure against an independent determination of the momentum distribution, based on the derivatives of the zero-temperature self-energy \cite{Rios2009}, and we have found quantitative agreement. 

Missing strength corrections are also relevant for the double convolution energy denominator of Eq.~(\ref{eq:chi}). We use the sum-rule of the lowest-order two-particle propagator,
\begin{align}
\int \frac{\text{d}\Omega}{2 \pi} \mathcal{G}_{II}^0(k,k'=k,\Omega) =1 - 2 n(k) \, ,
 \label{eq:sumrule}
\end{align}
to estimate the missing strength at a given momentum, $\varsigma_k$. The origin of this missing strength lies on the finite meshes in the calculation, and the difficulty of keeping track of narrow structures in energy space.
The missing strength correction in our ladder calculations is of the order of less than one percent away from the Fermi surface. $\varsigma_k$ is generally largest ($50\,\%$ or above) for momenta which are within $2-3\,\%$ of the Fermi surface. Hence, the energy denominator only needs corrections in the close vicinity of the Fermi surface. We implement this correction by means of the replacement,
\begin{align}
 \frac{1}{2 \chi_c(k)} \to  \frac{1}{2 \chi(k)} + \frac{\varsigma_k}{2 (\varepsilon_k - \mu)} \, .
 \label{eq:correction}
\end{align}
The resulting energy denominators are continuous, soft functions of momentum as a function of momentum (see Panel (b) in Fig.~\ref{fig:mom_chi}) and in a wide range of densities (see panels (d)-(f) in Fig.~\ref{fig:chi_den}). 

\section{Fits}
\label{app:fits}

The parameters for the gap fits of Eq.~(\ref{eq:gapfit}) in Tables~\ref{table:1S0} and \ref{table:3PF2} have been obtained from a robust bisquare non-linear least-square fit. The function is fit to all non-zero gap data plus the first zero value after the gap closure. For singlet gaps, in addition to the data points of Fig.~\ref{fig:gap1s0}, we have supplemented the fit with a point with zero gap at zero Fermi momentum. Other than the N3LO results, the fitting algorithm prefers a gap opening at $k_0=0.04-0.05$ fm$^{-1}$. No additional points were given in the triplet channel. The goodness of fit as measured by the $r^2$ coefficient of determination is above $0.99$ for singlet fits. Triplet fits, in contrast, have $r^2$ ranging between $0.76$ for CDBonn, $0.83$ for Av18 and $0.97$ for N3LO. 

\begin{acknowledgments}
This material is based upon work supported by the U.S. National Science Foundation under grant PHY-1304242; 
by the Consolider Ingenio 2010 Programme CPAN CSD2007-00042, Grant No. FIS2014-54672-P from MICINN (Spain), and Grant No. 2014SGR-401 from Generalitat de Catalunya (Spain); 
by STFC, through Grants No. ST/I005528/1, ST/L005816/1 and ST/L005743/1; 
and by NewCompstar, COST Action MP1304.
\end{acknowledgments}

\bibliographystyle{apsrev4-1}
\bibliography{biblio}

\begin{thebibliography}{86}%
\makeatletter
\providecommand \@ifxundefined [1]{%
 \@ifx{#1\undefined}
}%
\providecommand \@ifnum [1]{%
 \ifnum #1\expandafter \@firstoftwo
 \else \expandafter \@secondoftwo
 \fi
}%
\providecommand \@ifx [1]{%
 \ifx #1\expandafter \@firstoftwo
 \else \expandafter \@secondoftwo
 \fi
}%
\providecommand \natexlab [1]{#1}%
\providecommand \enquote  [1]{``#1''}%
\providecommand \bibnamefont  [1]{#1}%
\providecommand \bibfnamefont [1]{#1}%
\providecommand \citenamefont [1]{#1}%
\providecommand \href@noop [0]{\@secondoftwo}%
\providecommand \href [0]{\begingroup \@sanitize@url \@href}%
\providecommand \@href[1]{\@@startlink{#1}\@@href}%
\providecommand \@@href[1]{\endgroup#1\@@endlink}%
\providecommand \@sanitize@url [0]{\catcode `\\12\catcode `\$12\catcode
  `\&12\catcode `\#12\catcode `\^12\catcode `\_12\catcode `\%12\relax}%
\providecommand \@@startlink[1]{}%
\providecommand \@@endlink[0]{}%
\providecommand \url  [0]{\begingroup\@sanitize@url \@url }%
\providecommand \@url [1]{\endgroup\@href {#1}{\urlprefix }}%
\providecommand \urlprefix  [0]{URL }%
\providecommand \Eprint [0]{\href }%
\providecommand \doibase [0]{http://dx.doi.org/}%
\providecommand \selectlanguage [0]{\@gobble}%
\providecommand \bibinfo  [0]{\@secondoftwo}%
\providecommand \bibfield  [0]{\@secondoftwo}%
\providecommand \translation [1]{[#1]}%
\providecommand \BibitemOpen [0]{}%
\providecommand \bibitemStop [0]{}%
\providecommand \bibitemNoStop [0]{.\EOS\space}%
\providecommand \EOS [0]{\spacefactor3000\relax}%
\providecommand \BibitemShut  [1]{\csname bibitem#1\endcsname}%
\let\auto@bib@innerbib\@empty
\bibitem [{\citenamefont {Dean}\ and\ \citenamefont
  {Hjorth-Jensen}(2003)}]{Dean2003}%
  \BibitemOpen
  \bibfield  {author} {\bibinfo {author} {\bibfnamefont {D.~J.}\ \bibnamefont
  {Dean}}\ and\ \bibinfo {author} {\bibfnamefont {M.}~\bibnamefont
  {Hjorth-Jensen}},\ }\href {\doibase 10.1103/RevModPhys.75.607} {\bibfield
  {journal} {\bibinfo  {journal} {Rev. Mod. Phys.}\ }\textbf {\bibinfo {volume}
  {75}},\ \bibinfo {pages} {607} (\bibinfo {year} {2003})}\BibitemShut
  {NoStop}%
\bibitem [{\citenamefont {Steiner}\ and\ \citenamefont
  {Reddy}(2009)}]{Steiner2009}%
  \BibitemOpen
  \bibfield  {author} {\bibinfo {author} {\bibfnamefont {A.~W.}\ \bibnamefont
  {Steiner}}\ and\ \bibinfo {author} {\bibfnamefont {S.}~\bibnamefont
  {Reddy}},\ }\href {\doibase 10.1103/PhysRevC.79.015802} {\bibfield  {journal}
  {\bibinfo  {journal} {Phys. Rev. C}\ }\textbf {\bibinfo {volume} {79}},\
  \bibinfo {pages} {015802} (\bibinfo {year} {2009})}\BibitemShut {NoStop}%
\bibitem [{\citenamefont {Leinson}(2010)}]{Leinson2010}%
  \BibitemOpen
  \bibfield  {author} {\bibinfo {author} {\bibfnamefont {L.~B.}\ \bibnamefont
  {Leinson}},\ }\href {\doibase 10.1103/PhysRevC.81.025501} {\bibfield
  {journal} {\bibinfo  {journal} {Phys. Rev. C}\ }\textbf {\bibinfo {volume}
  {81}},\ \bibinfo {pages} {025501} (\bibinfo {year} {2010})}\BibitemShut
  {NoStop}%
\bibitem [{\citenamefont {Yakovlev}\ and\ \citenamefont
  {Pethick}(2004)}]{Yakovlev2004}%
  \BibitemOpen
  \bibfield  {author} {\bibinfo {author} {\bibfnamefont {D.~G.}\ \bibnamefont
  {Yakovlev}}\ and\ \bibinfo {author} {\bibfnamefont {C.~J.}\ \bibnamefont
  {Pethick}},\ }\href {\doibase 10.1146/annurev.astro.42.053102.134013}
  {\bibfield  {journal} {\bibinfo  {journal} {Annu. Rev. Astron. Astrophys.}\
  }\textbf {\bibinfo {volume} {42}},\ \bibinfo {pages} {169} (\bibinfo {year}
  {2004})}\BibitemShut {NoStop}%
\bibitem [{\citenamefont {Reddy}\ \emph {et~al.}(1997)\citenamefont {Reddy},
  \citenamefont {Prakash},\ and\ \citenamefont {Lattimer}}]{Reddy1997}%
  \BibitemOpen
  \bibfield  {author} {\bibinfo {author} {\bibfnamefont {S.}~\bibnamefont
  {Reddy}}, \bibinfo {author} {\bibfnamefont {M.}~\bibnamefont {Prakash}}, \
  and\ \bibinfo {author} {\bibfnamefont {J.~M.}\ \bibnamefont {Lattimer}},\
  }\href {\doibase 10.1103/PhysRevD.58.013009} {\bibfield  {journal} {\bibinfo
  {journal} {Phys. Rev. D}\ }\textbf {\bibinfo {volume} {58}},\ \bibinfo
  {pages} {41} (\bibinfo {year} {1997})}\BibitemShut {NoStop}%
\bibitem [{\citenamefont {Sedrakian}(2007)}]{Sedrakian2007}%
  \BibitemOpen
  \bibfield  {author} {\bibinfo {author} {\bibfnamefont {A.}~\bibnamefont
  {Sedrakian}},\ }\href {\doibase 10.1016/j.ppnp.2006.02.002} {\bibfield
  {journal} {\bibinfo  {journal} {Prog. Part. Nucl. Phys.}\ }\textbf {\bibinfo
  {volume} {58}},\ \bibinfo {pages} {168} (\bibinfo {year} {2007})}\BibitemShut
  {NoStop}%
\bibitem [{\citenamefont {Page}\ \emph {et~al.}(2011)\citenamefont {Page},
  \citenamefont {Prakash}, \citenamefont {Lattimer},\ and\ \citenamefont
  {Steiner}}]{Page2011}%
  \BibitemOpen
  \bibfield  {author} {\bibinfo {author} {\bibfnamefont {D.}~\bibnamefont
  {Page}}, \bibinfo {author} {\bibfnamefont {M.}~\bibnamefont {Prakash}},
  \bibinfo {author} {\bibfnamefont {J.~M.}\ \bibnamefont {Lattimer}}, \ and\
  \bibinfo {author} {\bibfnamefont {A.~W.}\ \bibnamefont {Steiner}},\ }\href
  {\doibase 10.1103/PhysRevLett.106.081101} {\bibfield  {journal} {\bibinfo
  {journal} {Phys. Rev. Lett.}\ }\textbf {\bibinfo {volume} {106}},\ \bibinfo
  {pages} {081101} (\bibinfo {year} {2011})}\BibitemShut {NoStop}%
\bibitem [{\citenamefont {Elshamouty}\ \emph {et~al.}(2013)\citenamefont
  {Elshamouty}, \citenamefont {Heinke}, \citenamefont {Sivakoff}, \citenamefont
  {Ho}, \citenamefont {Shternin}, \citenamefont {Yakovlev}, \citenamefont
  {Patnaude},\ and\ \citenamefont {David}}]{Elshamouty2013}%
  \BibitemOpen
  \bibfield  {author} {\bibinfo {author} {\bibfnamefont {K.~G.}\ \bibnamefont
  {Elshamouty}}, \bibinfo {author} {\bibfnamefont {C.~O.}\ \bibnamefont
  {Heinke}}, \bibinfo {author} {\bibfnamefont {G.~R.}\ \bibnamefont
  {Sivakoff}}, \bibinfo {author} {\bibfnamefont {W.~C.~G.}\ \bibnamefont {Ho}},
  \bibinfo {author} {\bibfnamefont {P.~S.}\ \bibnamefont {Shternin}}, \bibinfo
  {author} {\bibfnamefont {D.~G.}\ \bibnamefont {Yakovlev}}, \bibinfo {author}
  {\bibfnamefont {D.~J.}\ \bibnamefont {Patnaude}}, \ and\ \bibinfo {author}
  {\bibfnamefont {L.}~\bibnamefont {David}},\ }\href {\doibase
  10.1088/0004-637X/777/1/22} {\bibfield  {journal} {\bibinfo  {journal}
  {Astrophys. J.}\ }\textbf {\bibinfo {volume} {777}},\ \bibinfo {pages} {22}
  (\bibinfo {year} {2013})}\BibitemShut {NoStop}%
\bibitem [{\citenamefont {Ho}\ \emph {et~al.}(2015{\natexlab{a}})\citenamefont
  {Ho}, \citenamefont {Elshamouty}, \citenamefont {Heinke},\ and\ \citenamefont
  {Potekhin}}]{Ho2015}%
  \BibitemOpen
  \bibfield  {author} {\bibinfo {author} {\bibfnamefont {W.~C.~G.}\
  \bibnamefont {Ho}}, \bibinfo {author} {\bibfnamefont {K.~G.}\ \bibnamefont
  {Elshamouty}}, \bibinfo {author} {\bibfnamefont {C.~O.}\ \bibnamefont
  {Heinke}}, \ and\ \bibinfo {author} {\bibfnamefont {A.~Y.}\ \bibnamefont
  {Potekhin}},\ }\href {\doibase 10.1103/PhysRevC.91.015806} {\bibfield
  {journal} {\bibinfo  {journal} {Phys. Rev. C}\ }\textbf {\bibinfo {volume}
  {91}},\ \bibinfo {pages} {015806} (\bibinfo {year}
  {2015}{\natexlab{a}})}\BibitemShut {NoStop}%
\bibitem [{\citenamefont {Anderson}\ and\ \citenamefont
  {Itoh}(1975)}]{Anderson1975}%
  \BibitemOpen
  \bibfield  {author} {\bibinfo {author} {\bibfnamefont {P.~W.}\ \bibnamefont
  {Anderson}}\ and\ \bibinfo {author} {\bibfnamefont {N.}~\bibnamefont
  {Itoh}},\ }\href {\doibase 10.1038/256025a0} {\bibfield  {journal} {\bibinfo
  {journal} {Nature}\ }\textbf {\bibinfo {volume} {256}},\ \bibinfo {pages}
  {25} (\bibinfo {year} {1975})}\BibitemShut {NoStop}%
\bibitem [{\citenamefont {Pines}\ and\ \citenamefont
  {Alpar}(1985)}]{Pines1985}%
  \BibitemOpen
  \bibfield  {author} {\bibinfo {author} {\bibfnamefont {D.}~\bibnamefont
  {Pines}}\ and\ \bibinfo {author} {\bibfnamefont {M.~A.}\ \bibnamefont
  {Alpar}},\ }\href {\doibase 10.1038/316027a0} {\bibfield  {journal} {\bibinfo
   {journal} {Nature}\ }\textbf {\bibinfo {volume} {316}},\ \bibinfo {pages}
  {27} (\bibinfo {year} {1985})}\BibitemShut {NoStop}%
\bibitem [{\citenamefont {Haskell}\ and\ \citenamefont
  {Melatos}(2015)}]{Haskell2015}%
  \BibitemOpen
  \bibfield  {author} {\bibinfo {author} {\bibfnamefont {B.}~\bibnamefont
  {Haskell}}\ and\ \bibinfo {author} {\bibfnamefont {A.}~\bibnamefont
  {Melatos}},\ }\href {\doibase 10.1142/S0218271815300086} {\bibfield
  {journal} {\bibinfo  {journal} {Int. J. Mod. Phys. D}\ }\textbf {\bibinfo
  {volume} {24}},\ \bibinfo {pages} {1530008} (\bibinfo {year}
  {2015})}\BibitemShut {NoStop}%
\bibitem [{\citenamefont {Link}\ \emph {et~al.}(1992)\citenamefont {Link},
  \citenamefont {Epstein},\ and\ \citenamefont {{Van Riper}}}]{Link1992}%
  \BibitemOpen
  \bibfield  {author} {\bibinfo {author} {\bibfnamefont {B.}~\bibnamefont
  {Link}}, \bibinfo {author} {\bibfnamefont {R.~I.}\ \bibnamefont {Epstein}}, \
  and\ \bibinfo {author} {\bibfnamefont {K.~A.}\ \bibnamefont {{Van Riper}}},\
  }\href {\doibase 10.1038/359616a0} {\bibfield  {journal} {\bibinfo  {journal}
  {Nature}\ }\textbf {\bibinfo {volume} {359}},\ \bibinfo {pages} {616}
  (\bibinfo {year} {1992})}\BibitemShut {NoStop}%
\bibitem [{\citenamefont {Andersson}\ \emph {et~al.}(2012)\citenamefont
  {Andersson}, \citenamefont {Glampedakis}, \citenamefont {Ho},\ and\
  \citenamefont {Espinoza}}]{Andersson2012}%
  \BibitemOpen
  \bibfield  {author} {\bibinfo {author} {\bibfnamefont {N.}~\bibnamefont
  {Andersson}}, \bibinfo {author} {\bibfnamefont {K.}~\bibnamefont
  {Glampedakis}}, \bibinfo {author} {\bibfnamefont {W.~C.~G.}\ \bibnamefont
  {Ho}}, \ and\ \bibinfo {author} {\bibfnamefont {C.~M.}\ \bibnamefont
  {Espinoza}},\ }\href {\doibase 10.1103/PhysRevLett.109.241103} {\bibfield
  {journal} {\bibinfo  {journal} {Phys. Rev. Lett.}\ }\textbf {\bibinfo
  {volume} {109}},\ \bibinfo {pages} {241103} (\bibinfo {year}
  {2012})}\BibitemShut {NoStop}%
\bibitem [{\citenamefont {Chamel}(2013)}]{Chamel2013}%
  \BibitemOpen
  \bibfield  {author} {\bibinfo {author} {\bibfnamefont {N.}~\bibnamefont
  {Chamel}},\ }\href {\doibase 10.1103/PhysRevLett.110.011101} {\bibfield
  {journal} {\bibinfo  {journal} {Phys. Rev. Lett.}\ }\textbf {\bibinfo
  {volume} {110}},\ \bibinfo {pages} {011101} (\bibinfo {year}
  {2013})}\BibitemShut {NoStop}%
\bibitem [{\citenamefont {Ho}\ \emph {et~al.}(2015{\natexlab{b}})\citenamefont
  {Ho}, \citenamefont {Espinoza}, \citenamefont {Antonopoulou},\ and\
  \citenamefont {Andersson}}]{Ho2015b}%
  \BibitemOpen
  \bibfield  {author} {\bibinfo {author} {\bibfnamefont {W.~C.~G.}\
  \bibnamefont {Ho}}, \bibinfo {author} {\bibfnamefont {C.~M.}\ \bibnamefont
  {Espinoza}}, \bibinfo {author} {\bibfnamefont {D.}~\bibnamefont
  {Antonopoulou}}, \ and\ \bibinfo {author} {\bibfnamefont {N.}~\bibnamefont
  {Andersson}},\ }\href {\doibase 10.1126/sciadv.1500578} {\bibfield  {journal}
  {\bibinfo  {journal} {Sci. Adv.}\ }\textbf {\bibinfo {volume} {1}},\ \bibinfo
  {pages} {e1500578} (\bibinfo {year} {2015}{\natexlab{b}})}\BibitemShut
  {NoStop}%
\bibitem [{\citenamefont {Afanasjev}(2013)}]{Afanasjev2015}%
  \BibitemOpen
  \bibfield  {author} {\bibinfo {author} {\bibfnamefont {A.~V.}\ \bibnamefont
  {Afanasjev}},\ }\enquote {\bibinfo {title} {Isoscalar and isovector
  neutron-proton pairing},}\ in\ \href {\doibase 10.1142/9789814412490_0011}
  {\emph {\bibinfo {booktitle} {Fifty Years of Nuclear BCS}}}\ (\bibinfo
  {publisher} {World Scientific},\ \bibinfo {year} {2013})\ Chap.~\bibinfo
  {chapter} {11}, p.\ \bibinfo {pages} {138}\BibitemShut {NoStop}%
\bibitem [{\citenamefont {Frauendorf}\ and\ \citenamefont
  {Macchiavelli}(2014)}]{Frauendorf2014}%
  \BibitemOpen
  \bibfield  {author} {\bibinfo {author} {\bibfnamefont {S.}~\bibnamefont
  {Frauendorf}}\ and\ \bibinfo {author} {\bibfnamefont {A.}~\bibnamefont
  {Macchiavelli}},\ }\href {\doibase 10.1016/j.ppnp.2014.07.001} {\bibfield
  {journal} {\bibinfo  {journal} {Prog. Part. Nucl. Phys.}\ }\textbf {\bibinfo
  {volume} {78}},\ \bibinfo {pages} {24} (\bibinfo {year} {2014})}\BibitemShut
  {NoStop}%
\bibitem [{\citenamefont {Baldo}\ \emph {et~al.}(1992)\citenamefont {Baldo},
  \citenamefont {Bombaci},\ and\ \citenamefont {Lombardo}}]{Baldo1992}%
  \BibitemOpen
  \bibfield  {author} {\bibinfo {author} {\bibfnamefont {M.}~\bibnamefont
  {Baldo}}, \bibinfo {author} {\bibfnamefont {I.}~\bibnamefont {Bombaci}}, \
  and\ \bibinfo {author} {\bibfnamefont {U.}~\bibnamefont {Lombardo}},\ }\href
  {\doibase 10.1016/0370-2693(92)91416-7} {\bibfield  {journal} {\bibinfo
  {journal} {Phys. Lett. B}\ }\textbf {\bibinfo {volume} {283}},\ \bibinfo
  {pages} {8} (\bibinfo {year} {1992})}\BibitemShut {NoStop}%
\bibitem [{\citenamefont {M\"{u}ther}\ and\ \citenamefont
  {Dickhoff}(2005)}]{Muther2005}%
  \BibitemOpen
  \bibfield  {author} {\bibinfo {author} {\bibfnamefont {H.}~\bibnamefont
  {M\"{u}ther}}\ and\ \bibinfo {author} {\bibfnamefont {W.~H.}\ \bibnamefont
  {Dickhoff}},\ }\href {\doibase 10.1103/PhysRevC.72.054313} {\bibfield
  {journal} {\bibinfo  {journal} {Phys. Rev. C}\ }\textbf {\bibinfo {volume}
  {72}},\ \bibinfo {pages} {054313} (\bibinfo {year} {2005})}\BibitemShut
  {NoStop}%
\bibitem [{\citenamefont {Maurizio}\ \emph {et~al.}(2014)\citenamefont
  {Maurizio}, \citenamefont {Holt},\ and\ \citenamefont
  {Finelli}}]{Maurizio2014}%
  \BibitemOpen
  \bibfield  {author} {\bibinfo {author} {\bibfnamefont {S.}~\bibnamefont
  {Maurizio}}, \bibinfo {author} {\bibfnamefont {J.~W.}\ \bibnamefont {Holt}},
  \ and\ \bibinfo {author} {\bibfnamefont {P.}~\bibnamefont {Finelli}},\ }\href
  {\doibase 10.1103/PhysRevC.90.044003} {\bibfield  {journal} {\bibinfo
  {journal} {Phys. Rev. C}\ }\textbf {\bibinfo {volume} {90}},\ \bibinfo
  {pages} {044003} (\bibinfo {year} {2014})}\BibitemShut {NoStop}%
\bibitem [{\citenamefont {Gezerlis}\ \emph {et~al.}(2011)\citenamefont
  {Gezerlis}, \citenamefont {Bertsch},\ and\ \citenamefont
  {Luo}}]{Gezerlis2011}%
  \BibitemOpen
  \bibfield  {author} {\bibinfo {author} {\bibfnamefont {A.}~\bibnamefont
  {Gezerlis}}, \bibinfo {author} {\bibfnamefont {G.~F.}\ \bibnamefont
  {Bertsch}}, \ and\ \bibinfo {author} {\bibfnamefont {Y.~L.}\ \bibnamefont
  {Luo}},\ }\href {\doibase 10.1103/PhysRevLett.106.252502} {\bibfield
  {journal} {\bibinfo  {journal} {Phys. Rev. Lett.}\ }\textbf {\bibinfo
  {volume} {106}},\ \bibinfo {pages} {252502} (\bibinfo {year}
  {2011})}\BibitemShut {NoStop}%
\bibitem [{\citenamefont {Dickhoff}\ and\ \citenamefont {{Van
  Neck}}(2008)}]{Dickhoff08}%
  \BibitemOpen
  \bibfield  {author} {\bibinfo {author} {\bibfnamefont {W.~H.}\ \bibnamefont
  {Dickhoff}}\ and\ \bibinfo {author} {\bibfnamefont {D.}~\bibnamefont {{Van
  Neck}}},\ }\href@noop {} {\emph {\bibinfo {title} {Many-Body Theory Exposed!,
  2nd edition}}}\ (\bibinfo  {publisher} {World Scientific},\ \bibinfo
  {address} {New Jersey},\ \bibinfo {year} {2008})\BibitemShut {NoStop}%
\bibitem [{\citenamefont {Dickhoff}\ and\ \citenamefont
  {Barbieri}(2004)}]{Dickhoff2004a}%
  \BibitemOpen
  \bibfield  {author} {\bibinfo {author} {\bibfnamefont {W.~H.}\ \bibnamefont
  {Dickhoff}}\ and\ \bibinfo {author} {\bibfnamefont {C.}~\bibnamefont
  {Barbieri}},\ }\href {\doibase 10.1016/j.ppnp.2004.02.038} {\bibfield
  {journal} {\bibinfo  {journal} {Prog. Part. Nucl. Phys.}\ }\textbf {\bibinfo
  {volume} {52}},\ \bibinfo {pages} {377} (\bibinfo {year} {2004})}\BibitemShut
  {NoStop}%
\bibitem [{\citenamefont {Lapik{\'{a}}s}(1993)}]{Lapikas1993}%
  \BibitemOpen
  \bibfield  {author} {\bibinfo {author} {\bibfnamefont {L.}~\bibnamefont
  {Lapik{\'{a}}s}},\ }\href {\doibase 10.1016/0375-9474(93)90630-G} {\bibfield
  {journal} {\bibinfo  {journal} {Nucl. Phys. A}\ }\textbf {\bibinfo {volume}
  {553}},\ \bibinfo {pages} {297} (\bibinfo {year} {1993})}\BibitemShut
  {NoStop}%
\bibitem [{\citenamefont {Benhar}\ \emph {et~al.}(2008)\citenamefont {Benhar},
  \citenamefont {Sick},\ and\ \citenamefont {Day}}]{Benhar2008}%
  \BibitemOpen
  \bibfield  {author} {\bibinfo {author} {\bibfnamefont {O.}~\bibnamefont
  {Benhar}}, \bibinfo {author} {\bibfnamefont {I.}~\bibnamefont {Sick}}, \ and\
  \bibinfo {author} {\bibfnamefont {D.}~\bibnamefont {Day}},\ }\href {\doibase
  10.1103/RevModPhys.80.189} {\bibfield  {journal} {\bibinfo  {journal} {Rev.
  Mod. Phys.}\ }\textbf {\bibinfo {volume} {80}},\ \bibinfo {pages} {189}
  (\bibinfo {year} {2008})}\BibitemShut {NoStop}%
\bibitem [{\citenamefont {Arrington}\ \emph {et~al.}(2012)\citenamefont
  {Arrington}, \citenamefont {Higinbotham}, \citenamefont {Rosner},\ and\
  \citenamefont {Sargsian}}]{Arrington2011}%
  \BibitemOpen
  \bibfield  {author} {\bibinfo {author} {\bibfnamefont {J.}~\bibnamefont
  {Arrington}}, \bibinfo {author} {\bibfnamefont {D.}~\bibnamefont
  {Higinbotham}}, \bibinfo {author} {\bibfnamefont {G.}~\bibnamefont {Rosner}},
  \ and\ \bibinfo {author} {\bibfnamefont {M.}~\bibnamefont {Sargsian}},\
  }\href {\doibase 10.1016/j.ppnp.2012.04.002} {\bibfield  {journal} {\bibinfo
  {journal} {Prog. Part. Nucl. Phys.}\ }\textbf {\bibinfo {volume} {67}},\
  \bibinfo {pages} {898} (\bibinfo {year} {2012})},\ \Eprint
  {http://arxiv.org/abs/1104.1196} {1104.1196} \BibitemShut {NoStop}%
\bibitem [{\citenamefont {Rohe}\ \emph {et~al.}(2004)\citenamefont {Rohe},
  \citenamefont {Armstrong}, \citenamefont {Asaturyan}, \citenamefont {Baker},
  \citenamefont {Bueltmann}, \citenamefont {Carasco}, \citenamefont {Day},
  \citenamefont {Ent}, \citenamefont {Fenker}, \citenamefont {Garrow},
  \citenamefont {Gasparian}, \citenamefont {Gueye}, \citenamefont {Hauger},
  \citenamefont {Honegger}, \citenamefont {Jourdan}, \citenamefont {Keppel},
  \citenamefont {Kubon}, \citenamefont {Lindgren}, \citenamefont {Lung},
  \citenamefont {Mack}, \citenamefont {Mitchell}, \citenamefont {Mkrtchyan},
  \citenamefont {Mocelj}, \citenamefont {Normand}, \citenamefont {Petitjean},
  \citenamefont {Rondon}, \citenamefont {Segbefia}, \citenamefont {Sick},
  \citenamefont {Stepanyan}, \citenamefont {Tang}, \citenamefont
  {Tiefenbacher}, \citenamefont {Vulcan}, \citenamefont {Warren}, \citenamefont
  {Wood}, \citenamefont {Yuan}, \citenamefont {Zeier}, \citenamefont {Zhu},\
  and\ \citenamefont {Zihlmann}}]{Rohe2004}%
  \BibitemOpen
  \bibfield  {author} {\bibinfo {author} {\bibfnamefont {D.}~\bibnamefont
  {Rohe}}, \bibinfo {author} {\bibfnamefont {C.~S.}\ \bibnamefont {Armstrong}},
  \bibinfo {author} {\bibfnamefont {R.}~\bibnamefont {Asaturyan}}, \bibinfo
  {author} {\bibfnamefont {O.~K.}\ \bibnamefont {Baker}}, \bibinfo {author}
  {\bibfnamefont {S.}~\bibnamefont {Bueltmann}}, \bibinfo {author}
  {\bibfnamefont {C.}~\bibnamefont {Carasco}}, \bibinfo {author} {\bibfnamefont
  {D.}~\bibnamefont {Day}}, \bibinfo {author} {\bibfnamefont {R.}~\bibnamefont
  {Ent}}, \bibinfo {author} {\bibfnamefont {H.~C.}\ \bibnamefont {Fenker}},
  \bibinfo {author} {\bibfnamefont {K.}~\bibnamefont {Garrow}}, \bibinfo
  {author} {\bibfnamefont {A.}~\bibnamefont {Gasparian}}, \bibinfo {author}
  {\bibfnamefont {P.}~\bibnamefont {Gueye}}, \bibinfo {author} {\bibfnamefont
  {M.}~\bibnamefont {Hauger}}, \bibinfo {author} {\bibfnamefont
  {A.}~\bibnamefont {Honegger}}, \bibinfo {author} {\bibfnamefont
  {J.}~\bibnamefont {Jourdan}}, \bibinfo {author} {\bibfnamefont {C.~E.}\
  \bibnamefont {Keppel}}, \bibinfo {author} {\bibfnamefont {G.}~\bibnamefont
  {Kubon}}, \bibinfo {author} {\bibfnamefont {R.}~\bibnamefont {Lindgren}},
  \bibinfo {author} {\bibfnamefont {A.}~\bibnamefont {Lung}}, \bibinfo {author}
  {\bibfnamefont {D.~J.}\ \bibnamefont {Mack}}, \bibinfo {author}
  {\bibfnamefont {J.~H.}\ \bibnamefont {Mitchell}}, \bibinfo {author}
  {\bibfnamefont {H.}~\bibnamefont {Mkrtchyan}}, \bibinfo {author}
  {\bibfnamefont {D.}~\bibnamefont {Mocelj}}, \bibinfo {author} {\bibfnamefont
  {K.}~\bibnamefont {Normand}}, \bibinfo {author} {\bibfnamefont
  {T.}~\bibnamefont {Petitjean}}, \bibinfo {author} {\bibfnamefont
  {O.}~\bibnamefont {Rondon}}, \bibinfo {author} {\bibfnamefont
  {E.}~\bibnamefont {Segbefia}}, \bibinfo {author} {\bibfnamefont
  {I.}~\bibnamefont {Sick}}, \bibinfo {author} {\bibfnamefont {S.}~\bibnamefont
  {Stepanyan}}, \bibinfo {author} {\bibfnamefont {L.}~\bibnamefont {Tang}},
  \bibinfo {author} {\bibfnamefont {F.}~\bibnamefont {Tiefenbacher}}, \bibinfo
  {author} {\bibfnamefont {W.~F.}\ \bibnamefont {Vulcan}}, \bibinfo {author}
  {\bibfnamefont {G.}~\bibnamefont {Warren}}, \bibinfo {author} {\bibfnamefont
  {S.~A.}\ \bibnamefont {Wood}}, \bibinfo {author} {\bibfnamefont
  {L.}~\bibnamefont {Yuan}}, \bibinfo {author} {\bibfnamefont {M.}~\bibnamefont
  {Zeier}}, \bibinfo {author} {\bibfnamefont {H.}~\bibnamefont {Zhu}}, \ and\
  \bibinfo {author} {\bibfnamefont {B.}~\bibnamefont {Zihlmann}},\ }\href
  {\doibase 10.1103/PhysRevLett.93.182501} {\bibfield  {journal} {\bibinfo
  {journal} {Phys. Rev. Lett.}\ }\textbf {\bibinfo {volume} {93}},\ \bibinfo
  {pages} {182501} (\bibinfo {year} {2004})}\BibitemShut {NoStop}%
\bibitem [{\citenamefont {Rios}\ \emph {et~al.}(2009)\citenamefont {Rios},
  \citenamefont {Polls},\ and\ \citenamefont {Dickhoff}}]{Rios2009}%
  \BibitemOpen
  \bibfield  {author} {\bibinfo {author} {\bibfnamefont {A.}~\bibnamefont
  {Rios}}, \bibinfo {author} {\bibfnamefont {A.}~\bibnamefont {Polls}}, \ and\
  \bibinfo {author} {\bibfnamefont {W.~H.}\ \bibnamefont {Dickhoff}},\ }\href
  {\doibase 10.1103/PhysRevC.79.064308} {\bibfield  {journal} {\bibinfo
  {journal} {Phys. Rev. C}\ }\textbf {\bibinfo {volume} {79}},\ \bibinfo
  {pages} {064308} (\bibinfo {year} {2009})}\BibitemShut {NoStop}%
\bibitem [{\citenamefont {Rios}\ \emph {et~al.}(2014)\citenamefont {Rios},
  \citenamefont {Polls},\ and\ \citenamefont {Dickhoff}}]{Rios2014}%
  \BibitemOpen
  \bibfield  {author} {\bibinfo {author} {\bibfnamefont {A.}~\bibnamefont
  {Rios}}, \bibinfo {author} {\bibfnamefont {A.}~\bibnamefont {Polls}}, \ and\
  \bibinfo {author} {\bibfnamefont {W.~H.}\ \bibnamefont {Dickhoff}},\ }\href
  {\doibase 10.1103/PhysRevC.89.044303} {\bibfield  {journal} {\bibinfo
  {journal} {Phys. Rev. C}\ }\textbf {\bibinfo {volume} {89}},\ \bibinfo
  {pages} {044303} (\bibinfo {year} {2014})}\BibitemShut {NoStop}%
\bibitem [{\citenamefont {Alvioli}\ \emph {et~al.}(2013)\citenamefont
  {Alvioli}, \citenamefont {{Ciofi degli Atti}}, \citenamefont {Kaptari},
  \citenamefont {Mezzetti},\ and\ \citenamefont {Morita}}]{Alvioli2013}%
  \BibitemOpen
  \bibfield  {author} {\bibinfo {author} {\bibfnamefont {M.}~\bibnamefont
  {Alvioli}}, \bibinfo {author} {\bibfnamefont {C.}~\bibnamefont {{Ciofi degli
  Atti}}}, \bibinfo {author} {\bibfnamefont {L.~P.}\ \bibnamefont {Kaptari}},
  \bibinfo {author} {\bibfnamefont {C.~B.}\ \bibnamefont {Mezzetti}}, \ and\
  \bibinfo {author} {\bibfnamefont {H.}~\bibnamefont {Morita}},\ }\href
  {\doibase 10.1103/PhysRevC.87.034603} {\bibfield  {journal} {\bibinfo
  {journal} {Phys. Rev. C}\ }\textbf {\bibinfo {volume} {87}},\ \bibinfo
  {pages} {034603} (\bibinfo {year} {2013})}\BibitemShut {NoStop}%
\bibitem [{\citenamefont {{Ciofi degli Atti}}(2015)}]{Atti2015}%
  \BibitemOpen
  \bibfield  {author} {\bibinfo {author} {\bibfnamefont {C.}~\bibnamefont
  {{Ciofi degli Atti}}},\ }\href {\doibase 10.1016/j.physrep.2015.06.002}
  {\bibfield  {journal} {\bibinfo  {journal} {Phys. Rep.}\ }\textbf {\bibinfo
  {volume} {590}},\ \bibinfo {pages} {1} (\bibinfo {year} {2015})}\BibitemShut
  {NoStop}%
\bibitem [{\citenamefont {Wiringa}\ \emph {et~al.}(2014)\citenamefont
  {Wiringa}, \citenamefont {Schiavilla}, \citenamefont {Pieper},\ and\
  \citenamefont {Carlson}}]{Wiringa2014}%
  \BibitemOpen
  \bibfield  {author} {\bibinfo {author} {\bibfnamefont {R.~B.}\ \bibnamefont
  {Wiringa}}, \bibinfo {author} {\bibfnamefont {R.}~\bibnamefont {Schiavilla}},
  \bibinfo {author} {\bibfnamefont {S.~C.}\ \bibnamefont {Pieper}}, \ and\
  \bibinfo {author} {\bibfnamefont {J.}~\bibnamefont {Carlson}},\ }\href
  {\doibase 10.1103/PhysRevC.89.024305} {\bibfield  {journal} {\bibinfo
  {journal} {Phys. Rev. C}\ }\textbf {\bibinfo {volume} {89}},\ \bibinfo
  {pages} {024305} (\bibinfo {year} {2014})}\BibitemShut {NoStop}%
\bibitem [{\citenamefont {Neff}\ \emph {et~al.}(2015)\citenamefont {Neff},
  \citenamefont {Feldmeier},\ and\ \citenamefont {Horiuchi}}]{Neff2015}%
  \BibitemOpen
  \bibfield  {author} {\bibinfo {author} {\bibfnamefont {T.}~\bibnamefont
  {Neff}}, \bibinfo {author} {\bibfnamefont {H.}~\bibnamefont {Feldmeier}}, \
  and\ \bibinfo {author} {\bibfnamefont {W.}~\bibnamefont {Horiuchi}},\ }\href
  {\doibase 10.1103/PhysRevC.92.024003} {\bibfield  {journal} {\bibinfo
  {journal} {Phys. Rev. C}\ }\textbf {\bibinfo {volume} {92}},\ \bibinfo
  {pages} {024003} (\bibinfo {year} {2015})}\BibitemShut {NoStop}%
\bibitem [{\citenamefont {Ryckebusch}\ \emph {et~al.}(2015)\citenamefont
  {Ryckebusch}, \citenamefont {Vanhalst},\ and\ \citenamefont
  {Cosyn}}]{Ryckebusch2015}%
  \BibitemOpen
  \bibfield  {author} {\bibinfo {author} {\bibfnamefont {J.}~\bibnamefont
  {Ryckebusch}}, \bibinfo {author} {\bibfnamefont {M.}~\bibnamefont
  {Vanhalst}}, \ and\ \bibinfo {author} {\bibfnamefont {W.}~\bibnamefont
  {Cosyn}},\ }\href {\doibase 10.1088/0954-3899/42/5/055104} {\bibfield
  {journal} {\bibinfo  {journal} {J. Phys. G Nucl. Part. Phys.}\ }\textbf
  {\bibinfo {volume} {42}},\ \bibinfo {pages} {055104} (\bibinfo {year}
  {2015})}\BibitemShut {NoStop}%
\bibitem [{\citenamefont {M{\"u}ther}\ \emph {et~al.}(1995)\citenamefont
  {M{\"u}ther}, \citenamefont {Polls},\ and\ \citenamefont
  {Dickhoff}}]{Muther95}%
  \BibitemOpen
  \bibfield  {author} {\bibinfo {author} {\bibfnamefont {H.}~\bibnamefont
  {M{\"u}ther}}, \bibinfo {author} {\bibfnamefont {A.}~\bibnamefont {Polls}}, \
  and\ \bibinfo {author} {\bibfnamefont {W.~H.}\ \bibnamefont {Dickhoff}},\
  }\href {\doibase 10.1103/PhysRevC.51.3040} {\bibfield  {journal} {\bibinfo
  {journal} {Phys. Rev. C}\ }\textbf {\bibinfo {volume} {51}},\ \bibinfo
  {pages} {3040} (\bibinfo {year} {1995})}\BibitemShut {NoStop}%
\bibitem [{\citenamefont {Sargsian}(2014)}]{Sargsian2014}%
  \BibitemOpen
  \bibfield  {author} {\bibinfo {author} {\bibfnamefont {M.~M.}\ \bibnamefont
  {Sargsian}},\ }\href {\doibase 10.1103/PhysRevC.89.034305} {\bibfield
  {journal} {\bibinfo  {journal} {Phys. Rev. C}\ }\textbf {\bibinfo {volume}
  {89}},\ \bibinfo {pages} {034305} (\bibinfo {year} {2014})}\BibitemShut
  {NoStop}%
\bibitem [{\citenamefont {Hen}\ \emph {et~al.}(2014)\citenamefont {Hen} \emph
  {et~al.}}]{Hen2014}%
  \BibitemOpen
  \bibfield  {author} {\bibinfo {author} {\bibfnamefont {O.}~\bibnamefont
  {Hen}} \emph {et~al.},\ }\href {\doibase 10.1126/science.1256785} {\bibfield
  {journal} {\bibinfo  {journal} {Science}\ }\textbf {\bibinfo {volume}
  {346}},\ \bibinfo {pages} {614} (\bibinfo {year} {2014})}\BibitemShut
  {NoStop}%
\bibitem [{\citenamefont {Bo{\.z}ek}(1999)}]{Bozek1999}%
  \BibitemOpen
  \bibfield  {author} {\bibinfo {author} {\bibfnamefont {P.}~\bibnamefont
  {Bo{\.z}ek}},\ }\href {\doibase 10.1016/S0375-9474(99)00325-5} {\bibfield
  {journal} {\bibinfo  {journal} {Nucl. Phys. A}\ }\textbf {\bibinfo {volume}
  {657}},\ \bibinfo {pages} {187} (\bibinfo {year} {1999})}\BibitemShut
  {NoStop}%
\bibitem [{\citenamefont {Shen}\ \emph {et~al.}(2003)\citenamefont {Shen},
  \citenamefont {Lombardo}, \citenamefont {Schuck}, \citenamefont {Zuo},\ and\
  \citenamefont {Sandulescu}}]{Shen2003}%
  \BibitemOpen
  \bibfield  {author} {\bibinfo {author} {\bibfnamefont {C.}~\bibnamefont
  {Shen}}, \bibinfo {author} {\bibfnamefont {U.}~\bibnamefont {Lombardo}},
  \bibinfo {author} {\bibfnamefont {P.}~\bibnamefont {Schuck}}, \bibinfo
  {author} {\bibfnamefont {W.}~\bibnamefont {Zuo}}, \ and\ \bibinfo {author}
  {\bibfnamefont {N.}~\bibnamefont {Sandulescu}},\ }\href {\doibase
  10.1103/PhysRevC.67.061302} {\bibfield  {journal} {\bibinfo  {journal} {Phys.
  Rev. C}\ }\textbf {\bibinfo {volume} {67}},\ \bibinfo {pages} {061302}
  (\bibinfo {year} {2003})}\BibitemShut {NoStop}%
\bibitem [{\citenamefont {Som{\`a}}\ \emph {et~al.}(2011)\citenamefont
  {Som{\`a}}, \citenamefont {Duguet},\ and\ \citenamefont
  {Barbieri}}]{Soma2011}%
  \BibitemOpen
  \bibfield  {author} {\bibinfo {author} {\bibfnamefont {V.}~\bibnamefont
  {Som{\`a}}}, \bibinfo {author} {\bibfnamefont {T.}~\bibnamefont {Duguet}}, \
  and\ \bibinfo {author} {\bibfnamefont {C.}~\bibnamefont {Barbieri}},\ }\href
  {\doibase 10.1103/PhysRevC.84.064317} {\bibfield  {journal} {\bibinfo
  {journal} {Phys. Rev. C}\ }\textbf {\bibinfo {volume} {84}},\ \bibinfo
  {pages} {064317} (\bibinfo {year} {2011})}\BibitemShut {NoStop}%
\bibitem [{\citenamefont {Bo{\.z}ek}(2000)}]{Bozek2000}%
  \BibitemOpen
  \bibfield  {author} {\bibinfo {author} {\bibfnamefont {P.}~\bibnamefont
  {Bo{\.z}ek}},\ }\href {\doibase 10.1103/PhysRevC.62.054316} {\bibfield
  {journal} {\bibinfo  {journal} {Phys. Rev. C}\ }\textbf {\bibinfo {volume}
  {62}},\ \bibinfo {pages} {054316} (\bibinfo {year} {2000})}\BibitemShut
  {NoStop}%
\bibitem [{\citenamefont {Dong}\ \emph {et~al.}(2016)\citenamefont {Dong},
  \citenamefont {Lombardo}, \citenamefont {Zhang},\ and\ \citenamefont
  {Zuo}}]{Dong2015}%
  \BibitemOpen
  \bibfield  {author} {\bibinfo {author} {\bibfnamefont {J.~M.}\ \bibnamefont
  {Dong}}, \bibinfo {author} {\bibfnamefont {U.}~\bibnamefont {Lombardo}},
  \bibinfo {author} {\bibfnamefont {H.~F.}\ \bibnamefont {Zhang}}, \ and\
  \bibinfo {author} {\bibfnamefont {W.}~\bibnamefont {Zuo}},\ }\href
  {http://stacks.iop.org/0004-637X/817/i=1/a=6} {\bibfield  {journal} {\bibinfo
   {journal} {Astrophys. J.}\ }\textbf {\bibinfo {volume} {817}},\ \bibinfo
  {pages} {6} (\bibinfo {year} {2016})}\BibitemShut {NoStop}%
\bibitem [{\citenamefont {Bo{\.z}ek}(2003)}]{Bozek2003}%
  \BibitemOpen
  \bibfield  {author} {\bibinfo {author} {\bibfnamefont {P.}~\bibnamefont
  {Bo{\.z}ek}},\ }\href {\doibase 10.1016/S0370-2693(02)03007-1} {\bibfield
  {journal} {\bibinfo  {journal} {Phys. Lett. B}\ }\textbf {\bibinfo {volume}
  {551}},\ \bibinfo {pages} {93} (\bibinfo {year} {2003})}\BibitemShut
  {NoStop}%
\bibitem [{\citenamefont {Schnell}\ \emph {et~al.}(1999)\citenamefont
  {Schnell}, \citenamefont {R\"{o}pke},\ and\ \citenamefont
  {Schuck}}]{Schnell1999}%
  \BibitemOpen
  \bibfield  {author} {\bibinfo {author} {\bibfnamefont {A.}~\bibnamefont
  {Schnell}}, \bibinfo {author} {\bibfnamefont {G.}~\bibnamefont {R\"{o}pke}},
  \ and\ \bibinfo {author} {\bibfnamefont {P.}~\bibnamefont {Schuck}},\ }\href
  {\doibase 10.1103/PhysRevLett.83.1926} {\bibfield  {journal} {\bibinfo
  {journal} {Phys. Rev. Lett.}\ }\textbf {\bibinfo {volume} {83}},\ \bibinfo
  {pages} {1926} (\bibinfo {year} {1999})}\BibitemShut {NoStop}%
\bibitem [{\citenamefont {Thouless}(1960)}]{Thouless1960}%
  \BibitemOpen
  \bibfield  {author} {\bibinfo {author} {\bibfnamefont {J.}~\bibnamefont
  {Thouless}},\ }\href {\doibase 10.1016/0003-4916(60)90122-6} {\bibfield
  {journal} {\bibinfo  {journal} {Ann. Phys.}\ }\textbf {\bibinfo {volume}
  {10}},\ \bibinfo {pages} {553} (\bibinfo {year} {1960})}\BibitemShut
  {NoStop}%
\bibitem [{\citenamefont {Kadanoff}\ and\ \citenamefont
  {Martin}(1961)}]{Kadanoff1961}%
  \BibitemOpen
  \bibfield  {author} {\bibinfo {author} {\bibfnamefont {L.~P.}\ \bibnamefont
  {Kadanoff}}\ and\ \bibinfo {author} {\bibfnamefont {P.~C.}\ \bibnamefont
  {Martin}},\ }\href {\doibase 10.1103/PhysRev.124.670} {\bibfield  {journal}
  {\bibinfo  {journal} {Phys. Rev.}\ }\textbf {\bibinfo {volume} {124}},\
  \bibinfo {pages} {670} (\bibinfo {year} {1961})}\BibitemShut {NoStop}%
\bibitem [{\citenamefont {Alm}\ \emph {et~al.}(1993)\citenamefont {Alm},
  \citenamefont {Friman}, \citenamefont {R\"{o}pke},\ and\ \citenamefont
  {Schulz}}]{Alm1993}%
  \BibitemOpen
  \bibfield  {author} {\bibinfo {author} {\bibfnamefont {T.}~\bibnamefont
  {Alm}}, \bibinfo {author} {\bibfnamefont {B.}~\bibnamefont {Friman}},
  \bibinfo {author} {\bibfnamefont {G.}~\bibnamefont {R\"{o}pke}}, \ and\
  \bibinfo {author} {\bibfnamefont {H.}~\bibnamefont {Schulz}},\ }\href
  {\doibase 10.1016/0375-9474(93)90302-E} {\bibfield  {journal} {\bibinfo
  {journal} {Nucl. Phys. A}\ }\textbf {\bibinfo {volume} {551}},\ \bibinfo
  {pages} {45} (\bibinfo {year} {1993})}\BibitemShut {NoStop}%
\bibitem [{\citenamefont {Wambach}\ \emph {et~al.}(1993)\citenamefont
  {Wambach}, \citenamefont {Ainsworth},\ and\ \citenamefont
  {Pines}}]{Wambach1993}%
  \BibitemOpen
  \bibfield  {author} {\bibinfo {author} {\bibfnamefont {J.}~\bibnamefont
  {Wambach}}, \bibinfo {author} {\bibfnamefont {T.}~\bibnamefont {Ainsworth}},
  \ and\ \bibinfo {author} {\bibfnamefont {D.}~\bibnamefont {Pines}},\ }\href
  {\doibase 10.1016/0375-9474(93)90317-Q} {\bibfield  {journal} {\bibinfo
  {journal} {Nucl. Phys. A}\ }\textbf {\bibinfo {volume} {555}},\ \bibinfo
  {pages} {128} (\bibinfo {year} {1993})}\BibitemShut {NoStop}%
\bibitem [{\citenamefont {Schulze}\ \emph {et~al.}(2001)\citenamefont
  {Schulze}, \citenamefont {Polls},\ and\ \citenamefont {Ramos}}]{Schulze2001}%
  \BibitemOpen
  \bibfield  {author} {\bibinfo {author} {\bibfnamefont {H.-J.}\ \bibnamefont
  {Schulze}}, \bibinfo {author} {\bibfnamefont {A.}~\bibnamefont {Polls}}, \
  and\ \bibinfo {author} {\bibfnamefont {A.}~\bibnamefont {Ramos}},\ }\href
  {\doibase 10.1103/PhysRevC.63.044310} {\bibfield  {journal} {\bibinfo
  {journal} {Phys. Rev. C}\ }\textbf {\bibinfo {volume} {63}},\ \bibinfo
  {pages} {044310} (\bibinfo {year} {2001})}\BibitemShut {NoStop}%
\bibitem [{\citenamefont {Shen}\ \emph {et~al.}(2005)\citenamefont {Shen},
  \citenamefont {Lombardo},\ and\ \citenamefont {Schuck}}]{Shen2005}%
  \BibitemOpen
  \bibfield  {author} {\bibinfo {author} {\bibfnamefont {C.}~\bibnamefont
  {Shen}}, \bibinfo {author} {\bibfnamefont {U.}~\bibnamefont {Lombardo}}, \
  and\ \bibinfo {author} {\bibfnamefont {P.}~\bibnamefont {Schuck}},\ }\href
  {\doibase 10.1103/PhysRevC.71.054301} {\bibfield  {journal} {\bibinfo
  {journal} {Phys. Rev. C}\ }\textbf {\bibinfo {volume} {71}},\ \bibinfo
  {pages} {054301} (\bibinfo {year} {2005})}\BibitemShut {NoStop}%
\bibitem [{\citenamefont {Cao}\ \emph {et~al.}(2006)\citenamefont {Cao},
  \citenamefont {Lombardo},\ and\ \citenamefont {Schuck}}]{Cao2006}%
  \BibitemOpen
  \bibfield  {author} {\bibinfo {author} {\bibfnamefont {L.~G.}\ \bibnamefont
  {Cao}}, \bibinfo {author} {\bibfnamefont {U.}~\bibnamefont {Lombardo}}, \
  and\ \bibinfo {author} {\bibfnamefont {P.}~\bibnamefont {Schuck}},\ }\href
  {\doibase 10.1103/PhysRevC.74.064301} {\bibfield  {journal} {\bibinfo
  {journal} {Phys. Rev. C}\ }\textbf {\bibinfo {volume} {74}},\ \bibinfo
  {pages} {064301} (\bibinfo {year} {2006})}\BibitemShut {NoStop}%
\bibitem [{\citenamefont {Dong}\ \emph {et~al.}(2013)\citenamefont {Dong},
  \citenamefont {Lombardo},\ and\ \citenamefont {Zuo}}]{Dong2013}%
  \BibitemOpen
  \bibfield  {author} {\bibinfo {author} {\bibfnamefont {J.~M.}\ \bibnamefont
  {Dong}}, \bibinfo {author} {\bibfnamefont {U.}~\bibnamefont {Lombardo}}, \
  and\ \bibinfo {author} {\bibfnamefont {W.}~\bibnamefont {Zuo}},\ }\href
  {\doibase 10.1103/PhysRevC.87.062801} {\bibfield  {journal} {\bibinfo
  {journal} {Phys. Rev. C}\ }\textbf {\bibinfo {volume} {87}},\ \bibinfo
  {pages} {062801} (\bibinfo {year} {2013})}\BibitemShut {NoStop}%
\bibitem [{\citenamefont {Migdal}(1967)}]{Migdal1967}%
  \BibitemOpen
  \bibfield  {author} {\bibinfo {author} {\bibfnamefont {A.~B.}\ \bibnamefont
  {Migdal}},\ }\href@noop {} {\emph {\bibinfo {title} {{Theory of Finite Fermi
  Systems}}}}\ (\bibinfo  {publisher} {Interscience, New York},\ \bibinfo
  {year} {1967})\BibitemShut {NoStop}%
\bibitem [{\citenamefont {Schwenk}\ \emph {et~al.}(2003)\citenamefont
  {Schwenk}, \citenamefont {Friman},\ and\ \citenamefont
  {Brown}}]{Schwenk2003}%
  \BibitemOpen
  \bibfield  {author} {\bibinfo {author} {\bibfnamefont {A.}~\bibnamefont
  {Schwenk}}, \bibinfo {author} {\bibfnamefont {B.}~\bibnamefont {Friman}}, \
  and\ \bibinfo {author} {\bibfnamefont {G.~E.}\ \bibnamefont {Brown}},\ }\href
  {\doibase 10.1016/S0375-9474(02)01290-3} {\bibfield  {journal} {\bibinfo
  {journal} {Nucl. Phys. A}\ }\textbf {\bibinfo {volume} {713}},\ \bibinfo
  {pages} {191} (\bibinfo {year} {2003})},\ \Eprint
  {http://arxiv.org/abs/0207004} {0207004 [nucl-th]} \BibitemShut {NoStop}%
\bibitem [{\citenamefont {Schwenk}\ and\ \citenamefont
  {Friman}(2004)}]{Schwenk2004}%
  \BibitemOpen
  \bibfield  {author} {\bibinfo {author} {\bibfnamefont {A.}~\bibnamefont
  {Schwenk}}\ and\ \bibinfo {author} {\bibfnamefont {B.}~\bibnamefont
  {Friman}},\ }\href {\doibase 10.1103/PhysRevLett.92.082501} {\bibfield
  {journal} {\bibinfo  {journal} {Phys. Rev. Lett.}\ }\textbf {\bibinfo
  {volume} {92}},\ \bibinfo {pages} {082501} (\bibinfo {year}
  {2004})}\BibitemShut {NoStop}%
\bibitem [{\citenamefont {Frick}\ and\ \citenamefont
  {M\"{u}ther}(2003)}]{Frick2003}%
  \BibitemOpen
  \bibfield  {author} {\bibinfo {author} {\bibfnamefont {T.}~\bibnamefont
  {Frick}}\ and\ \bibinfo {author} {\bibfnamefont {H.}~\bibnamefont
  {M\"{u}ther}},\ }\href {\doibase 10.1103/PhysRevC.68.034310} {\bibfield
  {journal} {\bibinfo  {journal} {Phys. Rev. C}\ }\textbf {\bibinfo {volume}
  {68}},\ \bibinfo {pages} {034310} (\bibinfo {year} {2003})}\BibitemShut
  {NoStop}%
\bibitem [{\citenamefont {Machleidt}\ \emph {et~al.}(1996)\citenamefont
  {Machleidt}, \citenamefont {Sammarruca},\ and\ \citenamefont
  {Song}}]{Machleidt1995}%
  \BibitemOpen
  \bibfield  {author} {\bibinfo {author} {\bibfnamefont {R.}~\bibnamefont
  {Machleidt}}, \bibinfo {author} {\bibfnamefont {F.}~\bibnamefont
  {Sammarruca}}, \ and\ \bibinfo {author} {\bibfnamefont {Y.}~\bibnamefont
  {Song}},\ }\href {\doibase 10.1103/PhysRevC.53.R1483} {\bibfield  {journal}
  {\bibinfo  {journal} {Phys. Rev. C}\ }\textbf {\bibinfo {volume} {53}},\
  \bibinfo {pages} {R1483} (\bibinfo {year} {1996})}\BibitemShut {NoStop}%
\bibitem [{\citenamefont {Wiringa}\ \emph {et~al.}(1995)\citenamefont
  {Wiringa}, \citenamefont {Stoks},\ and\ \citenamefont
  {Schiavilla}}]{Wiringa1995}%
  \BibitemOpen
  \bibfield  {author} {\bibinfo {author} {\bibfnamefont {R.~B.}\ \bibnamefont
  {Wiringa}}, \bibinfo {author} {\bibfnamefont {V.~G.~J.}\ \bibnamefont
  {Stoks}}, \ and\ \bibinfo {author} {\bibfnamefont {R.}~\bibnamefont
  {Schiavilla}},\ }\href {\doibase 10.1103/PhysRevC.51.38} {\bibfield
  {journal} {\bibinfo  {journal} {Phys. Rev. C}\ }\textbf {\bibinfo {volume}
  {51}},\ \bibinfo {pages} {38} (\bibinfo {year} {1995})}\BibitemShut {NoStop}%
\bibitem [{\citenamefont {Entem}\ and\ \citenamefont
  {Machleidt}(2003)}]{Entem2003}%
  \BibitemOpen
  \bibfield  {author} {\bibinfo {author} {\bibfnamefont {D.~R.}\ \bibnamefont
  {Entem}}\ and\ \bibinfo {author} {\bibfnamefont {R.}~\bibnamefont
  {Machleidt}},\ }\href {\doibase 10.1103/PhysRevC.68.041001} {\bibfield
  {journal} {\bibinfo  {journal} {Phys. Rev. C}\ }\textbf {\bibinfo {volume}
  {68}},\ \bibinfo {pages} {041001} (\bibinfo {year} {2003})}\BibitemShut
  {NoStop}%
\bibitem [{\citenamefont {Holt}\ \emph {et~al.}(2010)\citenamefont {Holt},
  \citenamefont {Kaiser},\ and\ \citenamefont {Weise}}]{Holt2010}%
  \BibitemOpen
  \bibfield  {author} {\bibinfo {author} {\bibfnamefont {J.~W.}\ \bibnamefont
  {Holt}}, \bibinfo {author} {\bibfnamefont {N.}~\bibnamefont {Kaiser}}, \ and\
  \bibinfo {author} {\bibfnamefont {W.}~\bibnamefont {Weise}},\ }\href
  {\doibase 10.1103/PhysRevC.81.024002} {\bibfield  {journal} {\bibinfo
  {journal} {Phys. Rev. C}\ }\textbf {\bibinfo {volume} {81}},\ \bibinfo
  {pages} {024002} (\bibinfo {year} {2010})}\BibitemShut {NoStop}%
\bibitem [{\citenamefont {Carbone}\ \emph
  {et~al.}(2013{\natexlab{a}})\citenamefont {Carbone}, \citenamefont
  {Cipollone}, \citenamefont {Barbieri}, \citenamefont {Rios},\ and\
  \citenamefont {Polls}}]{Carbone2013a}%
  \BibitemOpen
  \bibfield  {author} {\bibinfo {author} {\bibfnamefont {A.}~\bibnamefont
  {Carbone}}, \bibinfo {author} {\bibfnamefont {A.}~\bibnamefont {Cipollone}},
  \bibinfo {author} {\bibfnamefont {C.}~\bibnamefont {Barbieri}}, \bibinfo
  {author} {\bibfnamefont {A.}~\bibnamefont {Rios}}, \ and\ \bibinfo {author}
  {\bibfnamefont {A.}~\bibnamefont {Polls}},\ }\href {\doibase
  10.1103/PhysRevC.88.054326} {\bibfield  {journal} {\bibinfo  {journal} {Phys.
  Rev. C}\ }\textbf {\bibinfo {volume} {88}},\ \bibinfo {pages} {054326}
  (\bibinfo {year} {2013}{\natexlab{a}})}\BibitemShut {NoStop}%
\bibitem [{\citenamefont {Carbone}\ \emph
  {et~al.}(2013{\natexlab{b}})\citenamefont {Carbone}, \citenamefont {Polls},\
  and\ \citenamefont {Rios}}]{Carbone2013b}%
  \BibitemOpen
  \bibfield  {author} {\bibinfo {author} {\bibfnamefont {A.}~\bibnamefont
  {Carbone}}, \bibinfo {author} {\bibfnamefont {A.}~\bibnamefont {Polls}}, \
  and\ \bibinfo {author} {\bibfnamefont {A.}~\bibnamefont {Rios}},\ }\href
  {\doibase 10.1103/PhysRevC.88.044302} {\bibfield  {journal} {\bibinfo
  {journal} {Phys. Rev. C}\ }\textbf {\bibinfo {volume} {88}},\ \bibinfo
  {pages} {044302} (\bibinfo {year} {2013}{\natexlab{b}})}\BibitemShut
  {NoStop}%
\bibitem [{\citenamefont {Carbone}\ \emph {et~al.}(2014)\citenamefont
  {Carbone}, \citenamefont {Rios},\ and\ \citenamefont {Polls}}]{Carbone2014}%
  \BibitemOpen
  \bibfield  {author} {\bibinfo {author} {\bibfnamefont {A.}~\bibnamefont
  {Carbone}}, \bibinfo {author} {\bibfnamefont {A.}~\bibnamefont {Rios}}, \
  and\ \bibinfo {author} {\bibfnamefont {A.}~\bibnamefont {Polls}},\ }\href
  {\doibase 10.1103/PhysRevC.90.054322} {\bibfield  {journal} {\bibinfo
  {journal} {Phys. Rev. C}\ }\textbf {\bibinfo {volume} {90}},\ \bibinfo
  {pages} {054322} (\bibinfo {year} {2014})}\BibitemShut {NoStop}%
\bibitem [{\citenamefont {Babu}\ and\ \citenamefont {Brown}(1973)}]{babu1973}%
  \BibitemOpen
  \bibfield  {author} {\bibinfo {author} {\bibfnamefont {S.}~\bibnamefont
  {Babu}}\ and\ \bibinfo {author} {\bibfnamefont {G.~E.}\ \bibnamefont
  {Brown}},\ }\href@noop {} {\bibfield  {journal} {\bibinfo  {journal} {Ann.
  Phys. (NY)}\ }\textbf {\bibinfo {volume} {78}},\ \bibinfo {pages} {1}
  (\bibinfo {year} {1973})}\BibitemShut {NoStop}%
\bibitem [{\citenamefont {Abrikosov}\ \emph {et~al.}(1965)\citenamefont
  {Abrikosov}, \citenamefont {Gorkov},\ and\ \citenamefont
  {Dzyaloshinskii}}]{Abrikosov1965}%
  \BibitemOpen
  \bibfield  {author} {\bibinfo {author} {\bibfnamefont {A.~A.}\ \bibnamefont
  {Abrikosov}}, \bibinfo {author} {\bibfnamefont {L.~P.}\ \bibnamefont
  {Gorkov}}, \ and\ \bibinfo {author} {\bibfnamefont {I.~Y.}\ \bibnamefont
  {Dzyaloshinskii}},\ }\href@noop {} {\emph {\bibinfo {title} {{Quantum Field
  Theoretical Methods in Statistical Physics}}}},\ \bibinfo {edition} {2nd}\
  ed.\ (\bibinfo  {publisher} {Pergamon Press},\ \bibinfo {year}
  {1965})\BibitemShut {NoStop}%
\bibitem [{\citenamefont {Haussmann}\ \emph {et~al.}(2007)\citenamefont
  {Haussmann}, \citenamefont {Rantner}, \citenamefont {Cerrito},\ and\
  \citenamefont {Zwerger}}]{Haussmann2007}%
  \BibitemOpen
  \bibfield  {author} {\bibinfo {author} {\bibfnamefont {R.}~\bibnamefont
  {Haussmann}}, \bibinfo {author} {\bibfnamefont {W.}~\bibnamefont {Rantner}},
  \bibinfo {author} {\bibfnamefont {S.}~\bibnamefont {Cerrito}}, \ and\
  \bibinfo {author} {\bibfnamefont {W.}~\bibnamefont {Zwerger}},\ }\href
  {\doibase 10.1103/PhysRevA.75.023610} {\bibfield  {journal} {\bibinfo
  {journal} {Phys. Rev. A}\ }\textbf {\bibinfo {volume} {75}},\ \bibinfo
  {pages} {023610} (\bibinfo {year} {2007})}\BibitemShut {NoStop}%
\bibitem [{\citenamefont {Chen}\ \emph {et~al.}(2005)\citenamefont {Chen},
  \citenamefont {Stajic}, \citenamefont {Tan},\ and\ \citenamefont
  {Levin}}]{Chen2005}%
  \BibitemOpen
  \bibfield  {author} {\bibinfo {author} {\bibfnamefont {Q.}~\bibnamefont
  {Chen}}, \bibinfo {author} {\bibfnamefont {J.}~\bibnamefont {Stajic}},
  \bibinfo {author} {\bibfnamefont {S.}~\bibnamefont {Tan}}, \ and\ \bibinfo
  {author} {\bibfnamefont {K.}~\bibnamefont {Levin}},\ }\href {\doibase
  10.1016/j.physrep.2005.02.005} {\bibfield  {journal} {\bibinfo  {journal}
  {Phys. Rep.}\ }\textbf {\bibinfo {volume} {412}},\ \bibinfo {pages} {1}
  (\bibinfo {year} {2005})}\BibitemShut {NoStop}%
\bibitem [{\citenamefont {Pines}(1962)}]{Pines1962}%
  \BibitemOpen
  \bibfield  {author} {\bibinfo {author} {\bibfnamefont {D.}~\bibnamefont
  {Pines}},\ }\href@noop {} {\emph {\bibinfo {title} {The many-body problem: a
  lecture note and reprint volume}}}\ (\bibinfo  {publisher} {W. A. Benjamin},\
  \bibinfo {year} {1962})\BibitemShut {NoStop}%
\bibitem [{\citenamefont {Frick}(2004)}]{FrickPhD}%
  \BibitemOpen
  \bibfield  {author} {\bibinfo {author} {\bibfnamefont {T.}~\bibnamefont
  {Frick}},\ }\emph {\bibinfo {title} {Self-consistent Green's Functions in
  Nuclear Matter at Finite Temperature}},\ \href@noop {} {Ph.D. thesis},\
  \bibinfo  {school} {University of T{\"u}bingen} (\bibinfo {year}
  {2004})\BibitemShut {NoStop}%
\bibitem [{\citenamefont {Rios}(2007)}]{RiosPhD}%
  \BibitemOpen
  \bibfield  {author} {\bibinfo {author} {\bibfnamefont {A.}~\bibnamefont
  {Rios}},\ }\emph {\bibinfo {title} {Thermodynamical Properties of Nuclear
  Matter from a Self-Consistent Green's Function Approach,}},\ \href@noop {}
  {Ph.D. thesis},\ \bibinfo  {school} {University of Barcelona} (\bibinfo
  {year} {2007})\BibitemShut {NoStop}%
\bibitem [{\citenamefont {Som\`{a}}\ and\ \citenamefont
  {Bo{\.z}ek}(2008)}]{Soma2008}%
  \BibitemOpen
  \bibfield  {author} {\bibinfo {author} {\bibfnamefont {V.}~\bibnamefont
  {Som\`{a}}}\ and\ \bibinfo {author} {\bibfnamefont {P.}~\bibnamefont
  {Bo{\.z}ek}},\ }\href {\doibase 10.1103/PhysRevC.78.054003} {\bibfield
  {journal} {\bibinfo  {journal} {Phys. Rev. C}\ }\textbf {\bibinfo {volume}
  {78}},\ \bibinfo {pages} {054003} (\bibinfo {year} {2008})}\BibitemShut
  {NoStop}%
\bibitem [{\citenamefont {Baldo}\ and\ \citenamefont
  {Grasso}(2000)}]{Baldo2000}%
  \BibitemOpen
  \bibfield  {author} {\bibinfo {author} {\bibfnamefont {M.}~\bibnamefont
  {Baldo}}\ and\ \bibinfo {author} {\bibfnamefont {A.}~\bibnamefont {Grasso}},\
  }\href {\doibase 10.1016/S0370-2693(00)00684-5} {\bibfield  {journal}
  {\bibinfo  {journal} {Phys. Lett. B}\ }\textbf {\bibinfo {volume} {485}},\
  \bibinfo {pages} {115} (\bibinfo {year} {2000})},\ \Eprint
  {http://arxiv.org/abs/0003039} {0003039} \BibitemShut {NoStop}%
\bibitem [{\citenamefont {Sedrakian}\ \emph {et~al.}(2006)\citenamefont
  {Sedrakian}, \citenamefont {Clark},\ and\ \citenamefont
  {Alford}}]{Sedrakian2006}%
  \BibitemOpen
  \bibfield  {author} {\bibinfo {author} {\bibfnamefont {A.}~\bibnamefont
  {Sedrakian}}, \bibinfo {author} {\bibfnamefont {J.~W.}\ \bibnamefont
  {Clark}}, \ and\ \bibinfo {author} {\bibfnamefont {M.~G.}\ \bibnamefont
  {Alford}},\ }\href@noop {} {\emph {\bibinfo {title} {Pairing in fermionic
  systems: basic concepts and modern applications}}},\ Vol.~\bibinfo {volume}
  {8}\ (\bibinfo  {publisher} {World Scientific},\ \bibinfo {year}
  {2006})\BibitemShut {NoStop}%
\bibitem [{\citenamefont {Dickhoff}\ and\ \citenamefont
  {M{\"u}ther}(1987)}]{dickhoff1987self}%
  \BibitemOpen
  \bibfield  {author} {\bibinfo {author} {\bibfnamefont {W.~H.}\ \bibnamefont
  {Dickhoff}}\ and\ \bibinfo {author} {\bibfnamefont {H.}~\bibnamefont
  {M{\"u}ther}},\ }\href {\doibase 10.1016/0375-9474(87)90133-3} {\bibfield
  {journal} {\bibinfo  {journal} {Nucl. Phys. A}\ }\textbf {\bibinfo {volume}
  {473}},\ \bibinfo {pages} {394} (\bibinfo {year} {1987})}\BibitemShut
  {NoStop}%
\bibitem [{\citenamefont {Friman}\ and\ \citenamefont
  {Dhar}(1979)}]{Friman1979}%
  \BibitemOpen
  \bibfield  {author} {\bibinfo {author} {\bibfnamefont {B.}~\bibnamefont
  {Friman}}\ and\ \bibinfo {author} {\bibfnamefont {A.}~\bibnamefont {Dhar}},\
  }\href {\doibase 10.1016/0370-2693(79)90763-9} {\bibfield  {journal}
  {\bibinfo  {journal} {Physics Letters B}\ }\textbf {\bibinfo {volume} {85}},\
  \bibinfo {pages} {1 } (\bibinfo {year} {1979})}\BibitemShut {NoStop}%
\bibitem [{\citenamefont {Dickhoff}\ \emph
  {et~al.}(1981{\natexlab{a}})\citenamefont {Dickhoff}, \citenamefont
  {Faessler}, \citenamefont {Meyer-Ter-Vehn},\ and\ \citenamefont
  {M\"uther}}]{Dickhoff1981x}%
  \BibitemOpen
  \bibfield  {author} {\bibinfo {author} {\bibfnamefont {W.~H.}\ \bibnamefont
  {Dickhoff}}, \bibinfo {author} {\bibfnamefont {A.}~\bibnamefont {Faessler}},
  \bibinfo {author} {\bibfnamefont {J.}~\bibnamefont {Meyer-Ter-Vehn}}, \ and\
  \bibinfo {author} {\bibfnamefont {H.}~\bibnamefont {M\"uther}},\ }\href
  {\doibase 10.1016/0375-9474(81)90767-3} {\bibfield  {journal} {\bibinfo
  {journal} {Nuclear Physics A}\ }\textbf {\bibinfo {volume} {368}},\ \bibinfo
  {pages} {445 } (\bibinfo {year} {1981}{\natexlab{a}})}\BibitemShut {NoStop}%
\bibitem [{\citenamefont {Dickhoff}\ \emph
  {et~al.}(1981{\natexlab{b}})\citenamefont {Dickhoff}, \citenamefont
  {Faessler}, \citenamefont {{Meyer-ter-Vehn}},\ and\ \citenamefont
  {M\"uther}}]{Dickhoff1981}%
  \BibitemOpen
  \bibfield  {author} {\bibinfo {author} {\bibfnamefont {W.~H.}\ \bibnamefont
  {Dickhoff}}, \bibinfo {author} {\bibfnamefont {A.}~\bibnamefont {Faessler}},
  \bibinfo {author} {\bibfnamefont {J.}~\bibnamefont {{Meyer-ter-Vehn}}}, \
  and\ \bibinfo {author} {\bibfnamefont {H.}~\bibnamefont {M\"uther}},\ }\href
  {\doibase 10.1103/PhysRevC.23.1154} {\bibfield  {journal} {\bibinfo
  {journal} {Phys. Rev. C}\ }\textbf {\bibinfo {volume} {23}},\ \bibinfo
  {pages} {1154} (\bibinfo {year} {1981}{\natexlab{b}})}\BibitemShut {NoStop}%
\bibitem [{\citenamefont {Pankratov}\ \emph {et~al.}(2015)\citenamefont
  {Pankratov}, \citenamefont {Baldo},\ and\ \citenamefont
  {Saperstein}}]{Pankratov2015}%
  \BibitemOpen
  \bibfield  {author} {\bibinfo {author} {\bibfnamefont {S.~S.}\ \bibnamefont
  {Pankratov}}, \bibinfo {author} {\bibfnamefont {M.}~\bibnamefont {Baldo}}, \
  and\ \bibinfo {author} {\bibfnamefont {E.~E.}\ \bibnamefont {Saperstein}},\
  }\href {\doibase 10.1103/PhysRevC.91.015802} {\bibfield  {journal} {\bibinfo
  {journal} {Phys. Rev. C}\ }\textbf {\bibinfo {volume} {91}},\ \bibinfo
  {pages} {015802} (\bibinfo {year} {2015})}\BibitemShut {NoStop}%
\bibitem [{\citenamefont {Sedrakian}(2003)}]{Sedrakian2003}%
  \BibitemOpen
  \bibfield  {author} {\bibinfo {author} {\bibfnamefont {A.}~\bibnamefont
  {Sedrakian}},\ }\href {\doibase 10.1103/PhysRevC.68.065805} {\bibfield
  {journal} {\bibinfo  {journal} {Phys. Rev. C}\ }\textbf {\bibinfo {volume}
  {68}},\ \bibinfo {pages} {065805} (\bibinfo {year} {2003})}\BibitemShut
  {NoStop}%
\bibitem [{\citenamefont {Margueron}\ \emph {et~al.}(2008)\citenamefont
  {Margueron}, \citenamefont {Sagawa},\ and\ \citenamefont
  {Hagino}}]{Margueron2008}%
  \BibitemOpen
  \bibfield  {author} {\bibinfo {author} {\bibfnamefont {J.}~\bibnamefont
  {Margueron}}, \bibinfo {author} {\bibfnamefont {H.}~\bibnamefont {Sagawa}}, \
  and\ \bibinfo {author} {\bibfnamefont {K.}~\bibnamefont {Hagino}},\ }\href
  {\doibase 10.1103/PhysRevC.77.054309} {\bibfield  {journal} {\bibinfo
  {journal} {Phys. Rev. C}\ }\textbf {\bibinfo {volume} {77}},\ \bibinfo
  {pages} {054309} (\bibinfo {year} {2008})}\BibitemShut {NoStop}%
\bibitem [{\citenamefont {De~Blasio}\ \emph {et~al.}(1997)\citenamefont
  {De~Blasio}, \citenamefont {Hjorth-Jensen}, \citenamefont {Elgar\o{}y},
  \citenamefont {Engvik}, \citenamefont {Lazzari}, \citenamefont {Baldo},\ and\
  \citenamefont {Schulze}}]{deBlasio1997}%
  \BibitemOpen
  \bibfield  {author} {\bibinfo {author} {\bibfnamefont {F.~V.}\ \bibnamefont
  {De~Blasio}}, \bibinfo {author} {\bibfnamefont {M.}~\bibnamefont
  {Hjorth-Jensen}}, \bibinfo {author} {\bibfnamefont {O.}~\bibnamefont
  {Elgar\o{}y}}, \bibinfo {author} {\bibfnamefont {L.}~\bibnamefont {Engvik}},
  \bibinfo {author} {\bibfnamefont {G.}~\bibnamefont {Lazzari}}, \bibinfo
  {author} {\bibfnamefont {M.}~\bibnamefont {Baldo}}, \ and\ \bibinfo {author}
  {\bibfnamefont {H.-J.}\ \bibnamefont {Schulze}},\ }\href {\doibase
  10.1103/PhysRevC.56.2332} {\bibfield  {journal} {\bibinfo  {journal} {Phys.
  Rev. C}\ }\textbf {\bibinfo {volume} {56}},\ \bibinfo {pages} {2332}
  (\bibinfo {year} {1997})}\BibitemShut {NoStop}%
\bibitem [{\citenamefont {Lombardo}\ \emph {et~al.}(2001)\citenamefont
  {Lombardo}, \citenamefont {Nozi{\`{e}}res}, \citenamefont {Schuck},
  \citenamefont {Schulze},\ and\ \citenamefont {Sedrakian}}]{Lombardo2001}%
  \BibitemOpen
  \bibfield  {author} {\bibinfo {author} {\bibfnamefont {U.}~\bibnamefont
  {Lombardo}}, \bibinfo {author} {\bibfnamefont {P.}~\bibnamefont
  {Nozi{\`{e}}res}}, \bibinfo {author} {\bibfnamefont {P.}~\bibnamefont
  {Schuck}}, \bibinfo {author} {\bibfnamefont {H.-J.}\ \bibnamefont {Schulze}},
  \ and\ \bibinfo {author} {\bibfnamefont {A.}~\bibnamefont {Sedrakian}},\
  }\href {\doibase 10.1103/PhysRevC.64.064314} {\bibfield  {journal} {\bibinfo
  {journal} {Phys. Rev. C}\ }\textbf {\bibinfo {volume} {64}},\ \bibinfo
  {pages} {064314} (\bibinfo {year} {2001})}\BibitemShut {NoStop}%
\bibitem [{\citenamefont {Stein}\ \emph {et~al.}(2014)\citenamefont {Stein},
  \citenamefont {Sedrakian}, \citenamefont {Huang},\ and\ \citenamefont
  {Clark}}]{Stein2014}%
  \BibitemOpen
  \bibfield  {author} {\bibinfo {author} {\bibfnamefont {M.}~\bibnamefont
  {Stein}}, \bibinfo {author} {\bibfnamefont {A.}~\bibnamefont {Sedrakian}},
  \bibinfo {author} {\bibfnamefont {X.-G.}\ \bibnamefont {Huang}}, \ and\
  \bibinfo {author} {\bibfnamefont {J.-W.}\ \bibnamefont {Clark}},\ }\href
  {\doibase 10.1103/PhysRevC.90.065804} {\bibfield  {journal} {\bibinfo
  {journal} {Phys. Rev. C}\ }\textbf {\bibinfo {volume} {90}},\ \bibinfo
  {pages} {065804} (\bibinfo {year} {2014})}\BibitemShut {NoStop}%
\bibitem [{\citenamefont {Ashcroft}\ and\ \citenamefont
  {Mermin}(1976)}]{Ashcroft1976}%
  \BibitemOpen
  \bibfield  {author} {\bibinfo {author} {\bibfnamefont {N.~W.}\ \bibnamefont
  {Ashcroft}}\ and\ \bibinfo {author} {\bibfnamefont {N.~D.}\ \bibnamefont
  {Mermin}},\ }\href {\doibase 10.1016/0038-1101(66)90069-4} {\emph {\bibinfo
  {title} {Solid State Physics}}},\ edited by\ \bibinfo {editor} {\bibfnamefont
  {F.}~\bibnamefont {Seitz}}\ and\ \bibinfo {editor} {\bibfnamefont
  {D.}~\bibnamefont {Turnbull}}\ (\bibinfo  {publisher} {Brooks Cole},\
  \bibinfo {year} {1976})\ p.\ \bibinfo {pages} {848}\BibitemShut {NoStop}%
\bibitem [{\citenamefont {Luttinger}(1961)}]{Luttinger1961}%
  \BibitemOpen
  \bibfield  {author} {\bibinfo {author} {\bibfnamefont {J.~M.}\ \bibnamefont
  {Luttinger}},\ }\href@noop {} {\bibfield  {journal} {\bibinfo  {journal}
  {Phys. Rev.}\ }\textbf {\bibinfo {volume} {121}},\ \bibinfo {pages} {942}
  (\bibinfo {year} {1961})}\BibitemShut {NoStop}%
\end{thebibliography}%

\end{document}